\documentclass[reprint,aps,amsmath,amssymb,prx, superscriptaddress,nofootinbib]{revtex4-2}

\usepackage{graphicx}
\usepackage{dcolumn}
\usepackage{bm}
\usepackage{xcolor}
\usepackage[normalem]{ulem}
\usepackage[USenglish]{babel}
\usepackage{float}
\usepackage{times}
\usepackage{physics}
\usepackage[colorlinks]{hyperref}
\usepackage[sep=5pt, offset=0.8em]{simpler-wick}
\usepackage{comment}
\usepackage{amssymb}
\usepackage{amsthm}
\usepackage{nccmath}
\usepackage{mathtools}
\usepackage{lipsum}
\usepackage{subcaption}
\usepackage[newcommands]{ragged2e}
\usepackage{caption}
\usepackage{tikz}
\usepackage{tikz-feynman}
\tikzfeynmanset{compat=1.0.0}
\usepackage{multirow}
\usetikzlibrary{decorations.pathreplacing,decorations.markings,shapes.misc,arrows.meta}
\tikzset{
	on each segment/.style={
		decorate,
		decoration={
			show path construction,
			moveto code={},
			lineto code={
				\path [#1]
				(\tikzinputsegmentfirst) -- (\tikzinputsegmentlast);
			},
			curveto code={
				\path [#1] (\tikzinputsegmentfirst)
				.. controls
				(\tikzinputsegmentsupporta) and (\tikzinputsegmentsupportb)
				..
				(\tikzinputsegmentlast);
			},
			closepath code={
				\path [#1]
				(\tikzinputsegmentfirst) -- (\tikzinputsegmentlast);
			},
		},
	},
	mid arrow/.style={postaction={decorate,decoration={
				markings,
				mark=at position .5 with {\arrow[#1]{stealth}}
	}}},
}

\definecolor{mthca_blue}{rgb}{0.368417, 0.506779, 0.709798}
\definecolor{mthca_orange}{rgb}{0.880722, 0.611041, 0.142051}
\definecolor{mthca_green}{rgb}{0.560181, 0.691569, 0.194885}

\newcommand{\cornerlineL}[3]{
  \par\noindent
  \hbox to \linewidth{%
    \vrule width #1 height #2 depth #3%
    \leaders\hrule height #1 depth 0pt\hfill
  }%
  \par
}

\newcommand{\cornerlineR}[3]{
  \par\noindent
  \hbox to \linewidth{%
    \leaders\hrule height #1 depth 0pt\hfill
    \vrule width #1 height #2 depth #3%
  }%
  \par
}

\usepackage{dashbox}

\makeatletter
\newcommand{\definelanguagealias}[2]{%
	\@namedef{languagealias@#1}{#2}%
}
\makeatother

\definelanguagealias{en}{english}
\definelanguagealias{EN}{english}

\newcommand{\Eq}[1]{Eq.\,(\ref{#1})}
\newcommand{\Sec}[1]{Section~\ref{#1}}
\newcommand{\Fig}[1]{Fig.\,\ref{#1}}

\newcommand{\OMBC}{\omega_0}

\newcommand\redsout{\bgroup\markoverwith{\textcolor{red}{\rule[0.5ex]{2pt}{0.8pt}}}\ULon}

\begin{document}
	


\title[]{Model Order Reduction for Open Quantum Systems Based on Measurement-adapted Time-coarse Graining}

\makeatletter

\date{\today}

\author{Wentao Fan}
\author{Hakan E. T\"{u}reci}
\affiliation{Department of Electrical and Computer Engineering, Princeton University, Princeton, NJ 08544, USA}
\begin{abstract}
Model order reduction encompasses mathematical techniques aimed at reducing the complexity of mathematical models in simulations while retaining the essential characteristics and behaviors of the original model. This is particularly useful in the context of large-scale dynamical systems, which can be computationally expensive to analyze and simulate. Here, we present a model order reduction technique to reduce the time complexity of open quantum systems, grounded in the principle of measurement-adapted coarse-graining. This method, governed by a coarse-graining time scale $\tau$ and the spectral band center $\omega_0$ of the measurement channel, organizes corrections to the lowest-order model which aligns with the RWA Hamiltonian in certain limits, and rigorously justifies the resulting effective quantum master equation (EQME). The focus on calculating to a high degree of accuracy only what can be resolved by the measurement introduces a principled regularization procedure to address singularities and generates low-stiffness models suitable for efficient long-time integration. The closed-form EQME parameters greatly enhance interpretability, while the predicted EQME structure provides a foundation for deriving stochastic master equations under continuous measurement and for constructing efficient graybox models for system identification and control. As a demonstration, we derive the fourth-order EQME for a challenging problem related to the dynamics of a superconducting qubit under high-power dispersive readout in the presence of a continuum of dissipative waveguide modes. This derivation shows that the lowest-order terms align with previous results, while higher-order corrections suggest new phenomena.

\end{abstract}

\maketitle


\section{Introduction}

When driven by oscillatory sources, non-linear quantum systems can exhibit dynamics at multiple time-scales. In an effort to ease the requirements on hardware, recent experimental efforts at building a quantum computer exploit this fact~\cite{Leghtas_2015,Reagor_2018,Villiers_2024,He2023} to engineer desired interactions between qubits by driving the associated parametric interactions through external oscillatory sources. The fidelity of quantum operations is often limited by the purity of these interactions, hence a general and systematic approach to deriving effective quantum models that describe the unitary and non-unitary processes alike is needed and has been the subject of recent studies~\cite{Bukov_2016,MPT_I,MPT_II,Hanai_McDonald_Clerk,Devoret_Kamiltonian}. In thinking about a systematic approach, one has several existing theoretical approaches at hand to choose from which in one way or another attempt to generalize and systematize corrections to the well-known rotating wave approximation (RWA) through the choice of an appropriately chosen set of unitary transformations, which we discuss in some detail below. 
However none of those approaches, to the extent analyzed by the authors, have sufficient generality and flexibility to organize the resulting series of effective corrections in a way that it simultaneously (i) delivers in equal measure all unitary and non-unitary corrections so that the truncation of the series is physically grounded, (ii) is by construction not subject to any divergences in the parameters of the series, and (iii) can be of direct practical utility in designing and analyzing experimental data obtained through a measurement apparatus with a finite bandwidth. 

Here we take the point of view that a physically grounded approach to the derivation of effective models should be informed by the available measurement processes, and explore its full implications to the dynamics of open quantum systems by treating the coherent and the driven-dissipative dynamics on the same footing. We acknowledge that the quantum systems one prepares and measures in experiments can never be fully isolated, which necessitates any measurement-adapted modeling of the system to account for its openness and the resulting dissipative dynamics. However, the coupling between the quantum system and the measurement channel cannot be arbitrarily strong either, which implies that there can never be ideal projective measurements taking place at any instant of time. In fact, the duration of the measurement process can be comparable to, or even exceed, certain internal system timescales -- this is typically the case in circuit QED systems~\cite{Blais_cQED_review}. As a result, fast system dynamics are inherently coarse-grained over the measurement timescale, leading to fluctuations/loss of information and coherence.
It is important to clarify that the results are broadly applicable and not limited to continuously monitored systems -- one only needs to assume that the system will \emph{ultimately} be measured with some finite time resolution. However, MaTCG does provide a natural and self-consistent prescription for deriving effective stochastic master equations in the presence of continuous measurements with finite time resolution, as discussed in \Sec{Subsec: SME}, \Sec{Subsec: spin-cavity SME}. We illustrate this point with a model for the high-power readout dynamics of a superconducting qubit. In this context, we note that the TCG method has also been applied to closed quantum systems with finite degrees of freedom~\cite{Leon_Wentao_STCG}, where the nature of the measurement process is unspecified.

\begin{figure}[h!]
\centering
\captionsetup{justification=Justified, font=footnotesize}
\includegraphics[width=0.95\textwidth]{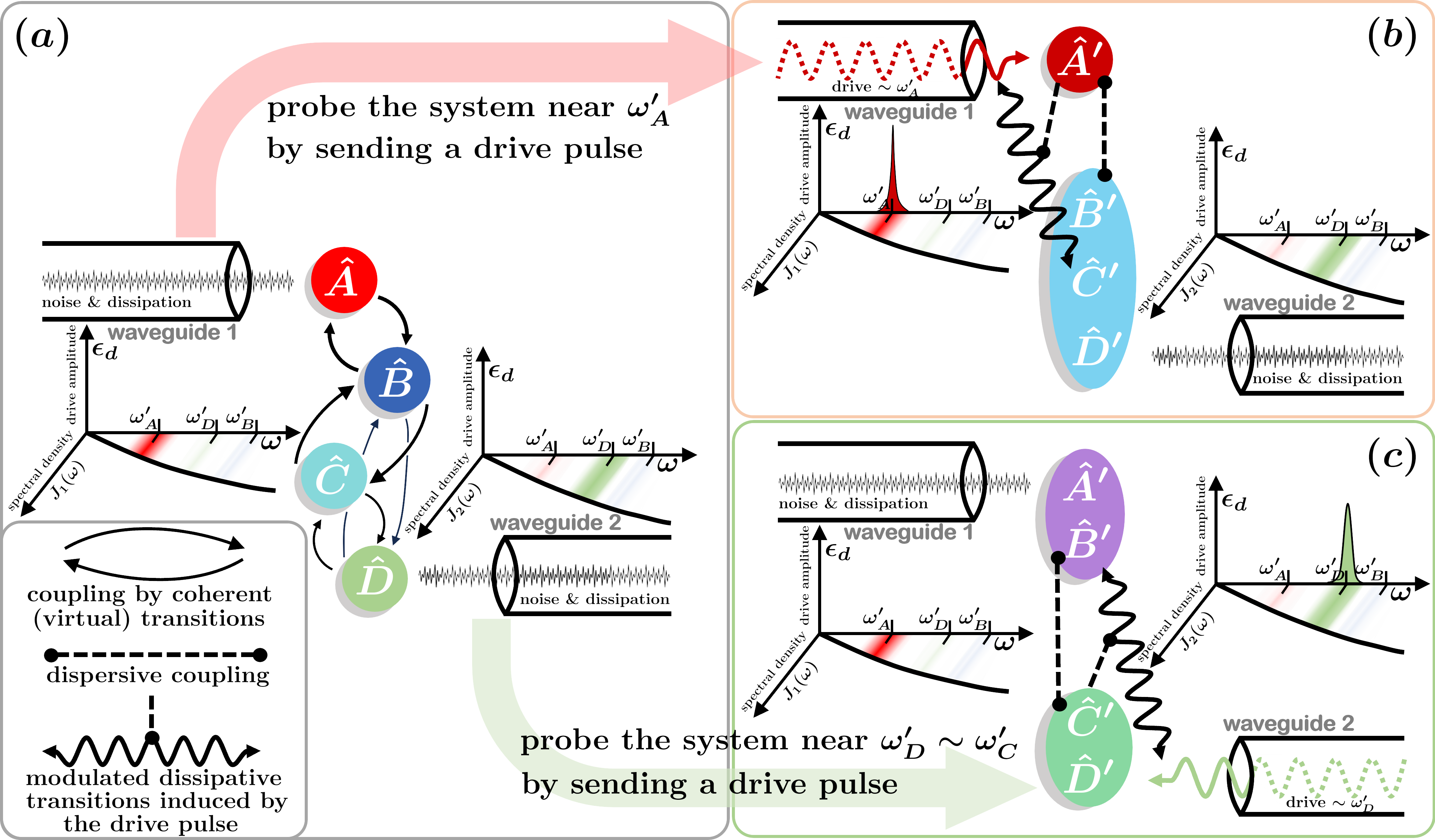}
\caption{A schematic showing measurement-adapted model reduction giving rise to different effective models depending on the measurement channel. During the readout of a quantum system represented by the coherently interacting modes in (a), one can send in drive pulses near frequency $\omega_{A}^{\prime}$ of the hybridized mode $\hat{A}^{\prime}$ or $\omega_{D}^{\prime}$ of mode $\hat{D}^{\prime}$, where the prime symbols indicate mode hybridization due to the couplings. Since readout pulses are sent in through channels which also function as sources of dissipation (here considered as bosonic modes in two waveguides), the spectral density and the drive amplitude are plotted together as functions of the mode frequency $\omega$ for each waveguide. Depending on choices of the readout channel and time resolution $\tau$, different effective models with reduced complexity can be obtained, as represented by (b) and (c). Considering the situation where differences between $\omega_{A}^{\prime}$, $\omega_{B}^{\prime}$, and $\omega_{D}^{\prime} \sim \omega_{C}^{\prime}$ are large in comparison with $\tau^{-1}$, one finds that the relevant degrees of freedom as well as their effective couplings and dissipation can be very different in the two different effective models. In particular, unresolvable (virtual) transitions are replaced by dispersive couplings, with the drive inducing dissipative transitions in some of the modes whose rates are modulated by the state of other parts of the system. Remarkably, these modulated dissipative transitions are different from direct or Purcell dissipation into the waveguide modes (which are denoted by the highlighted bands on the functions $J_{1,2}(\omega)$ in the schematic above) since they do not directly probe the corresponding spectral density; instead, they account for non-Markovian dissipative effects which, in the case of panel (b) for example, can manifest as dissipators of the form $\epsilon_{d} \langle 
f(\hat{A}^{\prime}) \rangle D_{g(\hat{B}^{\prime}, \hat{C}^{\prime}, \hat{D}^{\prime})}$ for some functions $f$ and $g$, where the corresponding rates are dependent on not only the drive amplitude $\epsilon_{d}$ but also the time-dependent expectation value $\langle 
f(\hat{A}^{\prime}) \rangle$ which depends on the entire dynamical history.}
\label{fig: MaTCG schematic}
\end{figure}

A physically grounded approach to deriving effective models must account for this inevitable information fluctuation/loss due to coarse-graining, and consider a non-unitary time evolution informed by the measurement process in order to accurately account for the information extractable from the system. In fact, the resulting dissipation and decoherence in such measurement-adapted effective models will in general depend on one's choice of the measurement channel, as illustrated by the schematic in Fig.\ref{fig: MaTCG schematic}.

Motivated by these observations, we introduce and analyze the measurement-adapted time-coarse graining (MaTCG) framework, a general scheme for deriving the effective dynamics of an open quantum system that can be resolved through a finite-bandwidth measurement chain defined by a certain band center $\OMBC$ and bandwidth $1/\tau$, with $\tau$ representing the coarse-grained information transfer through the measurement chain characterized by an overall response time scale $\tau$. The expansion of the resulting effective Liouvillian $\mathcal{L}$ governing the coarse-grained dynamics is then organized through these two time(energy) scales, with the band center $\OMBC$ defining an appropriate interaction picture (i.e., rotating frame), and the response time $\tau$ functioning as an additional free parameter characterizing the fundamental time resolution of the observable dynamics in that interaction picture. The $\tau\rightarrow 0$ limit corresponds to the microscopic dynamics defined by the von-Neumann equation, whereas finite values of $\tau$ correspond to the dynamics of the density matrix coarse-grained at the time scale $\tau$. This paper is devoted to exploring the implications of the general principle of measurement-adapted coarse-graining based on a minimal physical description of the measurement process, although the method can be straightforwardly generalized to more sophisticated measurement protocols by formulating the measurement chain in terms of carefully modeled filter functions.

\onecolumngrid
\vspace{8mm}

\vspace{2mm}
\twocolumngrid

From this perspective, the method's emphasis on measurable quantities inevitably requires a precise specification of the measurement process. While abstracting the measurement channel by the corresponding filter function is sufficient for exploring the general principle, assessing its practical utility demands a specific physical system and an extensively studied measurement setup. The superconducting qubit readout problem~\cite{Blais2004_PhysRevA, Wallraff2004, Blais_cQED_review} meets these criteria well. The readout problem has well-established theoretical foundations~\cite{Blais_etal_dispersive_cQED, Sank2016, MPT_I, MPT_II, Blais_etal_transmon_ionization, Hanai_McDonald_Clerk, Blais_etal_dispersive_cQED,Lescanne2019}, is of great practical importance for quantum computing~\cite{Blais_cQED_review, Heinsoo2018, Chen2023}, and has recently garnered increased interest in characterizing the readout dynamics under high-power readout pulses in an effort to increase the speed and accuracy of readout~\cite{Ginossar2010, Reed2010, Yeo2023}. We therefore consider the readout problem for the purpose of benchmarking the MaTCG method developed in this work. More specifically, we analyze the readout dynamics under a wide range of conditions, compare to existing results in the literature, and find analytical expressions for higher order processes that have not been captured before. 

MaTCG draws on the solution of a set of technical problems one encounters in describing the observable dynamics filtered through a measurement apparatus with a finite response time. Ideally, when the system is observed through a certain channel with finite time resolution $\tau$, the effective model should consist only of degrees of freedom that appear to be evolving slowly through that channel in comparison with the observation time scale $\tau$, while the secular effects of high-frequency (virtual) transitions are absorbed into the coefficients of a limited set of bandpass-filtered superoperators. This conceptual procedure is reminiscent of Wilsonian renormalization~\cite{Wilson1971_1, Wilson1971_2,Wilson_1983}, although the flow equation method~\cite{Wilson_flowEq, WEGNER2000141,Kehrein_flowEq} would be a more appropriate analogy to MaTCG since both involve the bandpass filtering of direct transitions without truncating the Hilbert space at any step. The Rotating Wave Approximation (RWA) along with its generalizations~\cite{Scully_Zubairy_1997, Zhang2015RWA}, as a venerable first line of attack in quantum optics, misses the impact of the high-frequency processes on the filtered, slow dynamics, whereas more sophisticated methods such as the Schrieffer-Wolff transformation and adiabatic elimination can be very laborious due to the lack of closed-form expressions at each order, and may encounter singularities in the presence of a resonant drive~\cite{MPT_I, MPT_II}. 

In order to tackle these difficulties, we adopt the time-coarse graining (TCG) approach which automatically filters out the fast dynamics in the observation channel with the time resolution of our choice. The TCG method has been introduced in the literature as an improved version of the RWA for multi-level atomic systems~\cite{Gamel_and_James,Lee_Noh_and_Kim}. However, without measurement-informed prescription for the choice of the interaction picture or the coarse-graining time scale, it is difficult to relate such approximation schemes to effective theories in experimentally relevant models. Furthermore, without closed-form expressions for the effective superoperators at each order, the difficulty of solving the noncommutative recurrence equation prohibits efficient calculation beyond the second order, and hence greatly limits the applicability of the method to open multi-oscillator systems subject to strong drive.

In this paper, we overcome these technical problems by systematically deriving the measurement-adapted time-coarse graining (MaTCG) method with an explicit formula for the TCG superoperators at arbitrary orders. In addition, we also develop a diagrammatic representation for each term in those superoperators, which is not only a convenient tool for visualizing the various corrections from MaTCG, but also a helpful conceptual technique for predicting the types of possible effective superoperators that can arise from a particular microscopic Hamiltonian. In the same way that Feynman diagrams can be considered as decomposition of the Dyson series into distinct (virtual) processes, each TCG diagram also represents a particular combination of (virtual) processes contributing to a superoperator in the effective quantum master equation (EQME). The resulting EQME has two added benefits: (i) it provides an interpretable series expansion for physical processes, ordered by their relevance to experimental preparation and measurement, and (ii) it is much less stiff in numerical simulations compared to the original microscopic von-Neumann equation, and can be integrated with considerably less computational resource. One important technical problem that we address here is that the coarse-graining procedure is applied starting with the original many-body system-bath Hamiltonian to derive an EQME that is informed by non-linear mixing processes that down- or up-convert excitations from the electromagnetic environment of the oscillator system. This stands in contrast to earlier studies that have analyzed coarse-graining directly on a Lindbladian~\cite{Lee_Noh_and_Kim}.

The sections of this paper are organized as follows: \Sec{Sec: TCG perturbation theory} introduces the TCG perturbation theory where we present a closed-form formula for the TCG master equation at each order assuming that the interaction-picture Hamiltonian admits a Fourier expansion in the frequency domain; we also propose a diagrammatic representation for each superoperator coefficient in the master equation. \Sec{Sec: TCG open quantum systems} discusses the time-coarse grained dynamics of open quantum systems in the system+bath formalism, and shows how a reduced Markovian master equation for the system can be derived from the EQME of the system+bath density matrix. We show how TCG provides natural justification for the Markov and secular approximations in the derivation of the Lindblad master equation (LME), and extend the LME to account for finite time resolutions. \Sec{Sec: Spin_Cavity} investigates a toy model of the dispersive readout of a spin with a linear cavity mode, and discusses in detail the effective corrections and emergent dynamics obtained at each order in the corresponding EQME. In particular we find measurement-induced spin energy jumps which have not been captured by any effective Hamiltonian methods in the literature. \Sec{Sec: the readout problem} applies the MaTCG to an experimentally-relevant model of the transmon readout problem, where we demonstrate the emergence of drive-induced dissipation of the artificial atom, along with other renormalization effects from MaTCG. Finally, we summarize our results in \Sec{Sec: Summary} and propose some directions for future studies. Details of the mathematical derivations and some lengthy formulas can be found in the Appendices.
	
\section{Dynamics of the time-coarse grained density matrix}
\label{Sec: TCG perturbation theory}

We first briefly review the notion of the time-coarse grained density matrix $\overline{\rho}(t)$ of a closed quantum system. It is helpful for the purposes of this section to think of the density matrix as a book-keeping device for the statistical information available about a system that in the limit of infinite time-resolution is described by $\rho(t)$. The time-evolution of $\rho(t)$ is described by the von Neumann equation
\begin{equation}
\partial_{t} \rho(t)
=
-i\big[ H_{I}(t), \rho(t) \big].
\end{equation}
where we like to think that there exists a Hamiltonian generator $H_{I}(t)$ that describes the perfectly resolved system dynamics in an interaction picture that we shall specify later. By considering time-coarse graining, we recognize the impossibility of the infinite time resolution of the dynamics of observables through a measurement. The finite time-resolution dynamics we actually observe can be described by the time-coarse grained density matrix $\overline{\rho}(t)$ in a certain interaction picture determined by the kind of measurement we perform. In order to describe the slow dynamics observed by realistic instruments, we need an effective dynamical equation for the time-coarse grained density matrix $\overline{\rho}(t)$ defined as
\begin{equation}
\overline{\rho}(t)
:=
\int_{-\infty}^{\infty} dt^{\prime} \cdot f(t^{\prime};\tau) \rho(t-t^{\prime})
\end{equation}
where $f(t^{\prime};\tau)$ is some window function centered around $t^{\prime}=0$ with the parameter $\tau$ indicating its width and providing a coarse-graining time scale. To ensure unit trace for $\overline{\rho}(t)$, we require that $\int_{-\infty}^{\infty}dt^{\prime}\cdot f(t^{\prime};\tau)=1$. We would also use
\[
\overline{O}(t):=\int_{-\infty}^{\infty} dt^{\prime} \cdot f(t^{\prime};\tau) O(t-t^{\prime})
\]
to denote the time average of any operator with respect to the window function $f(t;\tau)$.
	
Since the von Neumann equation and the coarse-graining operation $\rho(t) \mapsto \overline{\rho}(t)$ are both linear in $\rho(t)$, there also exists a linear dynamical equation for $\overline{\rho}(t)$ of the form
\begin{equation}
\label{TCG ME}
\partial_{t}\overline{\rho}(t)
=
\mathcal{L}(t) \overline{\rho}(t)
\end{equation}
which is local in time, with $\mathcal{L}$ being the TCG superoperator that describes the time-coarse grained evolution of $\overline{\rho}(t)$. We will refer to Eq.(\ref{TCG ME}) as the TCG master equation. Although it is difficult to obtain the exact TCG master equation explicitly in general, one may find a perturbative expansion of Eq.(\ref{TCG ME}) if the interaction Hamiltonian $H_{I}(t)$ can be considered as controlled by some small parameters. In most cases, the small parameters are either the ratios between the coupling strengths and the corresponding transition frequencies or the products of the coupling strengths and the coarse-graining time scale.
	
To perform the perturbative expansion, we expand the Liouvillian superoperator $\mathcal{L}(t)$ in powers of the Hamiltonian $H_{I}$:
\begin{equation}
\begin{split}
\mathcal{L}(t) = \sum_{k=1}^{\infty} \mathcal{L}_{k}(t).
\end{split}
\end{equation}
In the existing literature, closed-form formula for $\mathcal{L}_{k}$ has been recursively derived for $k = 1, 2, 3$, with explicit calculations done only up to the second order ($k=2$) due to the complexity of the third-order formula \cite{Gamel_and_James, Lee_Noh_and_Kim}. In this work, we report a closed-form formula for $\mathcal{L}_{k}$ at arbitrary order, and associate each term in the general formula with a diagrammatic representation. In fact, if the interaction-picture Hamiltonian can be written as $H_{I}(t) = \sum_{j} e^{i \omega_{j} t} h_{j}$, then one has the following general form for $\mathcal{L}_{k}$:

\onecolumngrid

\begin{equation}
\label{Eq: Lk}
\begin{split}
\boxed{
\mathcal{L}_{k}(t) \overline{\rho}
=
-
i \sum_{k_{1}=1}^{k} \sum_{ n_{1} \cdots n_{k} }
C_{k_{1}, k-k_{1}}\big( \omega_{n_{1}}, \omega_{n_{2}} \cdots, \omega_{n_{k}} \big)
e^{i ( \omega_{n_{1}}+\cdots+\omega_{n_{k}} ) t}
h_{n_{1}} \cdots h_{n_{k_{1}}} \overline{\rho} h_{n_{k_{1}+1}} \cdots h_{n_{k}}
-
h.c.
}
\end{split}
\end{equation}

\twocolumngrid
\noindent
where the numerical coefficient in Eq.(\ref{Eq: Lk}) is given by the function $C_{k_{1}, k-k_{1}}\big( \omega_{n_{1}}, \omega_{n_{2}} \cdots, \omega_{n_{k}} \big)$ of $k$ frequencies, which receives contribution from all possible loop diagrams containing $k_{1}$ left frequencies and $k-k_{1}$ right frequencies, as introduced with more details in Appendix B of the Supplementary Materials\cite{supplement}. To write down an explicit formula for the function $C_{l, r}$, we define $\tilde{f}(\omega)$ to be the Fourier transform of the time-coarse graining window function: $\tilde{f}(\omega) := \int_{-\infty}^{\infty} dt f(t) e^{-i \omega t}$ (typically, $\tilde{f}(\omega)$ is a low-pass filter, and $\tilde{f}(\omega) = e^{-\frac{\omega^{2}\tau^{2}}{2}}$ for a gaussian window of half width $\tau$), and define the factorial of a vector $\boldsymbol{\mu} = (\omega_{1}, \omega_{2}, \cdots, \omega_{n})$ to be
\begin{equation*}
\begin{split}
\boldsymbol{\mu} !
:=
(\omega_{1}+\omega_{2}+\cdots+\omega_{n})(\omega_{1}+\omega_{2}+\cdots+\omega_{n-1})\cdots\omega_{1}.
\end{split}
\end{equation*}
Then, as derived in Appendix B of the Supplementary Materials~\cite{supplement}, the following formula can be obtained for $C_{l, r}$ where $\norm{d}$ is the number of loops in the diagram while $\boldsymbol{\mu}_{i}$ and $\boldsymbol{\nu}_{i}$ are the ordered left, and right frequencies in $i$-th loop respectively:

\onecolumngrid

\begin{equation}
\label{Eq: contraction coefficient}
\begin{split}
\boxed{
C_{l,r}\big( \omega_{1}, \omega_{2}, \cdots, \omega_{l+r} \big)
=
\sum_{d\in \textrm{diagrams}}
( -1 )^{l+\norm{d}}
\frac{ \tilde{f}\big( \omega_{1} + \sum \boldsymbol{\mu}_{1} + \sum \boldsymbol{\nu}_{1} \big) \tilde{f}\big( \sum \boldsymbol{\mu}_{2} + \sum \boldsymbol{\nu}_{2} \big) \cdots \tilde{f}\big( \sum \boldsymbol{\mu}_{\norm{d}} + \sum \boldsymbol{\nu}_{\norm{d}} \big) }{ \boldsymbol{\mu}_{1} ! \boldsymbol{\nu}_{1} ! \boldsymbol{\mu}_{2} ! \boldsymbol{\nu}_{2} ! \cdots \boldsymbol{\mu}_{\norm{d}} ! \boldsymbol{\nu}_{\norm{d}} ! }
}
\end{split}
\end{equation}

\twocolumngrid

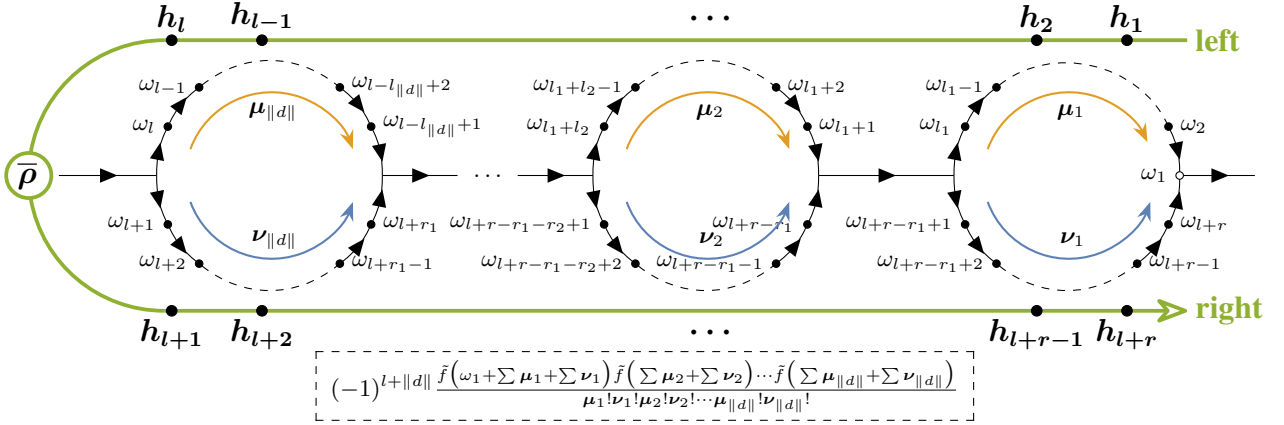
\begin{figure}[!h]
\captionsetup{justification=Justified, font=footnotesize}
\centering
\begin{tikzpicture}
\begin{feynman}

\vertex (a);

\vertex[label=left:\large$\boldsymbol{\overline{\rho}}$] at ($(a) + (-0.18cm,0cm)$) (al);
\vertex[right=1.3cm of a] (b);
\vertex[right=3cm of b] (c);
\vertex[right=1cm of c] (d);
\vertex[small, dot, label=left:$\omega_{l}$] at ($(b)!0.5!(c) + (-1.3515cm, 0.65083cm)$) (bc1){};
\vertex[small, dot, label=left:$\omega_{l-1}$] at ($(b)!0.5!(c) + (-0.93524cm, 1.1728cm)$) (bc2){};
\vertex[small, dot, label=right:$\omega_{l-l_{\norm{d}}+2}$] at ($(b)!0.5!(c) + (0.93524cm, 1.1728cm)$) (bc3){};
\vertex[small, dot, label=right:$\omega_{l-l_{\norm{d}}+1}$] at ($(b)!0.5!(c) + (1.3515cm, 0.65083cm)$) (bc4){};
\vertex[small, dot, label=left:$\omega_{l+1}$] at ($(b)!0.5!(c) + (-1.3515cm, -0.65083cm)$) (bc5){};
\vertex[small, dot, label=left:$\omega_{l+2}$] at ($(b)!0.5!(c) + (-0.93524cm, -1.1728cm)$) (bc6){};
\vertex[small, dot, label=right:$\omega_{l+r_{1}-1}$] at ($(b)!0.5!(c) + (0.93524cm, -1.1728cm)$) (bc7){};
\vertex[small, dot, label=right:$\omega_{l+r_{1}}$] at ($(b)!0.5!(c) + (1.3515cm, -0.65083cm)$) (bc8){};

\diagram* {
	(a) -- [fermion] (b) -- [fermion, out=90, in=-115.71] (bc1) -- [fermion, out=64.286, in=-141.43] (bc2) -- [scalar, out=38.571, in=141.43] (bc3) -- [fermion, out=-38.571, in=115.71] (bc4) -- [fermion, out=-64.286, in=90] (c) -- [fermion] (d),
	(b) -- [fermion, out=-90, in=115.71] (bc5) -- [fermion, out=-64.286, in=141.43] (bc6) -- [scalar, out=-38.571, in=-141.43] (bc7) -- [fermion, out=38.571, in=-115.71] (bc8) -- [fermion, out=64.286, in=-90] (c)
};

\vertex[label=left:$\boldsymbol{\mu}_{\norm{d}}$] at ($(bc1)!0.5!(bc4) + ( 0.6cm, 0.2cm )$) (mud){};
\vertex[label=left:$\boldsymbol{\nu}_{\norm{d}}$] at ($(bc5)!0.5!(bc8) + ( 0.6cm, -0.2cm )$) (nud){};

\vertex[] at ($(bc1) + ( 0.3cm, -0.3cm )$) (startud){};
\vertex[] at ($(bc5) + ( 0.3cm, 0.3cm )$) (startvd){};

\draw[mthca_orange, thick, -{Stealth[length=2.5mm]}] (startud) arc (-200:-340:1.15cm);
\draw[mthca_blue, thick, -{Stealth[length=2.5mm]}] (startvd) arc (-160:-20:1.15cm);

\vertex[right=0.1cm of d, label=right:$\ldots$] (e);

\vertex[right=0.7cm of e] (f);
\vertex[right=1cm of f] (g);
\vertex[right=3cm of g] (h);
\vertex[right=1cm of h] (i);
\vertex[small, dot, label=left:$\omega_{l_{1}+l_{2}}$] at ($(g)!0.5!(h) + (-1.3515cm, 0.65083cm)$) (gh1){};
\vertex[small, dot, label=left:$\omega_{l_{1}+l_{2}-1}$] at ($(g)!0.5!(h) + (-0.93524cm, 1.1728cm)$) (gh2){};
\vertex[small, dot, label=right:$\omega_{l_{1}+2}$] at ($(g)!0.5!(h) + (0.93524cm, 1.1728cm)$) (gh3){};
\vertex[small, dot, label=right:$\omega_{l_{1}+1}$] at ($(g)!0.5!(h) + (1.3515cm, 0.65083cm)$) (gh4){};
\vertex[small, dot, label=left:$\omega_{l+r-r_{1}-r_{2}+1}$] at ($(g)!0.5!(h) + (-1.3515cm, -0.65083cm)$) (gh5){};
\vertex[small, dot, label=left:$\omega_{l+r-r_{1}-r_{2}+2}$] at ($(g)!0.5!(h) + (-0.93524cm, -1.1728cm)$) (gh6){};
\vertex[small, dot, label=left:$\omega_{l+r-r_{1}-1}$] at ($(g)!0.5!(h) + (0.93524cm, -1.1728cm)$) (gh7){};
\vertex[small, dot, label=left:$\omega_{l+r-r_{1}}$] at ($(g)!0.5!(h) + (1.3515cm, -0.65083cm)$) (gh8){};

\vertex[label=left:$\boldsymbol{\mu}_{2}$] at ($(gh1)!0.5!(gh4) + ( 0.47cm, 0.2cm )$) (mu2){};
\vertex[label=left:$\boldsymbol{\nu}_{2}$] at ($(gh5)!0.5!(gh8) + ( 0.47cm, -0.2cm )$) (nu2){};

\vertex[] at ($(gh1) + ( 0.3cm, -0.3cm )$) (startu2){};
\vertex[] at ($(gh5) + ( 0.3cm, 0.3cm )$) (startv2){};

\draw[mthca_orange, thick, -{Stealth[length=2.5mm]}] (startu2) arc (-200:-340:1.15cm);
\draw[mthca_blue, thick, -{Stealth[length=2.5mm]}] (startv2) arc (-160:-20:1.15cm);

\diagram* {
	(f) -- [fermion] (g) -- [fermion, out=90, in=-115.71] (gh1) -- [fermion, out=64.286, in=-141.43] (gh2) -- [scalar, out=38.571, in=141.43] (gh3) -- [fermion, out=-38.571, in=115.71] (gh4) -- [fermion, out=-64.286, in=90] (h),
	(g) -- [fermion, out=-90, in=115.71] (gh5) -- [fermion, out=-64.286, in=141.43] (gh6) -- [scalar, out=-38.571, in=-141.43] (gh7) -- [fermion, out=38.571, in=-115.71] (gh8) -- [fermion, out=64.286, in=-90] (h)
};

\vertex[right=0.8cm of h] (k);
\vertex[right=1cm of k] (l);
\vertex[right=3cm of l] (m);
\vertex[right=1cm of m] (n);
\vertex[small, dot, label=left:$\omega_{l_{1}}$] at ($(l)!0.5!(m) + (-1.3515cm, 0.65083cm)$) (lm1){};
\vertex[small, dot, label=left:$\omega_{l_{1}-1}$] at ($(l)!0.5!(m) + (-0.93524cm, 1.1728cm)$) (lm2){};
\vertex[small, dot, label=right:$\omega_{2}$] at ($(l)!0.5!(m) + (1.3515cm, 0.65083cm)$) (lm3){};
\vertex[small, empty dot, label=left: $\omega_{1}$] at ($(l)!0.5!(m) + (1.5cm, 0)$) (lm4){};
\vertex[small, dot, label=left:$\omega_{l+r-r_{1}+1}$] at ($(l)!0.5!(m) + (-1.3515cm, -0.65083cm)$) (lm5){};
\vertex[small, dot, label=left:$\omega_{l+r-r_{1}+2}$] at ($(l)!0.5!(m) + (-0.93524cm, -1.1728cm)$) (lm6){};
\vertex[small, dot, label=right:$\omega_{l+r-1}$] at ($(l)!0.5!(m) + (0.93524cm, -1.1728cm)$) (lm7){};
\vertex[small, dot, label=right:$\omega_{l+r}$] at ($(l)!0.5!(m) + (1.3515cm, -0.65083cm)$) (lm8){};

\vertex[label=left:$\boldsymbol{\mu}_{1}$] at ($(lm1) + ( 1.82cm, 0.2cm )$) (mu1){};
\vertex[label=left:$\boldsymbol{\nu}_{1}$] at ($(lm5)!0.5!(lm8) + ( 0.47cm, -0.2cm )$) (nu1){};

\vertex[] at ($(lm1) + ( 0.3cm, -0.3cm )$) (startu1){};
\vertex[] at ($(lm5) + ( 0.3cm, 0.3cm )$) (startv1){};

\draw[mthca_orange, thick, -{Stealth[length=2.5mm]}] (startu1) arc (-200:-335:1.15cm);
\draw[mthca_blue, thick, -{Stealth[length=2.5mm]}] (startv1) arc (-160:-20:1.15cm);

\diagram* {
	(h) -- [fermion] (l) -- [fermion, out=90, in=-115.71] (lm1) -- [fermion, out=64.286, in=-141.43] (lm2) -- [scalar, out=38.571, in=115.71] (lm3) -- [fermion, out=-64.286, in=90] (lm4) -- [fermion] (n),
	(l) -- [fermion, out=-90, in=115.71] (lm5) -- [fermion, out=-64.286, in=141.43] (lm6) -- [scalar, out=-38.571, in=-141.43] (lm7) -- [fermion, out=38.571, in=-115.71] (lm8) -- [fermion, out=64.286, in=-90] (lm4)
};

\draw[mthca_green, line width=0.5mm, -] (1.4,1.8) -- (15,1.8);
\draw[mthca_green, line width=0.5mm] (1.4,1.8) arc[
		start angle=90,
		end angle=170,
		x radius=1.8cm,
		y radius=1.8cm
];
\draw[mthca_green, line width=0.5mm] (-0.37265,-0.31257) arc[
		start angle=190,
		end angle=270,
		x radius=1.8cm,
		y radius=1.8cm
];
\draw[mthca_green, line width=0.5mm](-0.4cm,0cm) circle (0.3);
\draw[mthca_green, line width=0.5mm, -{Stealth[length=5mm, open, round]}] (1.4,-1.8) -- (15,-1.8);

\vertex[label=right: \color{mthca_green}\large\textbf{left}\color{black}] at (15,1.8);
\vertex[label=right: \color{mthca_green}\large\textbf{right}\color{black}] at (15,-1.8);

\vertex[label=right:
\dashbox{
$
( -1 )^{l+\norm{d}}
\frac{ \tilde{f}\big( \omega_{1} + \sum \boldsymbol{\mu}_{1} + \sum \boldsymbol{\nu}_{1} \big) \tilde{f}\big( \sum \boldsymbol{\mu}_{2} + \sum \boldsymbol{\nu}_{2} \big) \cdots \tilde{f}\big( \sum \boldsymbol{\mu}_{\norm{d}} + \sum \boldsymbol{\nu}_{\norm{d}} \big) }{ \boldsymbol{\mu}_{1} ! \boldsymbol{\nu}_{1} ! \boldsymbol{\mu}_{2} ! \boldsymbol{\nu}_{2} ! \cdots \boldsymbol{\mu}_{\norm{d}} ! \boldsymbol{\nu}_{\norm{d}} ! }
$
}
] at (3.3,-2.8);

\filldraw[black] (14.2,1.8) circle (2pt) node[anchor=south]{\large$\boldsymbol{h_{1}}$};
\filldraw[black] (13,1.8) circle (2pt) node[anchor=south]{\large$\boldsymbol{h_{2}}$};
\filldraw[black] (1.5,1.8) circle (2pt) node[anchor=south]{\large$\boldsymbol{h_{l}}$};
\filldraw[black] (2.7,1.8) circle (2pt) node[anchor=south]{\large$\boldsymbol{h_{l-1}}$};
\filldraw[black] (1.5,-1.8) circle (2pt) node[anchor=north]{\large$\boldsymbol{h_{l+1}}$};
\filldraw[black] (2.7,-1.8) circle (2pt) node[anchor=north]{\large$\boldsymbol{h_{l+2}}$};
\filldraw[black] (14.2,-1.8) circle (2pt) node[anchor=north]{\large$\boldsymbol{h_{l+r}}$};
\filldraw[black] (13,-1.8) circle (2pt) node[anchor=north]{\large$\boldsymbol{h_{l+r-1}}$};

\vertex[label=above: \large$\boldsymbol{\cdots}$] at (8.7,1.88);
\vertex[label=above: \large$\boldsymbol{\cdots}$] at (8.7,-2.3);

\end{feynman}
\end{tikzpicture}
\caption{The structure of a generic diagram that contributes to the coefficient $C_{l,r}\big(\omega_{1},\omega_{2},\cdots,\omega_{l+r}\big)$ as in Eq.(\ref{Eq: contraction coefficient}). The green contour encircling the diagram indicates the order of the corresponding operators from left to right as they appear in the superoperator $h_{1} h_{2} \cdots h_{l} [ \bullet ] h_{l+1} h_{l+2} \cdots h_{l+r}$. The arrows inside each loop of the diagram indicate the order in which the frequencies are arranged in the corresponding vectors $\boldsymbol{\mu}_{i}$ and $\boldsymbol{\nu}_{i}$ used in the closed-form formula of the diagram's contribution.}
\label{fig: generic diagram}
\end{figure}

As shown in Fig.\ref{fig: generic diagram}, a generic diagram $d$ is an ordered set of $\norm{d}$ loops each of which consists of inserted operator frequencies that are arranged in a closed contour as one goes from the left-most operator (with frequency $\omega_{1}$) to the right-most one (with frequency $\omega_{l+r}$). In addition, we also require that the first loop (the one consisting of $(\boldsymbol{\mu}_{1},\boldsymbol{\nu}_{1})$) must contain the left-most frequency $\omega_{1}$. In other words, all the diagrams containing $l$ left frequencies and $r$ right frequencies can be generated by partitioning the 2-tuple $(l, r)$ into an ordered set of 2-tuples of non-negative integers $\big\{ (l_{1}, r_{1}), (l_{2}, r_{2}), \cdots, (l_{\norm{d}}, r_{\norm{d}}) \big\}$ so that
\begin{equation*}
\begin{split}
l_{1} + l_{2} + \cdots + l_{\norm{d}} = l
\quad
\textrm{and}
\quad
r_{1} + r_{2} + \cdots + r_{\norm{d}} = r
\end{split}
\end{equation*}
with $l_{1} \ge 1$.

Notice that it is possible for individual diagrams to diverge if certain frequencies sum up to zero in the denominator of the formula above. However, one can regularize the diverging diagrams by shifting all the frequencies by a small amount $\delta \omega$, summing up contributions from all the diagrams, and then take the $\delta\omega \rightarrow 0$ limit. Crucially, the singular contributions cancel out completely \emph{in all circumstances} regardless of one's assumption about the coarse-graining window function $f(t;\tau)$, leaving only the finite part which can also be calculated directly according to Eq.(B29) in Appendix B of the Supplementary Materials~\cite{supplement}. Therefore, effective master equations can be obtained straightforwardly without having to apply ad hoc regularizing techniques as is the case for many effective Hamiltonian methods in some circumstances (e.g. the Schrieffer-Wolff transformation used in \cite{MPT_II} is regularized based on a priori knowledge of the system state). We attribute the regularity of the MaTCG method to the fact that we are not trying to force the disappearance of certain predetermined types of terms in the effective Hamiltonian; rather, the effective dynamical equation is obtained from a physical procedure of coarse graining informed by the measurement channel.

\onecolumngrid

\twocolumngrid
	
In addition, we can organize the terms in $\mathcal{L}_{k}(t)$ at each order into the sum of a Hamiltonian term and a dissipator term (the dissipator term here may or may not dissipate energy over time):
\begin{equation}
\begin{split}
\mathcal{L}_{k}(t) \overline{\rho}
=
-
i \Big[
H_{\textrm{eff}}^{(k)}(t)
,
\overline{\rho}
\Big]
+
D_{\textrm{eff}}^{(k)}(t) \overline{\rho}
\end{split}
\end{equation}
where $H_{\textrm{eff}}^{(k)}(t)$ and $D_{\textrm{eff}}^{(k)}(t)$ are defined in Eq.(B21) and Eq.(B22) respectively as shown explicitly in Appendix B of the Supplementary Materials~\cite{supplement}.

In fact, these closed-form formulas allow for fully automated symbolic as well as numerical calculations on computers, which we address in an accompanying paper \cite{Leon_Wentao_STCG}.

Finally, TCG perturbation theory is grounded in the fundamental Hamiltonian formulation of quantum systems -- potentially with infinite degrees of freedom -- governed by the von Neumann equation. Quantum electrodynamical systems, in particular, fall into this category. We show in the next section that time-coarse graining justifies many of the standard approximations that is adopted in Quantum Optics. But in addition, with a suitable choice of the coarse-graining time scale, the MaTCG framework allows for a systematic derivation of the non-Markovian dynamics of a subsytem. In particular, it provides a rigorous justification for tracing out from $\overline{\rho}(t)$ parts of the system that evolve with a bath relaxation time shorter than the time-coarse-graining time $\tau$.

In the following section, we discuss in detail the dispersive readout of a spin-like degree of freedom, both as an illustration of the TCG method, and as a toy model towards an in-depth analysis of the readout problem of superconducting qubits. In order to obtain more explicit results, starting from here, we assume a gaussian window function $f(t;\tau) = \frac{1}{\sqrt{2\pi}\tau} e^{-\frac{t^{2}}{2\tau^{2}}}$ of half width $\tau$, which gives rise to the following low-pass filter function $\tilde{f}(\omega) = e^{-\frac{\omega^{2}\tau^{2}}{2}}$.

\section{TCG-based analysis of open quantum systems}
\label{Sec: TCG open quantum systems}

In this section, we extend the TCG framework to analyze quantum systems with infinite degrees of freedom, using a system-plus-bath Hamiltonian as our starting point. This choice does not limit generality, as any field theory can be cast in this form with appropriately defined system and bath operators.

When interactions between the system and the bath are sufficiently weak, one can obtain an effective master equation description for the system which is Markovian beyond the bath relaxation time scale. Using the MaTCG framework, we put ``separation of time scale'' arguments in more rigorous terms and propose a general strategy for deriving Lindblad-type master equations that are valid even when some of the internal dynamics compete over similar time scales as the bath relaxation; when such competitions are absent, our strategy reproduces the canonical Lindblad master equations.

The resulting Markovian TCG master equation for the system can be regarded as describing either (1) situations where projective measurements (with finite time resolution) only take place at the beginning (initialization) and the end (measurement) of an otherwise unmonitored evolution, or (2) situations where quantum trajectories are continuously recorded with finite time resolution and averaged over a large ensemble of repeated experiments \cite{Wiseman_Milburn_2009}.

When individual quantum trajectories are considered, one needs to include backaction from the continuous monitoring of the bath modes, which replaces the effective master equation (EME) by a stochastic master equation (SME) \cite{Carmichael_1993, Wiseman_Milburn_1993, Mabuchi_Wiseman_1999}. The MaTCG framework modifies both the smooth dynamics of the EME and the quantum noises in the SME \emph{in a self-consistent way} in formed by the time resolution $\tau$ of the measurements performed, as will be discussed in Subsection \ref{Subsec: spin_cavity_4th_order}.

\subsection{The effective master equation description}

Our derivation proceeds in two stages. First we obtain the TCG quantum Liouvillian for \emph{all} the system and bath degrees of freedom in the form of a perturbative expansion in the system-bath coupling strengths. At the second stage, we derive an EQME for the reduced system by tracing out degrees of freedom whose dynamics evolve at time scales significantly outside the bandwidth of one's measurement channel. This two-step approach to quantum systems with infinite degrees of freedom avoids imposing an a priori partitioning of system degrees of freedom. Instead, it relates different identification schemes to distinct limits of the full TCG master equation, enabling accurate treatment of non-Markovian dynamics.

As we will show in this section and the corresponding Supplementary Materials, the TCG framework generalizes the effective master equation approach to open quantum systems beyond the Caldeira-Leggett regime as informed by the interactions at the microscopic level. These generalizations are particularly important in low-temperature scenarios, where fast internal dynamics of the system take place around or below the thermal relaxation time scales of the baths. In particular, the TCG-based framework should be distinguished from effective theories which take a phenomenological Lindblad QME as the starting point and time-average it afterwards~\cite{MPT_II, Hanai_McDonald_Clerk}. This is because in the TCG approach to open quantum systems, the time-coarse graining process and thereby the measurement channel and its bandwidth play a central role in the justification of Markov and secular approximations involved in establishing the Lindblad QME in the first place.
Not surprisingly, the resulting effective quantum master equation for the system depends on the coarse-graining time scale $\tau$ in general, which is a feature that has been suggested by other analyses in the literature \cite{Cohen_Tannoudji_1986, Cohen_Tannoudji_1998, Cresser2017}, and may only be ignored when $\tau$ is limited to certain ranges in comparison with the various time scales of the system.

In order to simplify our presentation of the TCG approach to open quantum systems, we model the bath in this section by a continuum of bosonic modes described by the Schr\"{o}dinger-picture Hamiltonian
\begin{equation}
\begin{split}
\hat{H}_{B}
=
\int_{0}^{\infty} \frac{d\omega}{2\pi} \cdot \mathcal{D}_{\omega} \omega B_{\omega}^{\dagger} B_{\omega}
\end{split}
\end{equation}
with some density of states $\mathcal{D}_{\omega}$. It is straightforward to generalize to models with non-bosonic and/or multiple thermal baths, and discussions about such more general situations can be found in Appendix D of the Supplementary Materials~\cite{supplement}. Even in the absence of an accurate physical model of the system and its environment, experimentally observed non-Markovian dynamics in parts of the system can be effectively captured by embedding those components into an extended system comprising a Markovian bath and auxiliary degrees of freedom~\cite{Youssry_2023, Luchnikov_2020}.

For example, in the spin-readout toy model to be discussed in \Sec{Sec: Spin_Cavity}, if one regards the cavity mode as an auxiliary system, then the effective TCG dynamics of the spin can indeed be shown to demonstrate typical non-Markovian properties accurately captured by the MaTCG EQME. The MaTCG framework thus offers a versatile and broadly applicable approach for modeling a wide class of open quantum systems.

We assume that the interaction-picture Hamiltonian takes the form
\begin{equation}
\label{Eq: system+bath HI}
\begin{split}
&
H_{I}(t)
=
H_{s}(t)
+
H_{sB}(t)\\
\equiv&
H_{s}(t)
+
\int_{0}^{\infty} \frac{d\omega}{2\pi} \mathcal{D}_{\omega} g_{\omega} \Big( \sum_{k} e^{-i(\omega-\omega_{k})t} S_{k} B_{\omega} + h.c. \Big)
\end{split}
\end{equation}
where the coupling strength $g_{\omega}$ is much smaller than the other energy scales in the model as well as the TCG energy scale $\tau^{-1}$, which allows us to consider it as a small parameter for perturbative expansion. Following the recipe in the previous section, we can perturbatively calculate the TCG Liouvillian superoperator $\mathcal{L}$, and for the purpose of this section, we rewrite $\mathcal{L}$ as a power series in the system-bath coupling strengths $g_{\omega}$:
\begin{equation}
\begin{split}
\mathcal{L}
=
\mathcal{L}^{(0)}
+
\mathcal{L}^{(1)}
+
\mathcal{L}^{(2)}
+
\cdots
\end{split}
\end{equation}
where $\mathcal{L}^{(k)}$ is of order $\mathcal{O}\big( g_{\omega}^{k} \big)$. In particular, $\mathcal{L}^{(0)}$ acts as identity on the space of bath operators, whereas $\mathcal{L}^{(1)}$ is linear in $B_{\omega}$ and $B_{\omega}^{\dagger}$. Denoting the total density matrix of system+bath by $\overline{\rho}_{\textrm{tot}}(t)$, we can trace out the bath modes from the TCG master equation to obtain the following equation for the reduced density matrix of the system:
\begin{equation}
\label{Eq: formal drhobardt}
\begin{split}
&
\partial_{t} \overline{\rho}(t)
=
\textrm{Tr}_{B} \big[
\mathcal{L} (t) \overline{\rho}_{\textrm{tot}}(t)
\big]\\
=&
\textrm{Tr}_{B} \big[
\mathcal{L} (t) \big(
\overline{\rho}(t) \otimes \rho_{B}(0)
\big)
\big]
+
\textrm{Tr}_{B} \big[
\mathcal{L} (t) \overline{\chi}_{\textrm{corr}}(t)
\big]
\end{split}
\end{equation}
where $\overline{\rho}(t) \equiv \textrm{Tr}_{B}\big[ \overline{\rho}_{\textrm{tot}}(t) \big]$ is the TCG system density matrix, and the TCG system-bath correlation is defined as
\begin{equation}
\overline{\chi}_{\textrm{corr}}(t)
:=
\overline{\rho}_{\textrm{tot}}(t) - \overline{\rho}(t) \otimes \rho_{B}(0).
\end{equation}
Here $\rho_{B}(0)$ represents a thermal state of the bath, and we assume that $\overline{\chi}_{\textrm{corr}}(0) = 0$.
Now we define the projection operators $\mathcal{P}$ and $\mathcal{Q}$ as
\begin{equation}
\label{Eq: def P and Q}
\begin{split}
&
\mathcal{P} \overline{\rho}_{\textrm{tot}}(t)
:=
\textrm{Tr}_{B}\big[ \overline{\rho}_{\textrm{tot}}(t) \big] \otimes \rho_{B}(0)
\equiv
\overline{\rho}(t) \otimes \rho_{B}(0)\\
&
\mathcal{Q} := 1 - \mathcal{P}
\end{split}
\end{equation}
which allows us to write $\overline{\chi}_{\textrm{corr}}(t) = \mathcal{Q} \overline{\rho}_{\textrm{tot}}(t)$ and therefore
\begin{equation}
\label{Eq: dChidt}
\begin{split}
\partial_{t} \overline{\chi}_{\textrm{corr}}(t)
\equiv&
\partial_{t} \mathcal{Q} \overline{\rho}_{\textrm{tot}}(t)
=
\mathcal{Q} \mathcal{L}(t) \overline{\chi}_{\textrm{corr}}(t)
+
\mathcal{Q} \mathcal{L}(t) \mathcal{P} \overline{\rho}_{\textrm{tot}}(t).
\end{split}
\end{equation}
In addition, it is straightforward to verify from the definitions in Eq.(\ref{Eq: def P and Q}) that
\begin{equation}
\label{Eq: Q properties}
\begin{split}
\textrm{Tr}_{B} \big[ \mathcal{L}^{(0)}(t) \mathcal{Q} \rho \big]
=
0,
\quad\;
\textrm{Tr}_{B} \big[ \mathcal{L}^{(1)}(t) \mathcal{Q} \rho \big]
=
\textrm{Tr}_{B} \big[ \mathcal{L}^{(1)}(t) \rho \big]
\end{split}
\end{equation}
for any operator $\rho$.
Formally integrating Eq.(\ref{Eq: dChidt}), we obtain
\begin{equation}
\label{Eq: Chi_corr}
\begin{split}
&
\overline{\chi}_{\textrm{corr}}(t)\\
=&
\int_{0}^{t} dt_{1}
\mathcal{T} e^{\int_{t_{1}}^{t} dt_{2} \mathcal{Q} \mathcal{L}(t_{2})} \mathcal{Q}\big[
\mathcal{L}(t_{1})\big[
\overline{\rho}(t_{1}) \otimes \rho_{B}(0)
\big]
\big]\\
=&
\int_{0}^{t} dt_{1}
\mathcal{T} e^{\int_{t_{1}}^{t} dt_{2} \mathcal{L}^{(0)}(t_{2})}
\mathcal{L}^{(1)}(t_{1})\big[
\overline{\rho}(t_{1}) \otimes \rho_{B}(0)
\big]
+
\mathcal{O}\big( g^{2} \big)
\end{split}
\end{equation}
and therefore Eq.(\ref{Eq: formal drhobardt}) can be written as follows at order $\mathcal{O}\big( g^{2} \big)$:
\begin{equation}
\label{Eq: dRhodt_v1}
\begin{split}
&
\partial_{t} \overline{\rho}(t)
=
\textrm{Tr}_{B} \big[
\mathcal{L}(t) \big[
\overline{\rho}(t) \otimes \rho_{B}(0)
\big]
\big]
+
\textrm{Tr}_{B} \big[
\mathcal{L}(t) \overline{\chi}_{\textrm{corr}}(t)
\big]\\
=&
\mathcal{L}^{(0)}(t) \overline{\rho}(t)
+
\textrm{Tr}_{B} \Big[
\mathcal{L}^{(2)}(t) \big[
\overline{\rho}(t) \otimes \rho_{B}(0)
\big]
\Big]\\
&
+
\int_{0}^{t} dt_{1}
\mathcal{G}^{(2)}(t,t-t_{1})
\overline{\rho}(t-t_{1})
+
\mathcal{O}\big( g^{3} \big)
\end{split}
\end{equation}
where
\begin{equation}
\label{Eq: G(2)}
\begin{split}
&
\mathcal{G}^{(2)}(t,t-t_{1})\\
:=&
\textrm{Tr}_{B} \Big[
\mathcal{L}^{(1)}(t)
\big( \mathcal{T} e^{\int_{t-t_{1}}^{t} dt_{2} \mathcal{L}^{(0)}(t_{2})} \big)
\mathcal{L}^{(1)}(t-t_{1}) \rho_{B}(0)
\Big]
\end{split}
\end{equation}
and we have used Eq.(\ref{Eq: Chi_corr}) as well as the properties in Eq.(\ref{Eq: Q properties}). Since $\mathcal{L}^{(1)}$ is linear in $B_{\omega}$ and $B_{\omega}^{\dagger}$, it follows that $\mathcal{G}^{(2)}(t,t-t_{1})$ is on the same order of magnitude as the (time-coarse grained) two-point bath correlation functions which decay over the bath correlation time scale $\tau_{B}$. Under most realistic conditions, $\tau_{B}$ is much smaller than the coarse-graining time scale $\tau$ with which the system is observed\footnote{With the form of system-bath couplings in Eq.(\ref{Eq: system+bath HI}), the effective system-bath coupling coefficients produced by TCG are in general band-pass-filtered around certain system operator frequencies, which apparently reduces the width of the spectral density function and thereby increases the bath correlation time $\tau_{B}$. At first sight, this seems to suggest that the large-$\tau$ limit does not necessarily justify the Markov and secular approximations. However, we recognize that the increasing value of $\tau_{B}$ predicted by naive application of TCG essentially originates from taking into account some unrealistic Poincar\'e-recurrence-like phenomena in the bath which are in fact prevented by other thermalizing mechanisms of the bath that are not included in the simplistic model presented here. Therefore, we make the physical assumption that the bath correlation time $\tau_{B}$ is always limited below some very small value as we increase the coarse-graining time scale $\tau$.}, and consequently, the Markov approximation and the secular approximation are simultaneously justified with sufficiently low time resolution, which allows us to approximate Eq.(\ref{Eq: dRhodt_v1}) with a Markovian master equation:
\begin{equation}
\label{Eq: dRhodt_v2}
\begin{split}
\partial_{t} \overline{\rho}(t)
\approx&
\mathcal{L}^{(0)}(t) \overline{\rho}(t)
+
\textrm{Tr}_{B} \Big[
\mathcal{L}^{(2)}(t) \big[
\overline{\rho}(t) \otimes \rho_{B}(0)
\big]
\Big]\\
&
+
\int_{0}^{\infty} dt_{1}
\mathcal{G}^{(2)}(t,t-t_{1})
\overline{\rho}(t)
\end{split}
\end{equation}
where the first term represents the direct action of $\mathcal{L}(t)$ on $\overline{\rho}(t)$, the second term includes the effects of one virtual change with the bath, and the third term gives us the leading-order contribution from the accumulated system-bath correlation.
Furthermore, assuming the type of system-bath couplings in Eq.(\ref{Eq: system+bath HI}), we can in general rewrite Eq.(\ref{Eq: dRhodt_v2}) in the following form:
\begin{equation}
\label{Eq: open TCG master equation}
\begin{split}
\partial_{t} \overline{\rho}(t)
\approx&
\mathcal{L}^{(0)}(t) \overline{\rho}(t)
+
\textrm{Tr}_{B} \Big[
\mathcal{L}^{(2)}(t) \big[
\overline{\rho}(t) \otimes \rho_{B}(0)
\big]
\Big]\\
&
-
i \big[ H_{\textrm{corr}}^{(2)}(t), \overline{\rho}(t) \big]
+
D_{\textrm{corr}}^{(2)}(t) \overline{\rho}(t)
\end{split}
\end{equation}
where the analytical expressions for $H_{\textrm{corr}}^{(2)}(t)$ and $D_{\textrm{corr}}(t)$ are given in Appendix C of the Supplementary Materials\cite{supplement} at leading order of system interactions. In special situations where the Markovian and secular approximations are invalid, one can also develop TCG-based effective models starting from Eq.(\ref{Eq: dRhodt_v1}) and Eq.(\ref{Eq: G(2)}).

In the rest of this paper, all open quantum systems will be modeled using the approach presented in this section under the assumptions of weak system-bath couplings and sufficiently large coarse-graining time scale $\tau$, so that the dynamics of the reduced system density matrix $\overline{\rho}(t)$ can be described by a Markovian master equation in the form of Eq.(\ref{Eq: dRhodt_v2}) which is of second order in the system-bath coupling strengths $g_{\omega}$. However, we note that the derivation presented in this section can be straightforwardly generalized to obtain effective master equations at higher orders in $g_{\omega}$. In addition, even though we will see that some of the canonical Lindblad dissipators are exactly reproduced by the TCG-based analysis, we emphasize that they act on the coarse-grained density matrix $\overline{\rho}(t)$ rather than $\rho(t)$.

In situations where there are fast internal dynamics of the system taking place over time scales comparable to or smaller than the bath correlation time scale, we expect the TCG-based approach to be phenomenologically more accurate than methods which simply assume the canonical Lindblad dissipators to act on the instantaneous system density matrix $\rho(t)$. We refer the interested reader to Appendix D of the Supplementary Materials~\cite{supplement} for an explicit example.

Finally, we emphasize that the TCG approach to open quantum systems naturally extends beyond the system-bath dichotomy. In the ``second stage'' as discussed at the beginning of this section, instead of tracing out the environment degrees of freedom all at once, one can trace out degrees of freedom in a hierarchical manner where the dynamics at shorter time scales (e.g. the thermal baths) inform the effective dynamics at longer time scales (e.g. transient dynamics of the readout resonators), eventually leading to an effective model for the core degrees of freedom that are controlled and measured in an experiment (e.g. the qubits). As shown in \Sec{Sec: Spin_Cavity} and Appendix D of the Supplementary Materials~\cite{supplement}, the dynamics of such core degrees of freedom are generally non-Markovian and are well captured by reduced TCG master equations. These equations can be seamlessly integrated with system identification techniques, including ``gray-box" approaches~\cite{Genois_Quantum_Tailored_Machine_Learning, Youssry_2023, Youssry2024}, enabling MaTCG-guided machine learning models to infer intermediate dynamics not directly accessible through experimental measurement channels. Using the same approach, one can capture the effective impacts of impurities, defects, and quasiparticles in the environment, depending on their own dynamics which can be either explicitly modeled or studied using machine learning methods in combination with experimental probes.

\subsection{Continuous measurement and the effective stochastic master equation}
\label{Subsec: SME}

The finite time resolution of detectors not only affects the EMEs discussed previously corresponding to situations with no record of any information leaking out of the system through incoherent processes, but also has implications for the ``quantum trajectory'' picture where leaked photons are monitored during an extended period of time\footnote{Here we refer to the continuously detected degrees of freedom as ``photons''}.
One way to account for the finite time resolution of photocurrent detection is to start from the time-coarse grained master equation
\begin{equation}
\begin{split}
\partial_{t} \overline{\rho}(t;\tau)
=&
\mathcal{L}(t;\tau)
\overline{\rho}(t;\tau)
\equiv
\left[
\mathcal{J}
+
\mathcal{K}(t;\tau)
\right] \overline{\rho}(t;\tau)
\end{split}
\end{equation}
where the coarse-graining time scale $\tau$ reflects the finite resolution of the recorded photocurrent, and $\mathcal{J}$ is the quantum jump superoperator that can be related to photon detection events. For example, simple homodyne detections can be modeled with $\mathcal{J} \overline{\rho} = \gamma C \overline{\rho} C^{\dagger}$ where $C$ denoting the output field of the monitored mode, while
\begin{equation}
\begin{split}
\mathcal{K}(t;\tau)
\equiv
\mathcal{L}(t;\tau)
-
\mathcal{J}
\end{split}
\end{equation}
generates smooth evolution for the unmonitored part of the system.
The master equation can be formally solved by the propagator:

\cornerlineR{0.4pt}{6pt}{0pt}
\onecolumngrid

\begin{equation}
\begin{split}
&
\mathcal{U}(t;\tau)
\equiv
\mathcal{T} e^{\int_{0}^{t} dt_{1} \mathcal{L}(t_{1},\tau) }\\
=&
\sum_{m=0}^{\infty}
\int_{0}^{t} dt_{m} \int_{0}^{t_{m}} dt_{m-1} \cdots \int_{0}^{t_{2}} dt_{1}
\cdot
\mathcal{S}(t,t_{m-1};\tau)
\mathcal{J}
\mathcal{S}(t_{m},t_{m-1};\tau)
\cdots
\mathcal{J}
\mathcal{S}(t_{1},0;\tau)
\end{split}
\end{equation}
with 
\begin{equation}
\begin{split}
\mathcal{S}(t_{n},t_{n-1};\tau)
\equiv
\mathcal{T}
e^{\int_{t_{n-1}}^{t_{n}} dt^{\prime} \cdot \mathcal{K}(t^{\prime};\tau) }
.
\end{split}
\end{equation}
As demonstrated in \cite{Alsing_Carmichael_1991}, the probability density for a perfect detector with time resolution $\tau$ to record $m$ photon detection events in the time interval $[0,t]$ is
\begin{equation}
\begin{split}
&
p_{m}(t_{1}, \ldots, t_{m};[0,t],\tau)
=
\textrm{Tr}
\left[
\mathcal{S}(t,t_{m-1};\tau)
\mathcal{J}
\mathcal{S}(t_{m},t_{m-1};\tau)
\cdots
\mathcal{J}
\mathcal{S}(t_{1},0;\tau)
\overline{\rho}(0;\tau)
\right]
\end{split}
\end{equation}
if the detection events center around time
$
t
=
t_{1},t_{2},\cdots,t_{m}
$
respectively. Consequently, if one records a particular series of photon detection events centered around those moments in time, the conditional density operator is
\begin{equation}
\begin{split}
\overline{\rho}_{c}^{t_{1},\cdots,t_{m}}(t;\tau)
=
\frac{
\mathcal{S}(t,t_{m-1};\tau)
\mathcal{J}
\mathcal{S}(t_{m},t_{m-1};\tau)
\cdots
\mathcal{J}
\mathcal{S}(t_{1},0;\tau)
\overline{\rho}(0;\tau)
}{
\textrm{Tr}
\left[
\mathcal{S}(t,t_{m-1};\tau)
\mathcal{J}
\mathcal{S}(t_{m},t_{m-1};\tau)
\cdots
\mathcal{J}
\mathcal{S}(t_{1},0;\tau)
\overline{\rho}(0;\tau)
\right]
}
.
\end{split}
\end{equation}

\twocolumngrid

\cornerlineL{0.4pt}{0pt}{6pt}

Following the same analysis as in \cite{Wiseman_Milburn_1993}, we see that the time evolution of $\overline{\rho}_{c}(t;\tau)$, when subject to continuous homodyne measurements with efficiency $\eta = 1$, can be written as
\begin{equation}
\begin{split}
\partial_{t} \overline{\rho}(t;\tau)
=&
\mathcal{L}(t;\tau) \overline{\rho}(t;\tau)
+
\sqrt{\gamma} \xi(t;\tau)
\Big[
C \overline{\rho}(t;\tau) C^{\dagger}\\
&
-
\frac{1}{2}
\big(
C^{\dagger}C \overline{\rho}(t;\tau)
+
\overline{\rho}(t;\tau) C^{\dagger}C
\big)
\Big]
\end{split}
\end{equation}
where $\xi(t;\tau)$ denotes a stationary colored Gaussian noise process, arising from the Poisson statistic of the measuring local oscillator and determined by the coarse-graining window function $f(t;\tau)$ so that
$
\mathbb{E}
\left[
\xi(t;\tau)
\right]
=
0
$
while
\begin{equation}
\label{Eq: general homodyne SME}
\begin{split}
&
\textrm{Cov}
\left[
\xi(t;\tau),
\xi(t+\Delta t;\tau)
\right]\\
=&
\int dt_{1} dt_{2}
f^{\prime}(t_{1};\tau) f^{\prime}(t_{2};\tau)
\min\left(
t - t_{1}
,
t + \Delta t - t_{2}
\right)
.
\end{split}
\end{equation}

In situations where the time resolution of the continuous measurement is much higher than the rate of change in $\overline{\rho}$, one recovers the Lindblad master equation + Wiener white noise description of stochastic quantum trajectory by setting $\tau\rightarrow0$. For any non-negligible $\tau$, however, one should be more careful in general and use Eq.(\ref{Eq: general homodyne SME}) instead.

If one forces a set of finite-resolution measurement records on an infinite-resolution description of the measured quantum system, then the resulting SME would contain explicit non-Markovian terms \cite{Warszawski_Wiseman_2003}, leading to backactions that modify the average system dynamics and bring a number of subtly consistency issues \cite{Wiseman_Howard_Gambetta_2008}. We therefore contend that it is much more natural and self-consistent to carry out the whole discussion within the MaTCG framework where the non-Markovian phenomenology is consistently captured and straightforward to interpret.
For a more specific discussion, we refer the reader to \ref{Subsec: TCG spin dynamics} and Subsection \ref{Subsec: spin readout TCG}.

\section{Time-coarse grained dispersive spin readout}
\label{Sec: Spin_Cavity}

In this section we consider a model for the quantum nondemolition (QND) readout of a qubit through the lens of the TCG approach. We emphasize that QND measurement schemes, such as dispersive readouts, do not necessarily imply any direct continuous projective measurements on the measured subsystem. In fact, all ``readout-induced effects'' on the qubit in our model are due to coherent dynamics at the microscopic level, even though the effective dynamics at longer time scales may appear to be dissipative, as will be demonstrated in this section. If not record of the leaked photons is kept during the readout process or if a large ensemble of photocurrent records is averaged over, our EQME would provide a sufficient description; if the photocurrent is continuously monitored, then the MaTCG framework also provides a self-consistent prescription for deriving an effective stochastic master equation that takes into account the backaction due to the finite-resolution monitoring, where the standard Wiener process is recovered in the $\tau \rightarrow 0$ limit (see \Sec{Subsec: spin-cavity SME}).

In \Sec{Subsec: spin readout TCG} we derive the EQME to the fourth order in the dispersive limit. At that order, there are more than one thousand terms in the TCG series. Nonetheless, as we demonstrate in this section, the process of pruning these terms down to most significant terms, a process that depends on the parameters under consideration, can be automatized. We present analytical expressions for the most significant unitary and non-unitary corrections under assumed conditions, compare these terms to known expression obtained before using unitary transformation techniques and present corrections to them that have not been found before. We also find and discuss analytical expressions for processes that have not been noted before in literature. 

Next, in \Sec{Subsec: Spin-cavity resonance and tau}, we discuss a more difficult regime from the perspective of effective model synthesis, where the drive, cavity and the qubit are quasi-resonant with another and many terms in the TCG series are no longer negligibly small. This regime has been studied before using direct numerical techniques and does not present any special challenges unless the number of excitations becomes large~\cite{Blais_etal_transmon_ionization}. The goal of the analysis here is to demonstrate the flexibility and the broad applicability of the TCG method. The resulting terms are well-behaved and the series does not display the kind of divergences that one may encounter in unitary transformation techniques~\cite{MPT_II}, which have to be addressed using specialized techniques~\cite{KEHREIN19961}. 

We then proceed to numerically solve the resulting EQME in \Sec{Subsec: spin-cavity numerical solutions}. Although findings in this subsection are not entirely new, they provide us with an opportunity to examine the role of coarse-graining on qubit measurement through heterodyne readout of the cavity mode. We also point out a few new observations that are only possible through the TCG approach. In the section we also analyze the numerical stability and the computational resource requirements of EQME, and provide comparisons to directly solving the exact von-Neumann equation. One of the most compelling reasons why it may be meaningful to deal with thousands of terms analytically generated by TCG at high orders (once automatized through symbolic computation) becomes immediately apparent: greatly enhanced numerical stability for long-time simulation, and enormous savings in computational resources due to the resulting equations being far less stiffer than the original ones. 

Readers who are interested in the final expression for the effective spin-only QME to fourth order can skip to \Sec{Subsec: bath effects}. There we present the final formulas under approximations that have been previously employed in literature, where the TCG corrections can be cleanly delineated.

Finally, 

\subsection{The spin-cavity model for the dispersive readout of a qubit}
\label{Subsec: spin readout TCG}

In order to demonstrate the phenomenological predictions as well as the conceptual implications of the time-coarse grained effective master equation, we first study a toy model for the dispersive readout of a qubit. More specifically, we consider a system described by the following Schrodinger-picture Hamiltonian:
\begin{equation}
\begin{split}
\hat{H}
=&
\frac{\omega_{a}}{2} \sigma_{z}
+
\omega_{c} c^{\dagger} c
+
\hat{H}_{B}
+
g_{ac} \sigma_{x} \big( c + c^{\dagger} \big)\\
&
+
\epsilon_{d} \big( c + c^{\dagger} \big) \big( e^{-i \omega_{d} t} + e^{i \omega_{d} t} \big)
+
\hat{H}_{cB}
\end{split}
\end{equation}
where the bath is modeled by free bosonic modes:
\begin{equation}
\hat{H}_{B}
=
\int \frac{d\omega}{2\pi} \cdot \mathcal{D}_{\omega} \cdot \omega B_{\omega}^{\dagger} B_{\omega}
\end{equation}
and the cavity is coupled to the bath via the Hamiltonian term
\begin{equation}
\hat{H}_{cB}
=
\int \frac{d\omega}{2\pi} \cdot \mathcal{D}_{\omega} g_{\omega} \big( c + c^{\dagger} \big) \Big( B_{\omega} + B_{\omega}^{\dagger} \Big).
\end{equation}

In circuit QED experiments, the actual measurement process can hardly be modeled by an instantaneous ideal projection of the quantum state onto some predetermined basis. Instead, a readout pulse is sent to probe a particular component of the quantum device (e.g. the readout cavity), and the resulting output signal is then mixed with a local oscillator at some frequency $\omega_{\textrm{LO}}$, producing a down-converted signal which is then averaged over some time scale $\tau$ by an integrator to obtain a finite reading. As explained in Appendix H of the Supplementary Materials\cite{supplement}, quantum input-output theory suggests that measurement results obtained in such processes can be directly related to the time-coarse grained density matrix $\overline{\rho}(t)$ in the rotating frame where the mode coupled to the readout drive rotates at frequency $\omega_{\textrm{LO}}$. Therefore, for the spin-cavity readout system to be discussed in this section, we go to the rotating frame of the drive, i.e., the interaction picture defined by the free Hamiltonian
\begin{equation}
\begin{split}
\hat{H}_{0}
=
\frac{\omega_{a}}{2} \sigma_{z}
+
\omega_{d} c^{\dagger} c
+
\hat{H}_{B}
\end{split}
\end{equation}
where the interaction-picture Hamiltonian can be written as
\begin{equation}
\label{Eq: spin-cavity HI}
\begin{split}
H_{I}(t)
=&
\int \frac{d\omega}{2\pi} \cdot \mathcal{D}_{\omega} g_{\omega}
\Big(
e^{i(\omega-\omega_{d})t} B_{\omega}^{\dagger} c
+
e^{i(\omega+\omega_{d})t} B_{\omega}^{\dagger} c^{\dagger}\\
&
+
h.c.
\Big)
+
\epsilon_{d}
\big(
c
+
e^{-2i\omega_{d}t} c
+
h.c.
\big)
+
(\omega_{c} - \omega_{d}) c^{\dagger} c\\
&
+
g_{ac}
\big(
e^{i(\omega_{d}-\omega_{a})t} c^{\dagger} \sigma_{-}
+
e^{i(\omega_{d}+\omega_{a})t} c^{\dagger} \sigma_{+}
+
h.c.
\big).
\end{split}
\end{equation}

Following the recipe in \Sec{Sec: TCG perturbation theory}, we can write down the resulting TCG master equation for the full density matrix $\overline{\rho}_{\textrm{tot}}(t)$ describing the spin, the cavity mode, as well as the bath:
\begin{equation}
\begin{split}
\partial_{t} \overline{\rho}_{\textrm{tot}}(t)
=&
-
i \big[ H_{\textrm{TCG}}(t), \overline{\rho}_{\textrm{tot}}(t) \big]
+
D_{\textrm{TCG}}(t) \overline{\rho}_{\textrm{tot}}(t)
\end{split}
\end{equation}
where $H_{\textrm{TCG}}$ and $D_{\textrm{TCG}}$ receive contributions from each order in the perturbative expansion of the TCG Liouvillian $\mathcal{L}$
\begin{equation}
\begin{split}
&
H_{\textrm{TCG}}
=
H_{\textrm{TCG}}^{(1)}
+
H_{\textrm{TCG}}^{(2)}
+
H_{\textrm{TCG}}^{(3)}
+
\cdots\\
&
D_{\textrm{TCG}}
=
D_{\textrm{TCG}}^{(1)}
+
D_{\textrm{TCG}}^{(2)}
+
D_{\textrm{TCG}}^{(3)}
+
\cdots
\end{split}
\end{equation}

In the rest of this subsection, we discuss the most significant terms of the TCG master equation at different orders in the perturbative expansion. Unless stated otherwise, when numerical values need to be calculated we assume a typical superconducting qubit readout scenario with 
\begin{equation}
\label{Eq: dispersive parameters}
\begin{split}
&
\frac{\omega_{a}}{2\pi} = 5 \textrm{GHz};
\qquad\quad\;\;
\frac{\omega_{c}}{2\pi} = 7 \textrm{GHz};
\qquad\quad\;\;
\frac{\omega_{d}}{2\pi} = 7 \textrm{GHz};
\\
&
\frac{g_{ac}}{2\pi} = 48.0 \textrm{MHz};
\qquad\quad\,\,
\epsilon_{d}
=
12.1 \textrm{MHz}
\qquad\quad\,\,
\tau = 3 \textrm{ns};\\
&
J(\omega)
\equiv
\mathcal{D}_{\omega} g_{\omega}^{2}
=
\frac{\alpha \omega}{1 + (\frac{\omega}{\Lambda})^{2}}\\
&
\textrm{with}
\\
&
\kappa_{c}
\equiv
J(\omega_{d})
=
1 \textrm{MHz}
\quad\textrm{and}\quad
\frac{\Lambda}{2\pi}
=
50 \textrm{GHz}
\end{split}
\end{equation}

When the analytical expressions for the most significant terms are presented, we ignore superoperators that are at least exponentially suppressed by a factor of $e^{-\frac{\omega^{2}\tau^{2}}{2}}$ with $\omega > \frac{2\pi}{\tau}$, where all frequencies are assumed to be close to the numerical values listed above (close in comparison with $\frac{1}{\tau}$). In addition, we assume that $\omega_{d} = \omega_{c}$ for simplicity of the analytical expressions. After presenting the most significant TCG superoperators up to the third order, we discuss the effects of the thermal bath using the method presented in \Sec{Sec: TCG open quantum systems}.

\subsubsection{Rotating-wave approximation at the leading order}

At the leading order, TCG is equivalent to time-averaging the interaction-picture Hamiltonian, which reduces to the rotating-wave approximation (RWA) in the $\tau\rightarrow \infty$ limit.
In the spin-readout model studied here, we have
\begin{equation}
\begin{split}
H_{\textrm{TCG}}^{(1)}
\approx&
\int_{0}^{\infty} d\omega \cdot \mathcal{D}_{\omega} g_{\omega} e^{-\frac{(\omega-\omega_{d})^{2}\tau^{2}}{2}}
\Big(
e^{i(\omega-\omega_{d})t} c B_{\omega}^{\dagger}\\
&\qquad\quad\;
+
e^{-i(\omega-\omega_{d})t} c^{\dagger} B_{\omega}
\Big)
+
\epsilon_{d} \big( c + c^{\dagger} \big)
\end{split}
\end{equation}
and
\begin{equation}
\begin{split}
D_{\textrm{TCG}}^{(1)}
\approx
0.
\end{split}
\end{equation}
In particular, we see that the transverse coupling between the spin and the cavity mode is suppressed by TCG, and its effects on the slow dynamics show up at higher orders in the TCG expansion.
In addition, coupling coefficients between the cavity mode and the bath modes are band-filtered around the drive frequency $\omega_{d}$, and we will discuss their effects at the end of this subsection.

\subsubsection{Dispersive spin-cavity coupling and emergent spin decay channel at the second order}

At the second order, corrections to the effective Hamiltonian can be written as
\begin{equation}
\begin{split}
H_{\textrm{TCG}}^{(2)}
\approx
H_{g_{ac}^{2}}
+
H_{g_{ac} g}
\end{split}
\end{equation}
with
\begin{equation}
\begin{split}
\label{Eq: Hac}
H_{g_{ac}^{2}}
=&
-
2 g_{ac}^{2}
\Big(
\frac{1-e^{-(\omega_{d}-\omega_{a})^{2}\tau^{2}}}{\omega_{d} - \omega_{a}}
-
\frac{1-e^{-(\omega_{d}+\omega_{a})^{2}\tau^{2}}}{\omega_{d}+\omega_{a}}
\Big)\\
&\qquad\cdot
\frac{\sigma_{z}}{2} \big( c^{\dagger}c + \frac{1}{2} \big) + \textrm{const.}\\
\approx&
-
2g_{ac}^{2} \big(
\frac{1}{\omega_{d} - \omega_{a}}
-
\frac{1}{\omega_{d} + \omega_{a}}
\big)
\frac{\sigma_{z}}{2} \big( c^{\dagger}c + \frac{1}{2} \big)
+
\textrm{const.}
\end{split}
\end{equation}
and
\begin{equation}
\label{Eq: spin-bath couplings}
\begin{split}
H_{g_{ac} g}
=
\int_{0}^{\infty} \frac{\mathcal{D}_{\omega}}{2\pi}
\cdot
\tilde{g}^{(2)}_{\omega}
\Big(
\sigma_{-} B_{\omega}^{\dagger}
e^{i(\omega-\omega_{a})t}
+
h.c.
\Big)
\end{split}
\end{equation}
where
\begin{equation}
\label{Eq: gaB}
\begin{split}
\tilde{g}^{(2)}_{\omega}
=&
-
g_{\omega} g_{ac}
\Big[
e^{-\frac{(\omega-\omega_{a})^{2}\tau^{2}}{2}}
\Big(
\frac{ \omega_{d} }{ \omega_{d}^{2} - \omega_{a}^{2} }
+
\frac{ \omega_{d} }{ \omega_{d}^{2} - \omega^{2} }
\Big)\\
&
+
e^{-\frac{(\omega-\omega_{d})^{2}+(\omega_{d}-\omega_{a})^{2}}{2}\tau^{2}}
\Big(
\frac{1}{2(\omega-\omega_{d})}
+
\frac{1}{2(\omega_{a}-\omega_{d})}
\Big)
\Big].
\end{split}
\end{equation}
Notice that, starting from the second line of Eq.(\ref{Eq: Hac}), we have ignored terms that are exponentially suppressed by a factor of $e^{- \omega_{\textrm{eff}}^{2} \tau^{2}}$ with some $\frac{\omega_{\textrm{eff}}}{2\pi} > 1\textrm{GHz}$. For the purpose of concise presentation, we will always make this approximation in the rest of this paper unless otherwise stated. In fact, the coefficients of the TCG superoperators are $\tau$-dependent in general, and apparently $\tau$-independent expressions such as the second line of Eq.(\ref{Eq: Hac}) can only be valid when $\tau$ is assumed to be sufficiently large so that the superoperator coefficients have reached ``fixed points'' in the infrared (IR) limit of the renormalization flow due to coarse-graining.
We will later show that, once the bath modes are traced out, the effective direct couplings in Eq.(\ref{Eq: gaB}) between the spin and the bath modes lead to Purcell decay of the spin.

As for the second-order effective spin-cavity coupling $H_{g_{ac}^{2}}$, we notice that the term
\[
-\frac{2g_{ac}^{2}}{\omega_{d}-\omega_{a}} \frac{\sigma_{z}}{2} \big(c^{\dagger}c + \frac{1}{2}\big)
\]
reproduces the result in \cite{Blais_etal_dispersive_cQED} obtained from second-order perturbative diagonalization of the Jaynes-Cummings Hamiltonian; the additional term
\[
\frac{2g_{ac}^{2}}{\omega_{d}+\omega_{a}} \frac{\sigma_{z}}{2} \big(c^{\dagger}c + \frac{1}{2}\big)
\]
originates from the counter-rotating terms in the Rabi Hamiltonian, and is a first example of the effect of high-frequency virtual processes modifying the secular dynamics of a quantum system, which is not captured in \cite{Blais_etal_dispersive_cQED}. Here we emphasize that such high-frequency virtual processes are exceedingly cumbersome to be incorporated into well-known perturbative diagonalization methods as in \cite{Carbonaro_1979, Blais_etal_dispersive_cQED,C_F_Lo_1998}, because the unitary transformation that one has to postulate as an ansatz become much more complicated with those high-frequency (non-RWA) terms in the Hamiltonian, especially at high orders in the perturbative expansion. The time-coarse graining method developed in this work, on the other hand, puts both high- and low-frequency processes on the same footing, and efficiently calculates all the non-RWA corrections at arbitrary time resolution. Such non-RWA corrections are especially important in a systematic perturbation theory when $\frac{\omega_{d}-\omega_{a}}{\omega_{d}+\omega_{a}}$ is comparable with or greater than small parameters such as
\[
\frac{g_{ac}}{\omega_{d} - \omega_{c}}
\quad\textrm{or}\quad
g_{ac} \tau
\]
in the perturbative expansion. We will have more discussion on the corrections from non-RWA Hamiltonian terms in \Sec{Subsec: spin_cavity_4th_order}.

Bringing the discussion back to the second-order TCG corrections to the spin-cavity model, we notice that there is no significant TCG dissipator at this order, namely
\begin{equation}
\begin{split}
D_{\textrm{TCG}}^{(2)}
\approx
0.
\end{split}
\end{equation}
Hence we conclude that, at order $g_{ac}^{2}$, the effects of the spin-cavity coupling on the slow dynamics are:
\begin{itemize}
    \item A cavity-state independent shift of the spin transition frequency, as represented by
    \begin{equation}
    \delta^{(2)} \omega_{a} = - g_{ac}^{2} \big( \frac{1}{\omega_{d}-\omega_{a}} - \frac{1}{\omega_{d}+\omega_{a}} \big)
    \end{equation}
    \item A dispersive longitudinal coupling between the spin and the cavity, as represented by the effective Hamiltonian term
    \begin{equation}
    \delta H^{(2)}_{\textrm{disp}}
    =
    -
    g_{ac}^{2} \big(
    \frac{1}{\omega_{d} - \omega_{a}}
    -
    \frac{1}{\omega_{d} + \omega_{a}}
    \big)
    \sigma_{z} c^{\dagger}c
    \end{equation}
    which shifts the cavity frequency by
    \begin{equation}
    \begin{split}
    \qquad
    \delta\omega_{c}^{(2)}(n)
    =
    g_{ac}^{2} \big(
    \frac{1}{\omega_{d} - \omega_{a}}
    -
    \frac{1}{\omega_{d} + \omega_{a}}
    \big) (1 - 2n)
    \end{split}
    \end{equation}
    when the spin is at level $n\in \{ 0, 1\}$.
    \item Emergent direct couplings between the spin and the bath modes, as shown in Eq.(\ref{Eq: spin-bath couplings}). These effective spin-bath couplings function as decay channels for the spin and are responsible for its Purcell decay.
\end{itemize}

\subsubsection{Drive-induced spin level transitions at the third order}
At the third order, the only significant correction to the effective Hamiltonian can be written as
\begin{equation}
\label{Eq: HTCG3}
\begin{split}
&
H_{\textrm{TCG}}^{(3)}
\approx
H_{\epsilon_{d}g_{ac}^{2}}
+
H_{g_{ac}^{2}g}
\end{split}
\end{equation}
with
\begin{equation}
\begin{split}
&
H_{\epsilon_{d}g_{ac}^{2}}
=
\frac{\epsilon_{d} \lambda^{2}}{2}
\sigma_{z} (
c
+
c^{\dagger}
)
\end{split}
\end{equation}
and
\begin{equation}
\label{Eq: g3_Hamiltonian}
\begin{split}
H_{g_{ac}^{2}g}
=&
\int_{0}^{\infty} \frac{d\omega}{2\pi} \mathcal{D}_{\omega}
\cdot
\tilde{g}^{(3)}_{\omega}
\sigma_{z} \Big(
c B_{\omega}^{\dagger}
e^{i(\omega-\omega_{d})t}
+
h.c.
\Big)
\end{split}
\end{equation}
where
\begin{equation}
\label{Eq: lambda2}
\begin{split}
\lambda^{2}
=&
g_{ac}^{2}
\frac{
2\omega_{a}(3\omega_{d}^{2}-\omega_{a}^{2})
}{
\omega_{d}(\omega_{d}^{2}-\omega_{a}^{2})^{2}
}\\
=&
g_{ac}^{2}
\Big(
\frac{1}{(\omega_{d} - \omega_{a})^{2}}
+
\frac{2\omega_{a}}{\omega_{d}(\omega_{d}^{2}-\omega_{a}^{2})}
-
\frac{1}{(\omega_{d}+\omega_{a})^{2}}
\Big)
\end{split}
\end{equation}
while
\begin{equation}
\begin{split}
\tilde{g}^{(3)}_{\omega} \biggr\rvert_{\omega \rightarrow \omega_{d}}
=&
g_{\omega_{d}}\frac{\lambda^{2}}{2}.
\end{split}
\end{equation}

In particular, $H_{\epsilon_{d}g_{ac}^{2}}$ can be considered either as a spin-state dependent modification of the drive strength on the cavity, or as a cavity-state dependent shift in the spin transition frequency. In the second line of Eq.(\ref{Eq: lambda2}), we have split the frequency-dependent factor into three terms, where the first one is exactly the ``drive Hamiltonian'' correction in the first line of Eq.(3.15) in \cite{Blais_etal_dispersive_cQED}, and the following two terms are non-RWA corrections which are not captured by diagonalizing the Jaynes-Cummings Hamiltonian in \cite{Blais_etal_dispersive_cQED}. For the purpose of dispersive spin readout, this Hamiltonian correction provides another mechanism for the cavity state to distinguish between different spin states, in addition to the dispersive longitudinal coupling discussed previously. However, for moderate drive strengths, this correction is overwhelmed by the lower-order longitudinal coupling. On the other hand, $H_{g_{ac}^{2}g}$ is a spin-dependent modification to the cavity-bath coupling, which results in the following modification to the cavity thermal dissipator
\begin{equation}
\label{Eq: diss_def}
\begin{split}
\kappa_{c} D_{c, c^{\dagger}}
\longrightarrow
\kappa_{c} D_{c(1+\lambda^{2}\sigma_{z}/2),\; c^{\dagger}(1+\lambda^{2}\sigma_{z}/2)}
\end{split}
\end{equation}
as will be discussed in Subsection \ref{Subsec: bath effects}. Notice that here we have used the notation
\begin{equation}
\label{Eq: DLR_def}
\begin{split}
D_{L,R} \rho
\equiv
L \rho R
-
\frac{RL\rho + \rho RL}{2}
\end{split}
\end{equation}
for any operators $L$ and $R$, which generalizes the notation for Lindblad dissipators by allowing $R \neq L^{\dagger}$. So far, these results exactly reproduce the Hamiltonian corrections that can be obtained in principle  by perturbative diagonalization \cite{Blais_etal_dispersive_cQED}.

More interestingly, the leading-order significant TCG dissipators also appear at this order:
\begin{equation}
\label{Eq: spin D3}
\begin{split}
&
D_{\textrm{TCG}}^{(3)}
\approx
D_{\epsilon_{d}g_{ac}^{2}}\\
=&
i \epsilon_{d} \frac{g_{ac}^{2}}{(\omega_{d}-\omega_{a})^{2}}
\big(
D_{\sigma_{-},\sigma_{+}c}
+
D_{\sigma_{+}c,\sigma_{-}}
\big)\\
&
+
i \epsilon_{d} \frac{g_{ac}^{2}}{(\omega_{d}+\omega_{a})^{2}}
\big(
D_{\sigma_{-}c,\sigma_{+}}
+
D_{\sigma_{+},\sigma_{-}c}
\big)
+
h.c.
\end{split}
\end{equation}
which has not been predicted by any effective Hamiltonian method. Noticeably, these dissipators represent non-coherent processes in the effective coarse-grained dynamics that emerge with or without coupling to a bath. In fact, one can tell from the coefficients of these dissipators that they are induced by the cavity drive as well as the spin-cavity coupling. Focusing on their effects on the spin dynamics, we trace out the cavity mode and write the effective spin dissipators as
\begin{equation}
\label{Eq: TCG spin dissipator}
\begin{split}
D_{\epsilon_{d}g_{ac}^{2}}^{\textrm{spin}}(t)
=
K(t) D_{\sigma_{-},\sigma_{+}}
+
\Gamma(t) D_{\sigma_{+},\sigma_{-}}
\end{split}
\end{equation}
with the drive-induced spin transition rates being
\begin{equation}
\label{Eq: formula for K and Gamma}
\begin{split}
&
K(t)
:=
-
2 \epsilon_{d} g_{ac}^{2} \Big(
\frac{1}{(\omega_{d} - \omega_{a})^{2}}
+
\frac{1}{(\omega_{d} + \omega_{a})^{2}}
\Big)
\textrm{Im}\langle c(t) \rangle_{1}\\
&
\Gamma(t)
:=
-
2 \epsilon_{d} g_{ac}^{2} \Big(
\frac{1}{(\omega_{d} - \omega_{a})^{2}}
+
\frac{1}{(\omega_{d} + \omega_{a})^{2}}
\Big)
\textrm{Im}\langle c(t) \rangle_{0}
\end{split}
\end{equation}
where we use $\langle c(t) \rangle_{l}$ to denote the conditional expectation value of $c$ when the spin is at level $l$:
\begin{equation}
\begin{split}
\langle c(t) \rangle_{l}
:=
\frac{
\textrm{Tr}\big( 
c | l\rangle\langle l| \overline{\rho}(t) \big)
}{
\textrm{Tr}\big( 
| l\rangle\langle l| \overline{\rho}(t) \big)
}.
\end{split}
\end{equation}
In the scenario of dispersive spin readout, the lifetime $\frac{1}{\kappa_{c}}$ of the cavity mode is much shorter than that of the spin, and therefore the quasi-steady state of the cavity mode will be close to a spin-state dependent coherent state with
\begin{equation}
\begin{split}
\lim_{\kappa_{c} t \rightarrow \infty} \langle c(t)\rangle_{n}
\approx
\frac{\epsilon_{d}}{
- \delta\omega_{c}^{(2)}(n)
+
\frac{\kappa_{c}}{2} i
}
\end{split}
\end{equation}
at low temperatures where energy absorption from the bath is much smaller than the energy emission into the bath. In such cases, we find the drive-induced relaxation and excitation rates to be approximately
\begin{equation}
\begin{split}
&
\lim_{\kappa_{c} t \rightarrow \infty}
K(t)
\approx
\kappa_{c} g_{ac}^{2} \Big(
\frac{1}{(\omega_{d} - \omega_{a})^{2}}
+
\frac{1}{(\omega_{d} + \omega_{a})^{2}}
\Big)\\
&\qquad\qquad\qquad\quad\cdot
\frac{
\abs{\epsilon_{d}}^{2}
}{
| \delta\omega_{c}^{(2)}(1) |^{2}
+
\frac{\kappa_{c}^{2}}{4}
}\\
&
\lim_{\kappa_{c} t \rightarrow \infty}
\Gamma(t)
\approx
\kappa_{c} g_{ac}^{2} \Big(
\frac{1}{(\omega_{d} - \omega_{a})^{2}}
+
\frac{1}{(\omega_{d} + \omega_{a})^{2}}
\Big)\\
&\qquad\qquad\qquad\quad\cdot
\frac{
\abs{\epsilon_{d}}^{2}
}{
| \delta\omega_{c}^{(2)}(0) |^{2}
+
\frac{\kappa_{c}^{2}}{4}
}
\end{split}
\end{equation}
after the cavity mode settles to its (quasi-)steady state. Notice that $K(\infty)$ and $\Gamma(\infty)$ are proportional to the (steady-state) cavity photon occupation number
\begin{equation}
\begin{split}
\lim_{\quad t \gg \kappa_{c}^{-1}}
\langle n_{c}(t) \rangle_{l}
\approx
\frac{
\abs{\epsilon_{d}}^{2}
}{
| \delta\omega_{c}^{(2)}(l) |^{2}
+
\frac{\kappa_{c}^{2}}{4}
}.
\end{split}
\end{equation}
And by the end of this subsection, we will show that these drive-induced transition rates become greater than the Purcell decay rate of the spin at low temperatures once the drive is strong enough to maintain a few photons in the readout cavity. Therefore, for strong readout drives, these drive-induced transitions can become the dominant source of spin relaxation/excitation, which eventually destroys the quantum non-demolition feature of the dispersive readout scheme.

We note that the TCG master equation is not positive-definite in general for arbitrary density matrix. In fact, for certain extreme $\overline{\rho}$, the effective qubit transition rates can be negative. However, as explained in Appendix E of the Supplementary Materials\cite{supplement}, if physically achievable initial conditions for $\overline{\rho}(t)$ are chosen, then the TCG master equation will not lead to unphysical states with negative probabilities, even if some of the effective transition rates are temporarily negative during parts of the evolution. In fact, by comparing with the dynamics obtained from coarse-graining the exact numerical solution in Fig.2 of Appendix E in the Supplementary Materials\cite{supplement}, we are able to demonstrate that the appearance of negative effective transition rates does \textit{not} signal failure of the TCG perturbation theory. Instead, it is a faithful representation of physical phenomena at finite time resolution which cannot be explained by Lindblad master equations with positive relaxation/excitation rates.

\subsubsection{RWA and non-RWA corrections at the fourth order}
\label{Subsec: spin_cavity_4th_order}
At the fourth order in the TCG expansion, non-RWA corrections are more prevalent. For instance, the most significant terms in $H_{\textrm{TCG}}^{(4)}$ can be written as
\begin{equation}
\begin{split}
H_{\textrm{TCG}}^{(4)}
\approx&
H_{g_{ac}^{4}}
+
H_{\epsilon_{d}^{2} g_{ac}^{2}}
\end{split}
\end{equation}
with
\begin{equation}
\begin{split}
H_{g_{ac}^{4}}
=&
\zeta^{\prime} c^{\dagger} c
+
\zeta \Big( c^{\dagger}c + \frac{1}{2} \Big) \sigma_{z}
+
\zeta \big(c^{\dagger}c\big)^{2} \sigma_{z}
\end{split}
\end{equation}
and
\begin{equation}
\begin{split}
H_{\epsilon_{d}^{2} g_{ac}^{2}}
=&
-
\xi c^{\dagger}c
+
\mu \sigma_{z}
\end{split}
\end{equation}
where the coefficients in the formulas above are defined as
\begin{equation}
\begin{split}
&
\zeta
=
g_{ac}^{4}
\Big[
\frac{1}{(\omega_{d} - \omega_{a})^{3}}
-
\frac{2(3\omega_{a}^{2}+\omega_{d}^{2})}{(\omega_{d}+\omega_{a})^{3}(\omega_{d}-\omega_{a})^{2}}\\
&\qquad\qquad
+
\frac{1}{(\omega_{d}+\omega_{a})^{3}}
\Big];\\
&
\zeta^{\prime}
=
g_{ac}^{4}
\Big[
\frac{1}{(\omega_{d} - \omega_{a})^{3}}
-
\frac{2(\omega_{a}^{2}+\omega_{d}^{2})}{\omega_{d}(\omega_{d}+\omega_{a})^{2}(\omega_{d}-\omega_{a})^{2}}\\
&\qquad\qquad
+
\frac{1}{(\omega_{d}+\omega_{a})^{3}}
\Big];\\
&
\xi
=
\frac{g_{ac}^{2}\epsilon_{d}^{2}}{\omega_{d}(\omega_{d}^{2}-\omega_{a}^{2})};
\qquad
\mu
=
-
g_{ac}^{2}\epsilon_{d}^{2}
\frac{\omega_{a}(3\omega_{d}^{2}-\omega_{a}^{2})}{\omega_{d}^{2}(\omega_{d}^{2}-\omega_{a}^{2})^{2}}.
\end{split}
\end{equation}
In particular, we notice that the first terms in $\zeta$ and $\zeta^{\prime}$ reproduce the highest-order corrections in the effective system Hamiltonian in \cite{Blais_etal_dispersive_cQED}, while the rest of the terms therein correspond to non-RWA corrections; the effective AC-Stark shifts to the spin and cavity frequencies in $H_{\epsilon_{d}^{2} g_{ac}^{2}}$, on the other hand, are not captured by the effective Hamiltonian in \cite{Blais_etal_dispersive_cQED} since there the drive is excluded from the diagonalization procedure. However, it is already remarkable that the effective Hamiltonian obtained by perturbative diagonalization is exactly identical to the IR limit of the TCG Hamiltonian up to high orders in the coupling constants, considering that the physical pictures are different in the two formalisms. Nevertheless, since the terms in the diagonalized Hamiltonian are conserved by the time evolution, it is not unreasonable to expect phenomenological convergence to the TCG Hamiltonian in the IR limit.

The TCG dissipators, on the other hand, account for corrections to the effective dynamics which cannot be captured by any effective Hamiltonian method. In particular, the most significant fourth-order TCG dissipators for the spin-cavity model can be written as
\begin{equation}
\begin{split}
&
D_{\textrm{TCG}}^{(4)}
\approx
D_{g_{ac}^{4}}
+
D_{\epsilon_{d}^{2}g_{ac}^{2}}
\end{split}
\end{equation}
with
\begin{equation}
\begin{split}
D_{g_{ac}^{4}}
=&
\tilde{\kappa}_{+} \big(
D_{\sigma_{-}c^{\dagger} c^{2}, \sigma_{+}c^{\dagger}}
-
D_{\sigma_{+}c^{\dagger 2} c, \sigma_{-}c}
\big)\\
&
+
\tilde{\kappa}_{-} \big(
D_{\sigma_{-}c^{\dagger 2} c, \sigma_{+}c}
-
D_{\sigma_{+}c^{\dagger} c^{2}, \sigma_{-}c^{\dagger}}
\big)
+
h.c.
\end{split}
\end{equation}
and
\begin{equation}
\begin{split}
D_{\epsilon_{d}^{2}g_{ac}^{2}}
=&
\tilde{\gamma} \big(
D_{\sigma_{-}c^{\dagger}c, \sigma_{+}}
+
D_{\sigma_{+}c^{\dagger}c, \sigma_{-}}
\big)\\
&
+
\tilde{\gamma}_{z} \big(
D_{\sigma_{z} c, c^{\dagger}}
-
D_{\sigma_{z} c^{\dagger}, c}
\big)
+
h.c.
\end{split}
\end{equation}
where $\tilde{\kappa}_{\pm}$, $\tilde{\gamma}$, and $\tilde{\gamma}_{z}$ are defined as
\begin{equation}
\begin{split}
&
\tilde{\kappa}_{\pm}
=
\frac{4i g_{ac}^{4}\omega_{a}}{(\omega_{d} \mp \omega_{a})(\omega_{d}\pm\omega_{a})^{3}};\\
&
\tilde{\gamma}
=
-
i
\frac{\epsilon_{d}^{2}g_{ac}^{2}}{\omega_{d}(\omega_{d}^{2}-\omega_{a}^{2})};
\qquad
\tilde{\gamma}_{z}
=
i
\frac{\epsilon_{d}^{2}g_{ac}^{2}\omega_{a}}{\omega_{d}^{2}(\omega_{d}^{2}-\omega_{a}^{2})}
.
\end{split}
\end{equation}

\subsubsection{Effects of the bath and the reduced spin master equation}
\label{Subsec: bath effects}

Following the recipe in \Sec{Sec: TCG open quantum systems}, we know that the bath-independent superoperators at order $g^{0}$ acts directly on the reduced spin+cavity density matrix $\overline{\rho}(t) \equiv \textrm{Tr}_{B} \overline{\rho}_{\textrm{tot}}(t)$, whereas the bath-induced corrections at order $g^{2}$ can be calculated from $\mathcal{L}^{(1)}$ and $\mathcal{L}^{(2)}$ according to Eq.(\ref{Eq: dRhodt_v1}). Based on the numerical values of the parameters assumed at the beginning of this subsection, we find the most significant bath-related effects to be described by the following terms in Eq.(\ref{Eq: open TCG master equation}):
\begin{equation}
\begin{split}
&
\partial_{t} \overline{\rho}(t)
=
\cdots
-
i
\Big[
H_{g^{2}}
+
H_{\textrm{corr}}^{(2)}
,
\overline{\rho}(t)
\Big]
+
D_{\textrm{corr}}^{(2)}(t) \overline{\rho}(t)
\end{split}
\end{equation}
with
\begin{equation}
\begin{split}
&
H_{g^{2}}\\
=&
-
\int_{0}^{\infty} \frac{d\omega}{2\pi} \cdot \mathcal{D}_{\omega} g_{\omega}^{2}
\Big(
\frac{1 - e^{-(\omega-\omega_{c})^{2}\tau^{2}}}{\omega-\omega_{c}}
+
\frac{1}{\omega+\omega_{c}}
\Big) c^{\dagger} c,
\end{split}
\end{equation}

\begin{equation}
\begin{split}
H_{\textrm{corr}}^{(2)}
=
-
\int_{0}^{\infty} \frac{d\omega}{2\pi} \cdot \mathcal{P}
\Big(
\mathcal{D}_{\omega} g_{\omega}^{2}
\frac{e^{-(\omega-\omega_{c})^{2}\tau^{2}}}{\omega-\omega_{c}}
\Big) c^{\dagger} c
,
\end{split}
\end{equation}
and
\begin{equation}
\begin{split}
D_{\textrm{corr}}^{(2)}
=&
J(\omega_{d}) \big( \overline{n}(\omega_{d}) + 1 \big) D_{c,c^{\dagger}}
+
J(\omega_{d}) \overline{n}(\omega_{d}) D_{c^{\dagger},c}\\
&
+
\tilde{J}(\omega_{a}) \big( \overline{n}(\omega_{a}) + 1 \big) D_{\sigma_{-},\sigma_{+}}
+
\tilde{J}(\omega_{a}) \overline{n}(\omega_{a}) D_{\sigma_{+},\sigma_{-}}
\end{split}
\end{equation}
where
\begin{equation}
\begin{split}
&
\overline{n}(\omega)
=
\textrm{Tr}_{B} \big(
B^{\dagger}_{\omega} B_{\omega_{d}} \rho_{B}(0)
\big);
\qquad
\tilde{J}(\omega)
=
\mathcal{D}_{\omega} \abs{ \tilde{g}^{(2)}_{\omega} }^{2}.
\end{split}
\end{equation}
In particular, the sum of $H_{g^{2}}$ and $H_{\textrm{corr}}^{(2)}$ gives us the following formula for the Lamb shift of the cavity frequency
\begin{equation}
\begin{split}
&
H_{g^{2}}
+
H_{\textrm{corr}}^{(2)}\\
=&
-
\int_{0}^{\infty} \frac{d\omega}{2\pi} \cdot
\mathcal{P}
\Big[
\mathcal{D}_{\omega} g_{\omega}^{2}
\Big(
\frac{1}{\omega-\omega_{c}}
+
\frac{1}{\omega+\omega_{c}}
\Big)
\Big],
\end{split}
\end{equation}
which reproduces the standard formula for the Lamb shift of a harmonic oscillator coupled to a bosonic thermal bath.
As for dissipators, we find the zero-temperature cavity decay rate to be 
\begin{equation}
\begin{split}
\kappa_{c}
=
J(\omega_{d})
\equiv
\mathcal{D}_{\omega_{d}} g_{\omega_{d}}^{2}
\end{split}
\end{equation}
at the second order in the coupling constants. Incorporating the additional coupling terms in Eq.(\ref{Eq: g3_Hamiltonian}), we find the spin-dependent cavity decay rates to be
\begin{equation}
\kappa_{c}^{e}
=
(1+\lambda^{2}/2)^{2} \kappa_{c}
\quad\textrm{and}\quad
\kappa_{c}^{g}
=
(1-\lambda^{2}/2)^{2} \kappa_{c}
\end{equation}
for the excited and ground spin states respectively.
Similarly, for an Ohmic spectral density $J(\omega)$, the low-temperature Purcell decay rate of the spin is found to be
\begin{equation}
\begin{split}
\kappa_{s}
=&
\tilde{J}(\omega_{a})
\equiv
\mathcal{D}_{\omega_{a}} \big( \tilde{g}^{(2)}_{\omega_{a}} \big)^{2}
\approx
\mathcal{D}_{\omega_{a}} g_{\omega_{a}}^{2}
\frac{ 4g_{ac}^{2}\omega_{d}^{2} }{ \big( \omega_{d}^{2} - \omega_{a}^{2} \big)^{2} }\\
=&
\mathcal{D}_{\omega_{a}} g_{\omega_{a}}^{2}
\Big(\frac{1}{(\omega_{d} - \omega_{a})^{2}}
+
\frac{4\omega_{d}}{(\omega_{d}+\omega_{a})^{2}(\omega_{d}-\omega_{a})}\\
&\qquad\qquad
-
\frac{1}{(\omega_{d}+\omega_{a})^{2}}
\Big)
=
\frac{4\kappa_{c} g_{ac}^{2}\omega_{a}\omega_{d}}{(\omega_{d}^{2}-\omega_{a}^{2})^{2}}.
\end{split}
\end{equation}
Again, the first term in the second line reproduces the Purcell decay rate calculated in \cite{Blais_etal_dispersive_cQED}, while the rest of the terms account for the non-RWA corrections.

Similar to the system superoperators, the effective TCG spectral density also depends on the coarse-graining time scale $\tau$. For example, if we assume the infinite-resolution spectral density to be Ohmic, then we can lump together all the TCG superoperators that are proportional to $B_{\omega}^{\dagger}$ by adding up their coefficients and obtain the following effective spectral densities for $\tau = 0, 0.1\,\textrm{ns},$ and $1\,\textrm{ns}$:
\begin{figure}[h!]
\captionsetup{justification=Justified, font=footnotesize}
\centering
\captionsetup{justification=Justified, font=footnotesize}
\includegraphics[width=0.47\textwidth]{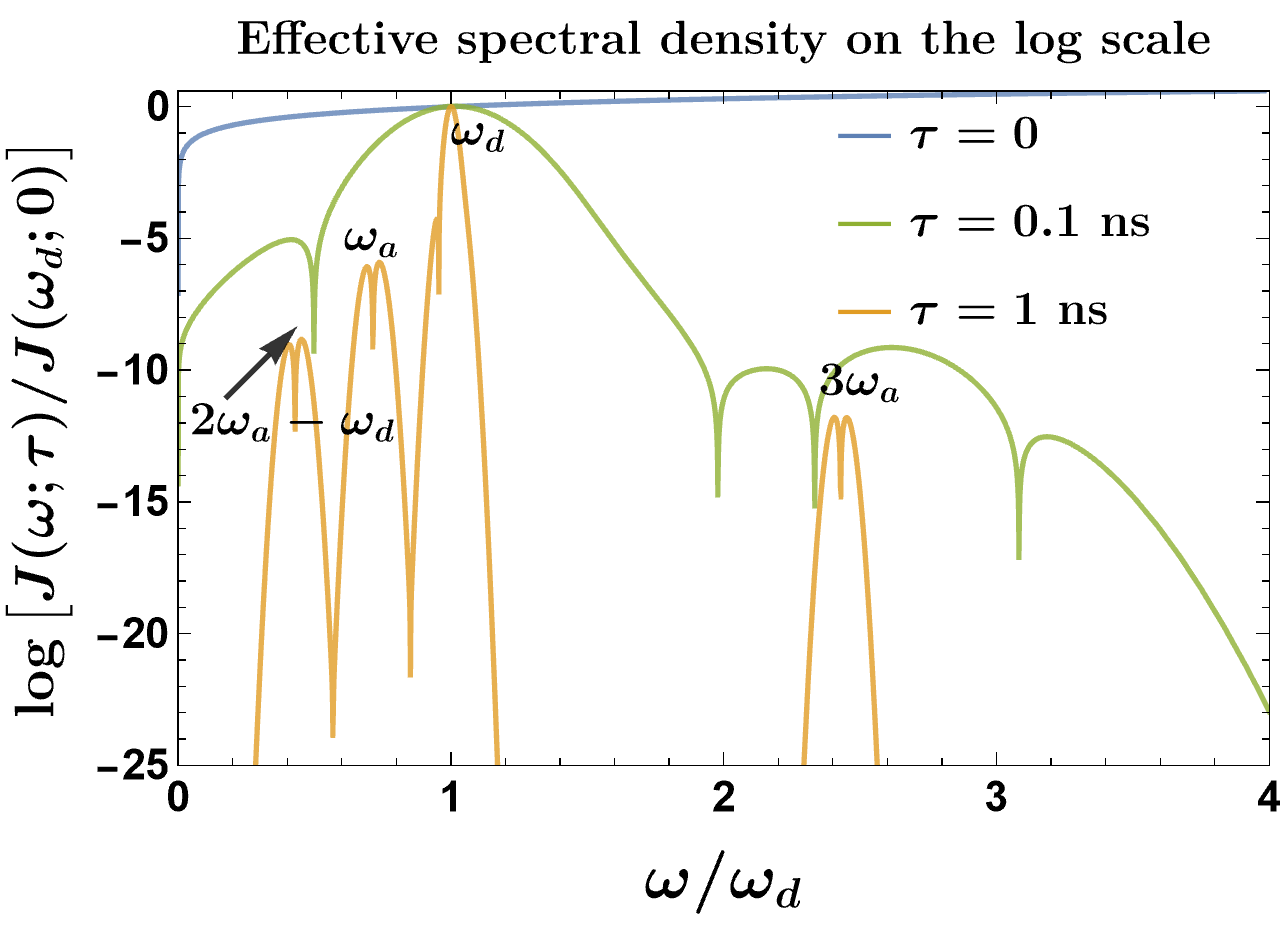}
\caption{The effective spectral densities at different coarse-graining time scale $\tau$ as obtained from the third-order TCG master equation. The vertical axis is on the log scale, and the reference point $J(\omega_{d};0)$ is equal to the leading-order cavity decay rate $\kappa_{c}$. In this figure, $\tau=0$ corresponds to the infinite-resolution Ohmic spectral density, whereas $\tau=1\,\textrm{ns}$ is close to the IR limit.} 
\label{fig: effective spectral density}
\end{figure}

We note in Fig.\ref{fig: effective spectral density} that, as we approach the IR limit, the effective spectral density peaks more and more sharply around a discrete set of system resonant frequencies (in this example the peaks are at $2\omega_{a}-\omega_{d},$ $\omega_{a},$ $\omega_{d},$ and $3\omega_{a}$). In general, these system resonant frequencies would include nonlinear (e.g., the peaks at $2\omega_{a}-\omega_{d}$ and $3\omega_{a}$) as well as linear processes (e.g., the peaks at $\omega_{a}$ and $\omega_{d}$). In addition, the finite band width at feasible values of $\tau$ makes the ultraviolet (UV) cutoff $\Lambda$ irrelevant to the experimentally measured dynamics in most if not all situations.

Comparing this low-temperature Purcell decay rate with the drive-induced transition rates at the (conditional) steady state of the cavity, we find that
\begin{equation}
\label{Eq: KvsPurcell1}
\begin{split}
\lim_{\kappa_{c} t \rightarrow \infty}
\frac{K(t)}{\kappa_{s}}
\approx&
\frac{J(\omega_{d})}{J(\omega_{a})}
\cdot
\frac{ \omega_{d}^{2} + \omega_{a}^{2} }{ 2\omega_{d}^{2} }
\langle n_{c}(\infty) \rangle_{1}\\
\approx&
\frac{ \omega_{d}^{2} + \omega_{a}^{2} }{ 2\omega_{d}\omega_{a} }
\langle n_{c}(\infty) \rangle_{1}
\end{split}
\end{equation}
\begin{equation}
\label{Eq: KvsPurcell2}
\begin{split}
\lim_{\kappa_{c} t \rightarrow \infty}
\frac{\Gamma(t)}{\kappa_{s}}
\approx&
\frac{J(\omega_{d})}{J(\omega_{a})}
\cdot
\frac{ \omega_{d}^{2} + \omega_{a}^{2} }{ 2\omega_{d}^{2} }
\langle n_{c}(\infty) \rangle_{0}\\
\approx&
\frac{ \omega_{d}^{2} + \omega_{a}^{2} }{ 2\omega_{d}\omega_{a} }
\langle n_{c}(\infty) \rangle_{0}.
\end{split}
\end{equation}
Since $\omega_{d}$ is on the same order of magnitude as $\omega_{a}$, the equations above suggest that the drive-induced transition rates are comparable with the low-temperature Purcell decay rate when the corresponding conditional cavity occupation numbers are on the order of $1$.

Although the TCG superoperators obtained so far imply intricate effective interactions between the spin and the cavity, one can significantly simplify the master equation by focusing on the spin degrees of freedom alone. Following the assumptions and approximations made in \cite{Blais_etal_dispersive_cQED}, we can trace out the cavity degrees of freedom assuming that the cavity is approximately in conditional coherent states depending on the spin state. In particular, if the (approximate) coherent-state amplitudes are $\alpha_{g}$ and $\alpha_{e}$ for relaxed and excited spin states respectively, then we can use the results in Appendix F of the Supplementary Materials\cite{supplement} to obtain the following fourth-order TCG effective master equation for the reduced spin density matrix $\overline{\rho}_{\textrm{S}}(t) = \textrm{Tr}_{\textrm{C}} \overline{\rho}(t)$:
\begin{equation}
\label{Eq: spin EQME}
\begin{split}
\partial_{t} \overline{\rho}_{\textrm{S}}
\approx&
-
i \frac{\omega_{\textrm{S}}}{2}
\big[
\sigma_{z}
, \overline{\rho}_{\textrm{S}} \big]
+
\frac{\gamma_{\varphi_{\textrm{eff}}}}{2} D_{\sigma_{z}, \sigma_{z}} \overline{\rho}_{\textrm{S}}
+
\gamma_{\downarrow} D_{\sigma_{-}, \sigma_{+}} \overline{\rho}_{\textrm{S}}\\
&
+
\gamma_{\uparrow} D_{\sigma_{+}, \sigma_{-}} \overline{\rho}_{\textrm{S}}
\end{split}
\end{equation}
where
\begin{equation}
\label{Eq: omega_S}
\begin{split}
\omega_{\textrm{S}}
=&
\chi
+
2\big[
\chi
+
\zeta(1 + \abs{\alpha_{e}}^{2} + \abs{\alpha_{g}}^{2})
+
\frac{\lambda^{4}\kappa_{c}}{4}
\big] \textrm{Re}(\alpha_{g} \alpha_{e}^{\ast})\\
&
-
\zeta (\abs{\alpha_{g}}^{4} + \abs{\alpha_{e}}^{4})
+
\lambda^{2} \textrm{Re}\big(\epsilon_{d} (\alpha_{e}^{\ast}+\alpha_{g}^{\ast})\big)\\
&
+
2\mu
+
\textrm{Im}(\kappa^{d}_{-}+\kappa^{d}_{+})
\textrm{Re}(\alpha_{e}-\alpha_{g})
\end{split}
\end{equation}
\begin{equation}
\label{Eq: gamma_pure}
\begin{split}
\gamma_{\varphi_{\textrm{eff}}}
=&
2\big[
\chi
+
\zeta (1+\abs{\alpha_{e}}^{2}+\abs{\alpha_{g}}^{2})
-
\frac{\lambda^{4}\kappa_{c}}{4}
\big]
\textrm{Im}(\alpha_{g}\alpha_{e}^{\ast})\\
&
+
\lambda^{2} \textrm{Im}\big(
\epsilon_{d} (\alpha_{e}^{\ast} - \alpha_{g}^{\ast})
\big)
\end{split}
\end{equation}
\begin{equation}
\label{Eq: gamma_jump}
\begin{split}
\gamma_{\downarrow}
=
\kappa_{s}
+
K(t)
\qquad\qquad\qquad\qquad\qquad\quad
\gamma_{\uparrow}
=
\Gamma(t)
\end{split}
\end{equation}
with
\begin{equation}
\begin{split}
\chi
=
-
g_{ac}^{2} \big(
\frac{1}{\omega_{d} - \omega_{a}}
-
\frac{1}{\omega_{d} + \omega_{a}}
\big)
+
\zeta
\end{split}
\end{equation}
and
\begin{equation}
\begin{split}
\kappa^{d}_{\pm}
=
i \frac{\epsilon_{d} g_{ac}^{2}}{(\omega_{d}\pm \omega_{a})^{2}}
.
\end{split}
\end{equation}
Details of the derivation can be found in Appendix F of the Supplementary Materials\cite{supplement}.
In particular, we see that the results above not only reproduce all the relevant corrections in Eq.(5.4) and Eq.(5.6) of \cite{Blais_etal_dispersive_cQED} but also include all the non-RWA corrections and, most importantly, additional measurement-induced corrections to the qubit frequency $\omega_{\textrm{S}}$ and the qubit transition rates $\gamma_{\uparrow \downarrow}$. We emphasize that the measurement-induced corrections $K$ and $\Gamma$ to the effective spin transition rates have not been captured by any effective Hamiltonian methods in the literature. In fact, more sophisticated analyses based on the method in \cite{Blais_etal_dispersive_cQED} mostly focus on the effects of non-RWA terms or the introduction of additional noise or heat bath \cite{Zueco_beyond_RWA, MPT_I,Blais_dissipation_ultrastrong}, and cannot account for spin dissipation of the type of Eq.(\ref{Eq: TCG spin dissipator}) to the best of our knowledge. Therefore, the drive-induced TCG dissipators on the qubit are potentially very important for explaining measurement-induced enhancement of spin decay rates that has been observed in experiments with relatively strong readout strength \cite{Devoret_March_Meeting,Minev_Nature}.

\subsection{Near-resonant behavior and parametric dependence on the coarse-graining time scale \texorpdfstring{$\boldsymbol{\tau}$}{TEXT}}
\label{Subsec: Spin-cavity resonance and tau}
Unlike the dispersive case, if one brings the cavity mode frequency $\omega_{d}$ close to the spin transition frequency $\omega_{a}$, the TCG effective superoperators will become sensitive to the coarse-graining time scale $\tau$, and more corrections from TCG will become relevant. Alternatively, when $\tau$ is sufficiently small to be comparable with some of the system (virtual) transition frequencies, the effective EQME will also depend strongly on $\tau$, even in the dispersive limit of the model parameters.

\onecolumngrid
\vspace{8mm}

\begin{figure}[h]
\centering
\captionsetup{justification=Justified, font=footnotesize}
\includegraphics[width=0.99\textwidth]{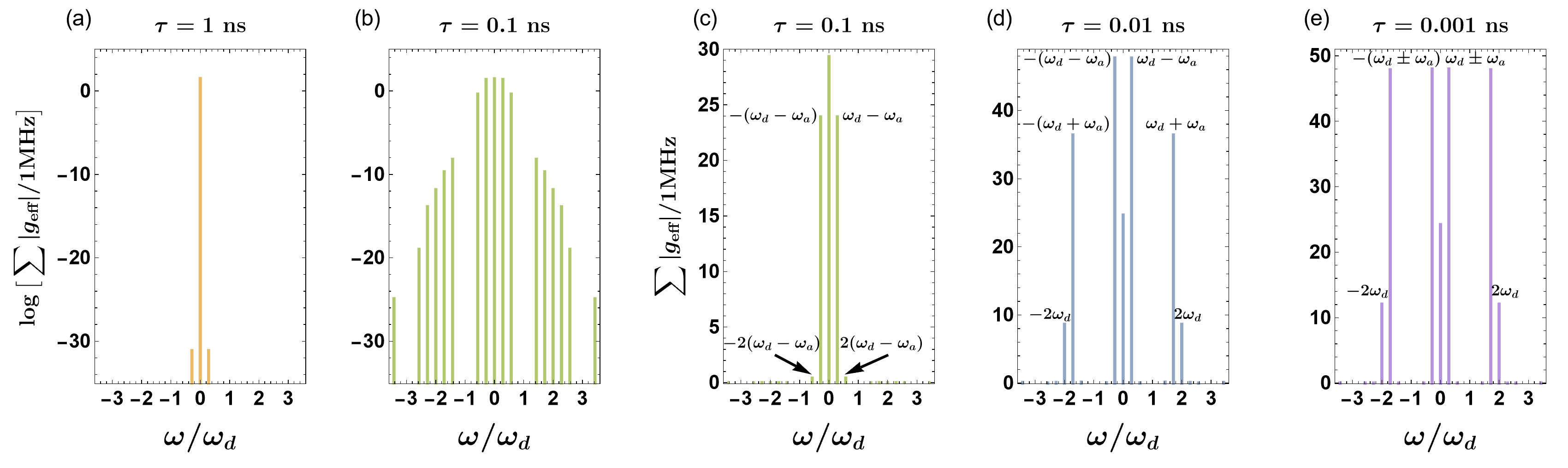}
\caption{The lumped discrete spectra of effective coupling strengths at different coarse-graining time scales $\tau$, as obtained from the third-order TCG master equation of the spin-cavity model. The vertical axes of panel (a) and panel (b) are on the log scale, whereas those of the other panels are on the linear scale. Superoperators involving the bath modes are not included in this figure, and the TCG effective spectral density will be discussed in the next subsection. These discrete spectra lump together all the superoperators at a certain frequency, and therefore does not correspond to any physical observables. Instead, they only provide a simple illustration of how the TCG effective master equation ``flows'' from one fixed point to another as one changes the value fo $\tau$.}
\label{fig: discrete spectrum}
\end{figure}

\vspace{2mm}
\twocolumngrid

For example, the spin-cavity coupling can no longer be ignored in the leading-order TCG Hamiltonian $H_{\textrm{TCG}}^{(1)}$
\begin{equation}
\begin{split}
H_{\textrm{TCG}}^{(1)}
\approx&
\int \frac{d\omega}{2\pi} \cdot \mathcal{D}_{\omega} g_{\omega} e^{-\frac{(\omega-\omega_{d})\tau^{2}}{2}}
\Big(
e^{i(\omega-\omega_{d})t} c B_{\omega}^{\dagger}\\
&
+
e^{-i(\omega-\omega_{d})t} c^{\dagger} B_{\omega}
\Big)
+
\epsilon_{d} \big( c + c^{\dagger} \big)\\
&
+
g_{ac} e^{-\frac{(\omega_{d}-\omega_{a})^{2}\tau^{2}}{2}}
\big(
e^{i(\omega_{d}-\omega_{a})t} c^{\dagger} \sigma_{-}
+
h.c.
\big)
\end{split}
\end{equation}
whereas the spin-cavity hybridization correction $H_{g_{ac}^{2}}$ becomes
\begin{equation}
\begin{split}
H_{g_{ac}^{2}}
=&
-
g_{ac}^{2} \big(
\frac{1 - e^{-(\omega_{d}-\omega_{a})^{2}\tau^{2}}}{\omega_{d} - \omega_{a}}
-
\frac{1}{\omega_{d} + \omega_{a}}
\big)
\frac{\sigma_{z}}{2} \big( 1 + 2c^{\dagger}c \big)\\
&
+
\textrm{const.},
\end{split}
\end{equation}
and the TCG dissipator $D_{\textrm{TCG}}^{(2)}$ is no longer negligible:
\begin{equation}
\begin{split}
D_{\textrm{TCG}}^{(2)}(t)
=&
-
2i g_{ac}^{2} e^{-(\omega_{d}-\omega_{a})^{2}\tau^{2}}
\frac{1 - e^{-(\omega_{d}-\omega_{a})^{2}\tau^{2}}}{\omega_{d}-\omega_{a}}\\
&\qquad\cdot
e^{2i(\omega_{a}-\omega_{d})t} D_{\sigma_{+} c, \sigma_{+} c}
+
h.c.
\end{split}
\end{equation}

Noticeably, $H_{g_{ac}^{2}}$ and $D_{\textrm{TCG}}^{(2)}(t)$ remain regular in the $\omega_{a} \rightarrow \omega_{d}$ limit:
\begin{equation}
\begin{split}
&
\lim_{\omega_{a} \rightarrow \omega_{d}}
H_{g_{ac}^{2}}\\
=&
-
g_{ac}^{2} \Big(
(\omega_{d} - \omega_{a}) \tau^{2}
-
\frac{1}{\omega_{d} + \omega_{a}}
\Big)
\frac{\sigma_{z}}{2} \big( 1 + 2c^{\dagger}c \big)\\
&
+
\mathcal{O}\big( (\omega_{d} - \omega_{a})^{2} \big)
+
\textrm{const.}
\end{split}
\end{equation}
\begin{equation}
\begin{split}
&
\lim_{\omega_{a} \rightarrow \omega_{d}}
D_{\textrm{TCG}}^{(2)}(t)\\
=&
-
2i g_{ac}^{2} (\omega_{d}-\omega_{a}) \tau^{2}
e^{2i(\omega_{a}-\omega_{d})t} D_{\sigma_{+} c, \sigma_{+} c}
+
h.c.\\
&
+
\mathcal{O}\big( (\omega_{d} - \omega_{a})^{2} \big)
\end{split}
\end{equation}
In fact, we emphasize that the superoperators generated in the perturbative expansion of a TCG master equation are always regular and do not suffer from near-resonant divergences that often plague other method such as the Schrieffer–Wolff transformation \cite{MPT_II}.

In general, all coefficients in the TCG effective master equation depend smoothly on the coarse-graining time scale $\tau$. Usually, the value of a particular coefficient will approach certain constants (i.e., the UV and IR fixed points) when $\tau$ is much smaller or much greater than certain threshold scales. As a result, there typically exist a few effective models that depend weakly on $\tau$ within their respective range of validity, and those effective models are bridged by relatively rapid transitions from one to another when the time resolution is close to one of the threshold time scales. For instance, if we sum up the absolute values of the superoperator coefficients obtained from the third-order TCG master equation of the spin-cavity model, then the resulting lumped discrete spectra of coupling strengths at various coarse-graining time scales $\tau$ can be seen in Fig.\ref{fig: discrete spectrum}.
In particular, we note that
\begin{itemize}
    \item With large $\tau$, as shown in panel (a) of Fig.\ref{fig: discrete spectrum}, we would be in the IR limit where only the superoperators that are static in the rotating frame remain significant. However, these include effective dissipators as well as effective Hamiltonian corrections in general, and account for various virtual processes that cannot be resolved by the time resolution.
    \item When $\tau$ decreases sufficiently to resolve $(\omega_{d}-\omega_{a})^{-1} = 0.080\,\textrm{ns}$, as shown in panel (b) and (c), superoperators at frequencies $\pm(\omega_{d}-\omega_{a})$ and $\pm2(\omega_{d}-\omega_{a})$ begin to have noticeable effects on the observed dynamics, and the coefficients of TCG superoperators drop quickly as the corresponding frequencies become much greater in magnitude than $\tau^{-1}$.
    \item When $\tau$ begins to resolve $(\omega_{d}+\omega_{a})^{-1} = 0.013\,\textrm{ns}$ and $(2\omega_{d})^{-1} = 0.011\,\textrm{ns}$, as shown in panel (d), superoperators at frequencies $\pm(\omega_{d}+\omega_{a})^{-1}$ and $(2\omega_{d})^{-1}$ become important, whereas those at frequencies $\pm2(\omega_{d}-\omega_{a})$ get suppressed back down again. Both $\tau=0.1\,\textrm{ns}$ and $\tau=0.01\,\textrm{ns}$ are close to some ``threshold time scales'' discussed earlier, and the effective dynamics are relatively sensitive to $\tau$ near those thresholds.
    \item Finally, with sufficiently small $\tau$, as shown in panel (e), we approach the UV limit which is exactly described by the infinite-resolution Hamiltonian $H_{I}$.
\end{itemize}

As stated before, we will focus on the IR limit of the TCG dynamics in the remaining of this work.
In addition, for arbitrary values of the spin transition frequency $\omega_{a}$, the TCG perturbative expansion remains valid as long as the coarse-graining time scale $\tau$ is small compared to the inverse of the coefficients of the terms in $H_{I}$. For example, we have
\begin{equation}
\begin{split}
&
g_{ac}^{2} \frac{1 - e^{-(\omega_{d}-\omega_{a})^{2}\tau^{2}}}{\omega_{d} - \omega_{a}}\\
\le&
g_{ac}^{2} \tau \cdot \max\Big[
2 (\omega_{d}-\omega_{a})\tau e^{-(\omega_{d}-\omega_{a})^{2}\tau^{2}}
\Big]
=
\sqrt{\frac{2}{e}} g_{ac}^{2} \tau 
\end{split}
\end{equation}
which shows that the magnitude of the prefactor at order $g_{ac}^{2}$ is bounded from above by $\sqrt{\frac{2}{e}} g_{ac} \tau$ times the prefactor at order $g_{ac}^{1}$. And in general, if we denote the largest coupling (drive) strength in $H_{I}$ by $g_{\max}$, then $\tau g_{\max}$ can be used in the TCG perturbative expansion as the small control parameter.

\subsection{Numerical solutions to the EQME}
\label{Subsec: spin-cavity numerical solutions}

To solve the TCG master equation for the driven dissipative spin-cavity system, we use the conditional Husimi representation $\overline{Q}^{ab}_{\mu}(\phi, n)$ of the density matrix $\overline{\rho}$ discussed in Appendix G of the Supplementary Materials\cite{supplement}, where for any $\mu \in \{x, y, z\}$ there is a complete set of three independent functions for $(a,b) \in \{(0,0),(1,1),(0,1)\}$. The TCG master equation then gets translated into a set of partial differential equations for $\overline{Q}^{ab}_{z}(\phi, n)$ which we solve numerically in this subsection. The same QME can also be solved by truncating the cavity mode Hilbert space or applying a cumulant expansion in the cavity quadrature variables, as we do in \Sec{Sec: the readout problem}. Here we choose to numerically solve for the evolution of the conditional Husimi Q functions, to allow us to get physical insight into measured observables in a heterodyne measurement. $\overline{Q}^{aa}_{\mu}(\phi, n)$ can be interpreted as the joint probability distribution of the spin being found in state $|\sigma_{\mu} = a\rangle$ and the quadrature variables $(\phi, n)$ measured in an an ideal heterodyne record~\cite{Muller_Evading_Vacuum_Noise, STENHOLM1992233, Shapiro_quantum_theory_optical_communication}. The resulting solutions are closely related to experimental observables and can be conveniently used to infer the mutual information between the spin state and the cavity mode pointers.

In the rest of this subsection, we first demonstrate the convergence of the TCG perturbation theory by comparing solutions to the EQME at different orders with that obtained by directly time-coarse graining the exact numerical solution to the microscopic von-Neumann equation. Once the TCG method has been validated, we continue to discuss details of the spin dynamics, relating numerical results to the analytical formulas presented in Subsection \ref{Subsec: bath effects}. In particular, we focus on how the cavity state affects the dephasing, relaxation, and excitation of the spin. Finally, we take a closer look at the cavity mode dynamics and how it is conditioned by the spin state. The impact of higher-order TCG corrections on the cavity mode cumulants is emphasized.
All the simulations presented in this subsection are performed assuming $\tau = 3\textrm{ns}$ as well as all the other parameter values in Eq.(\ref{Eq: dispersive parameters}).

\subsubsection{Comparison with direct coarse-graining of the exact numerical solution}

As discussed previously, we numerically solve the TCG effective master equation at each order in terms of the conditional Husimi functions $\overline{Q}_{z}^{ab}(\phi, n)$.
In order to demonstrate convergence of the TCG perturbation theory, we also numerically solve the following microscopic master equation for the exact interaction-picture density matrix $\rho(t)$ of the spin-cavity system:
\begin{equation}
\label{Eq: exact ME}
\begin{split}
\partial_{t} \rho(t)
=
\big[
H_{I}(t)
,
\rho(t)
\big]
+
\kappa_{c} D_{c, c^{\dagger}} \rho(t).
\end{split}
\end{equation}
In particular, we assume the initial condition
\begin{equation}
\label{Eq: ICforTCG}
\begin{split}
&
\rho\big(-9 \textrm{ ns}\big)\\
=&
\Big(
\sqrt{\frac{2}{5}} |0\rangle
+
\sqrt{\frac{3}{5}} |1\rangle \Big)
\Big(
\sqrt{\frac{2}{5}} \langle 0|
+
\sqrt{\frac{3}{5}} \langle 1|
\Big)
\otimes
|0\rangle_{c}\langle 0|_{c}
\end{split}
\end{equation}
at time $t_{i} = -9 \textrm{ ns}$ so that the solution to Eq.(\ref{Eq: exact ME}) can be manually time-coarse grained to provide the initial conditions for the TCG master equations at time $t = 0$.

\begin{figure}[h]
\centering
\captionsetup{justification=Justified, font=footnotesize}
\includegraphics[width=0.98\textwidth]{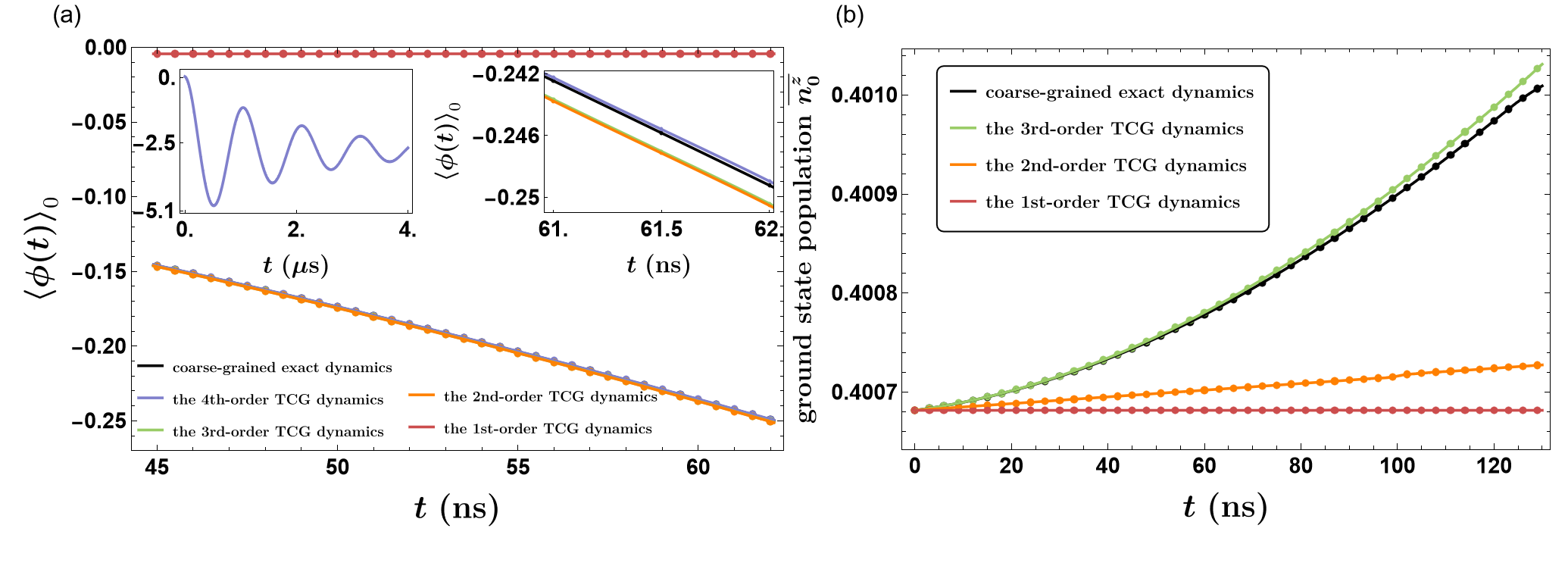}
\caption{Numerical results obtained from solving the PDE for the conditional Husimi functions $\overline{Q}^{ab}_{\mu}(\phi, n)$ assuming the parameters in Eq.(\ref{Eq: dispersive parameters}). In particular, TCG is performed with a time resolution of $\tau = 3\textrm{ns}$. Panel (a) shows the time evolution of the conditional expectation value of the quadrature variable $\phi$ if the spin is in the ground state $|0\rangle$; the inset in the upper-right corner shows the zoomed-in view of all the simulated dynamics around $t=61.5\textrm{ns}$ whereas the inset in the upper-left corner shows the zoomed-out view of the 4th-order TCG dynamics for $t$ from $0$ to $4\mu\textrm{s}$. Panel (b) shows the time evolution of the ground state population of the spin. For the parameters in Eq.(\ref{Eq: dispersive parameters}), increasing order in the TCG perturbative expansion gives rise to solutions that gradually converge to the exact coarse-grained dynamics. For different dynamical variables, however, different orders of the TCG expansion are required to attain the same level of accuracy.}
\label{fig: TCG convergence}
\end{figure}

In addition, directly coarse-graining the UV-limit ($\tau=0$) solution $Q_{z}^{aa}(t)$ gives us a target of convergence against which solutions to the TCG master equations at different orders can be compared. In Fig.\ref{fig: TCG convergence} we plot the time-evolution of different observables. Panel (a) of Fig.\ref{fig: TCG convergence} displays the dynamics of the conditional expectation value of the cavity quadrature variable $\phi$ when the spin is in the ground state. We see in the main plot that the coarse-grained exact cavity dynamics is reproduced reasonably well with the second order TCG master equation, and the accuracy is even better if we go to the 4th order, as shown in the inset in the upper-right corner. The inset in the upper-left corner shows a zoomed-out view of $\langle\phi(t)\rangle_{0}$ over longer time scales, which confirms one's expectation of a typical damped oscillation (the averaged exact dynamics has not been calculated over this longer time period since that is too time-consuming in the Husimi-PDE formalism). On the other hand, drive-induced spin transitions from the third-order TCG are indispensable for understanding the spin dynamics. As shown in panel (b) of Fig.\ref{fig: TCG convergence}, the time evolution of the coarse-grained ground state population $\overline{n_{0}^{z}}(t)$ of the spin cannot be accurately described in terms of the Purcell decay of the spin which appears at the second order, and drive-induced spin level transitions (\Eq{Eq: formula for K and Gamma}) are necessary for capturing its dynamics (see also \Eq{Eq: KvsPurcell1} and \Eq{Eq: KvsPurcell2}).

\onecolumngrid

\twocolumngrid

For the examples shown in Fig.\ref{fig: TCG convergence}, the total time of simulation is limited by the maximum time step size required to numerically solve the infinite-resolution master equation. 
To quote some performance numbers for examples studied here, depending on the model parameters and the time period to simulate, the speedup by EQME can range from about $100$-fold to over $1000$-fold. Perhaps more importantly, in the analysis in Appendix A of the Supplementary Materials\cite{supplement} we argue that it is precisely the coarse-grained cavity state represented by $\overline{Q}_{z}^{00} + \overline{Q}_{z}^{11}$ that describes more faithfully the distributions that can be expected to be observed in realistic readout experiments with finite time resolution, even though further digital integration over the classical readout signal is often applied after the quantum measurement to increase the SNR.

\subsubsection{The EQME and the cavity state}

\begin{figure}[h!]
\captionsetup{justification=Justified, font=footnotesize}
\centering
\captionsetup{justification=Justified, font=footnotesize}
\includegraphics[width=0.48\textwidth]{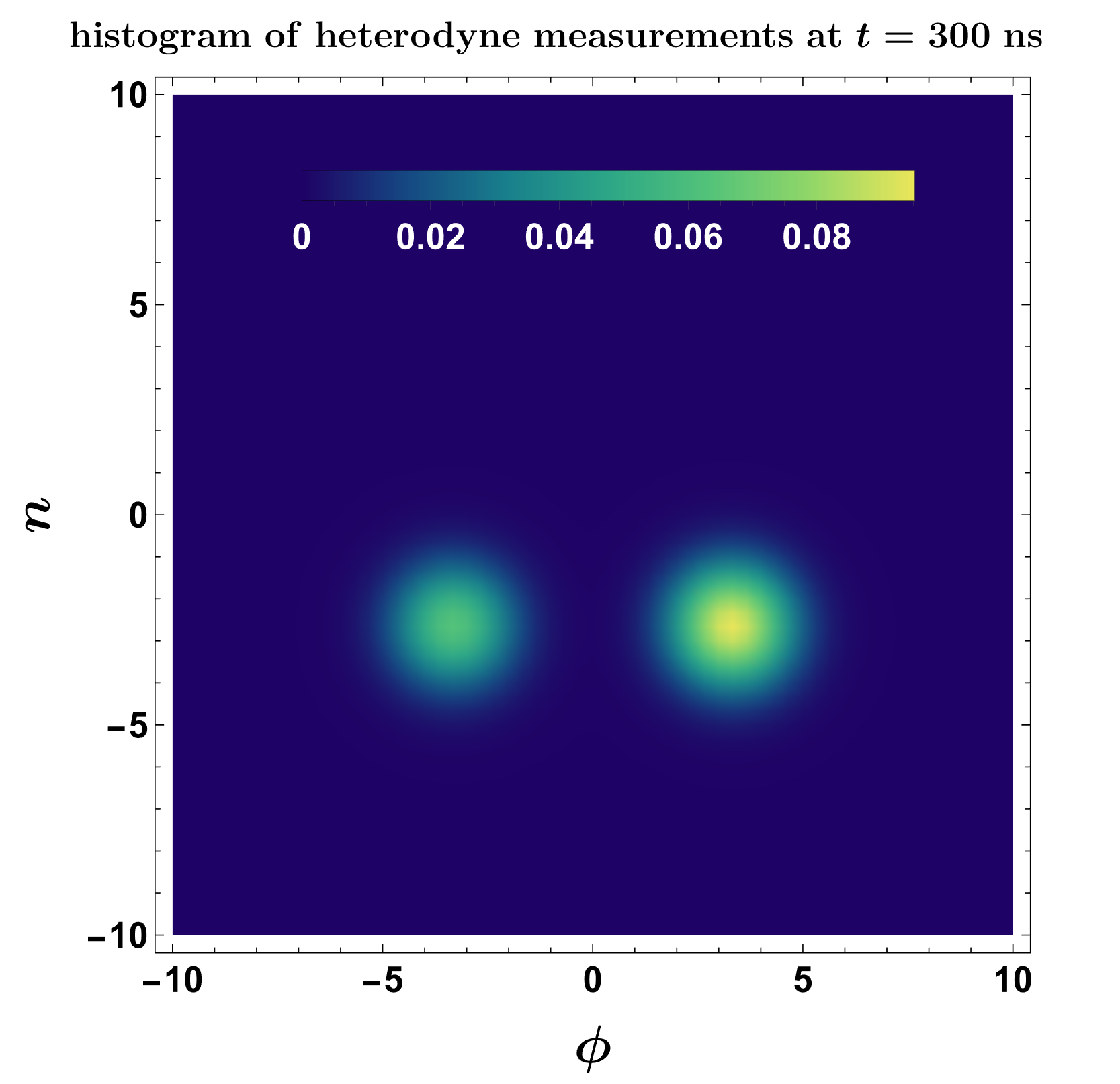}
\caption{The simulated histogram of heterodyne measurement results for the dispersive spin readout system at time $t = 300\textrm{ns}$ assuming the set of parameters in Eq.(\ref{Eq: dispersive parameters}). The left and right high-probability clusters correspond predominantly to the ground and the excited states of the spin respectively.}
\label{fig: histogram T300}
\end{figure}

\begin{figure}[!h]
\centering
\captionsetup{justification=Justified, font=footnotesize}
\includegraphics[width=0.48\textwidth]{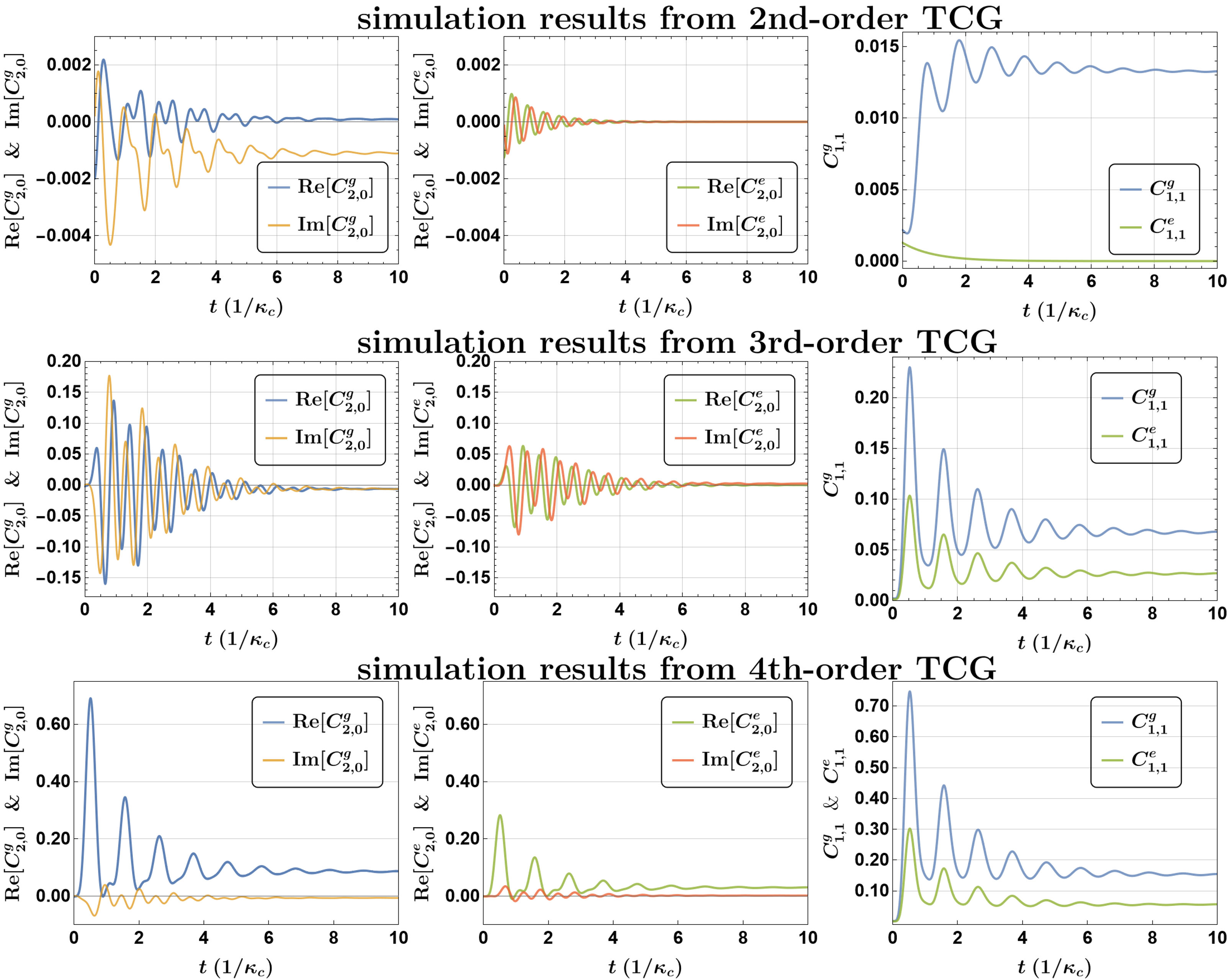}
\caption{The second-order cumulants of the cavity pointer states calculated from EQME at different orders. In particular, the fourth-order superoperators provide important corrections for the pointer state deformation from a perfect coherent state, although they do not have much effects on the spin dynamics or the position of the cavity state in the phase space.}
\label{fig: SC_Cumulants}
\end{figure}

For the initial conditions in Eq.(\ref{Eq: ICforTCG}) and the model parameters in Eq.(\ref{Eq: dispersive parameters}), a typical histogram of simulated heterodyne measurement results at $t=300\textrm{ns}$ is shown in Fig.\ref{fig: histogram T300}, where the cavity pointer states are both close to coherent states, as one would expect in a typical readout experiment. In particular, the positions of the pointer states on the $\phi-n$ plane can be predicted relatively accurately by the second-order EQME which accounts for the cavity drive and the dispersive shift of the cavity mode frequency. The precise shapes of the pointer states, however, can receive considerable corrections from the third- and the fourth-order EQME, depending on parameters. This can be seen by directly visualizing the evolution of second-order cumulants
\[
C_{2,0}
\equiv
\langle a^{2} \rangle
-
\langle a \rangle^{2};
\qquad
C_{1,1}
\equiv
\langle a^{\dagger}a \rangle
-
\abs{\langle a \rangle}^{2}
\]
whose time evolution is shown in Fig.\ref{fig: SC_Cumulants}.

These second-order cumulants characterize the squeezing of the pointer states, and will be sufficient for describing the shapes of the pointer states as long as their Husimi functions are roughly of elliptical shapes in the phase space (i.e., the cavity stays close to some gaussian state depending on the spin state).

Once we have solved for the positions and shapes of the cavity pointer states in the phase space, we can use them to calculate the mutual information between the heterodyne observables and the spin energy levels.
More precisely, with the joint probability distributions $\overline{Q}^{00}_{z}$ and $\overline{Q}^{11}_{z}$, we can calculate the mutual information shared by the random variables $\sigma_{\mu}$ and $(\phi, n)$ according to the equation
\begin{equation}
\begin{split}
&
I\big( \sigma_{\mu}, (\phi,n) \big)\\
=&
\sum_{a = 0, 1}
\int d\phi dn \cdot
\overline{Q}_{\mu}^{aa}(\phi, n) \log_{2} \frac{ \overline{Q}_{\mu}^{aa}(\phi, n) }{ \overline{Q}_{\mu} (\phi, n) \cdot \overline{n_{a}^{z}} }
\end{split}
\end{equation}
where $\overline{Q}_{\mu} (\phi, n) \equiv \overline{Q}_{\mu}^{00}(\phi, n) + \overline{Q}_{\mu}^{11}(\phi, n)$ and the log is taken with base $2$ so that $I\big( \sigma_{\mu}, (\phi,n) \big)$ is bounded between $0$ and $1$, with $I = 0$ meaning complete independence of the two variables and $I = 1$ representing the limit where ideal heterodyne measurements of the cavity state is equivalent to projective measurements of the $\mu$-component of the spin. Conceptually, one can think of $I\big( \sigma_{\mu}, (\phi,n) \big)$ as indicating the amount of information about the $\mu$-component of the spin stored in the cavity state, and we will see in the next subsection that this amount of information is closely related to the dephasing of the spin.

In other words, although the high-order terms in the EQME are not necessary for simulating the evolution of the cavity quadrature variables during their transient period. We would nonetheless need them for obtaining more accurate estimates of the amount of spin-state information that one can in principle learn from measuring the state of the cavity mode. And the same is true if we would like to more accurately calculate the corresponding backaction to the spin state in the form of a time-dependent dephasing rate. On the other hand, over the much longer time period of spin dissipation, the higher-order terms in the QME also has noticeable impact on the expectation values of the cavity quadrature variables, as will be discussed in more detail in \Sec{Sec: the readout problem}.

\begin{figure}[h]
\centering
\captionsetup{justification=Justified, font=footnotesize}
\includegraphics[width=0.92\textwidth]{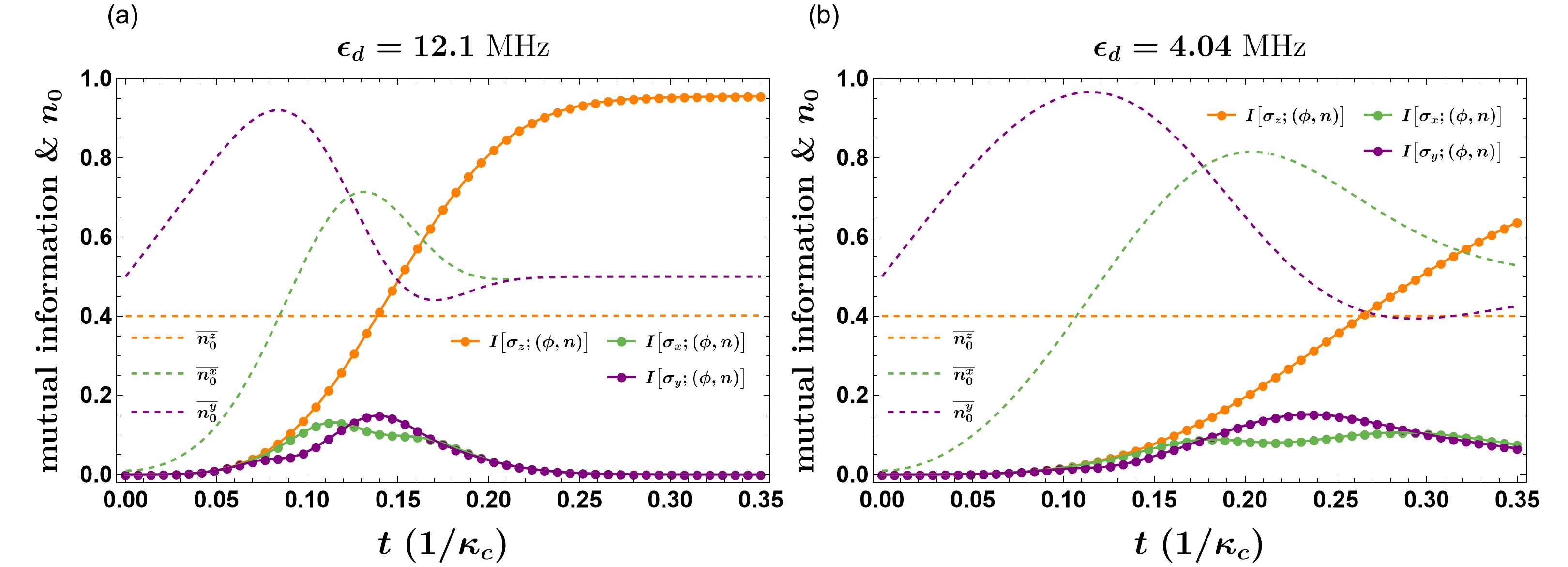}
\caption{The time evolution of the spin state population and spin-cavity mutual information in different spin basis for the dispersive readout model, as obtained from the third-order TCG master equation. The cavity is initially empty whereas the spin at the initial time $t=-9\textrm{ns}$ is a coherent superposition of the two pointer states: $|\Psi(-9\textrm{ns})\rangle = \sqrt{0.4} |0\rangle + \sqrt{0.6} |1\rangle$. Panel (a) shows the simulated dynamics with the parameters given in Eq.(\ref{Eq: dispersive parameters}), whereas panel (b) shows the same same variables for a smaller readout drive.}
\label{fig: mutual information}
\end{figure}

As shown in Fig.\ref{fig: mutual information}, the mutual information with $\sigma_{z}$ rises quickly above $0.9$, whereas those with $\sigma_{x}$ and $\sigma_{y}$ first increase together with $I\big( \sigma_{z}, (\phi,n) \big)$ before they get suppressed back down to near zero during the same period of time
\begin{equation}
\label{Eq: tau_info}
\begin{split}
\tau_{\textrm{info}}
\approx
\frac{1}{\abs{\chi}} \arccos\big( 1 - \frac{\abs{\chi}}{\epsilon_{d}} \big)
\xrightarrow[\epsilon_{d} \gg \abs{\chi}]{}
\sqrt{\frac{2}{\epsilon_{d}\abs{\chi}}}
\end{split}
\end{equation}
in the $\abs{\chi} \gg \kappa_{c}$ limit. As will be explained in the next subsection, $\tau_{\textrm{info}}$ is on the same order of magnitude as the spin dephasing time scale $T_{2}$, i.e., the time scale for $\overline{n_{0}^{x}}(t)$ and $\overline{n_{0}^{y}}(t)$ to stabilize around $0.5$.
In fact, while the cavity mode acquires information about the spin state in a certain direction, the spin also receives backaction from the cavity mode which reduces its coherence (manifest in Fig.\ref{fig: mutual information} as the damping of the oscillations in $\overline{n_{0}^{x}}(t)$ and $\overline{n_{0}^{y}}(t)$). However, as discussed in the next subsection as well as in the past literature \cite{Blais_etal_dispersive_cQED, Gambetta_measurement_induced_dephasing}, the measurement-induced dephasing can be negative during some periods of time, signaling the backflow of spin information from the cavity. We will revisit this point after discussing the spin dynamics in Subsection \ref{Subsec: TCG spin dynamics}. In particular, we will find that there are also measurement-induced spin energy jumps whose rates can be temporarily negative, which is a phenomenon that has not been captured by any effective Hamiltonian methods in the literature.

\onecolumngrid

\twocolumngrid

\subsubsection{The EQME and the spin dynamics}
\label{Subsec: TCG spin dynamics}

\begin{figure}[!h]
\centering
\captionsetup{justification=Justified, font=footnotesize}
\includegraphics[width=0.48\textwidth]{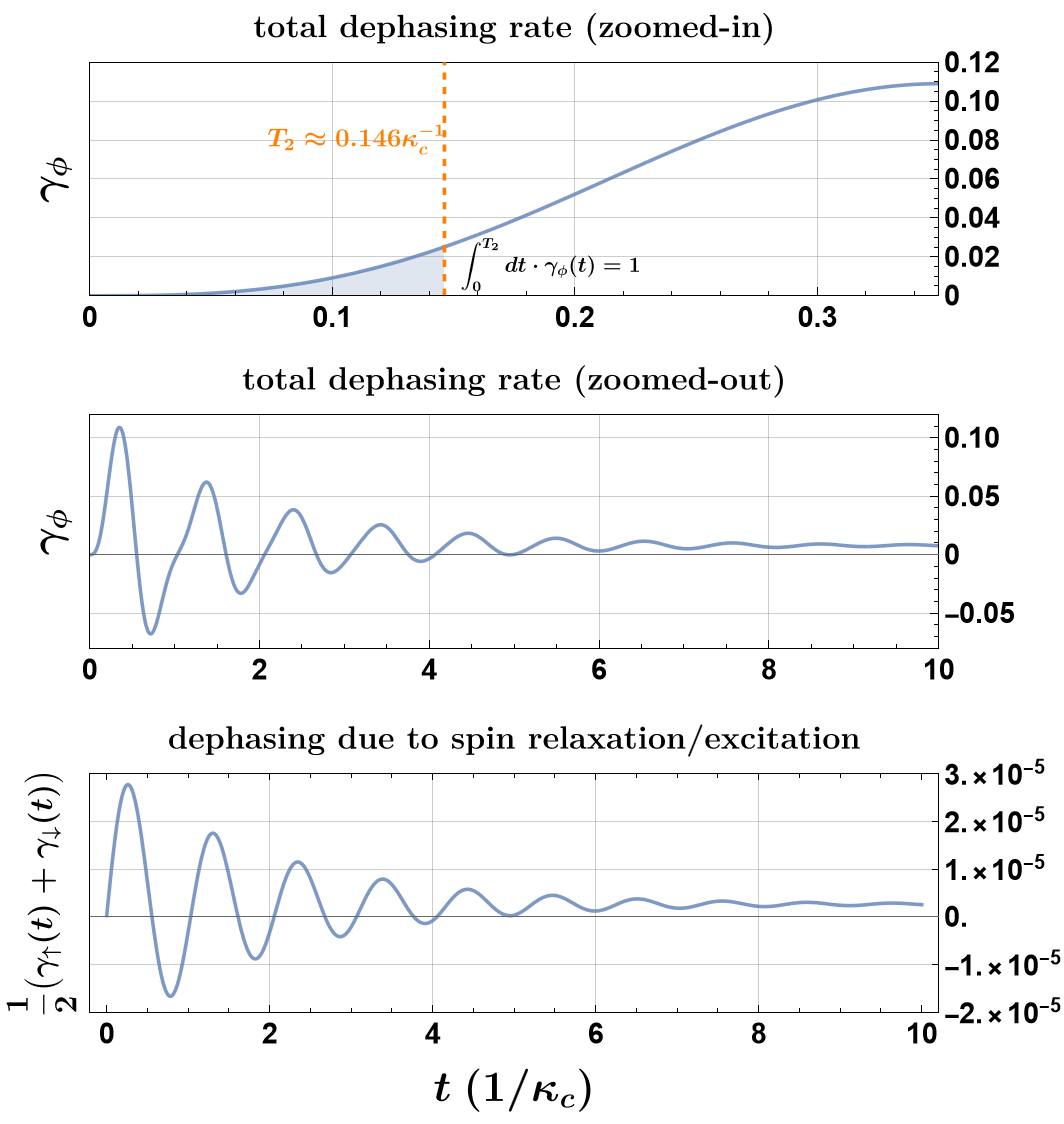}
\caption{Time dependence of the total dephasing rate $\gamma_{\phi}(t)$ as well as the quantum-jump-induced dephasing rate $\frac{1}{2}\big(\gamma_{\uparrow}(t) + \gamma_{\downarrow}(t)\big)$ calculated from Eq.(\ref{Eq: gamma_pure}) and Eq.(\ref{Eq: gamma_jump}) with numerically solved conditional cavity states assuming the initial condition in Eq.(\ref{Eq: ICforTCG}) and the parameters in Eq.(\ref{Eq: dispersive parameters}). The vertical axes assume the unit of GHz.}
\label{fig: GammaPhiVsTime}
\end{figure}

As shown in Subsection \ref{Subsec: bath effects}, the cavity degrees of freedom can be traced out from the EQME, which gives us the following effective master equation for the spin in the interaction picture:
\begin{equation*}
\begin{split}
\partial_{t} \overline{\rho}_{\textrm{S}}
\approx&
-
i \frac{\omega_{\textrm{S}}(t)}{2}
\big[
\sigma_{z}
, \overline{\rho}_{\textrm{S}} \big]
+
\frac{\gamma_{\varphi_{\textrm{eff}}}(t)}{2} D_{\sigma_{z}, \sigma_{z}} \overline{\rho}_{\textrm{S}}\\
&
+
\gamma_{\downarrow}(t) D_{\sigma_{-}, \sigma_{+}} \overline{\rho}_{\textrm{S}}
+
\gamma_{\uparrow}(t) D_{\sigma_{+}, \sigma_{-}} \overline{\rho}_{\textrm{S}}
\end{split}
\end{equation*}
if we ignore the superoperators that are exponentially suppressed by the TCG operation. Notice in particular that the spin frequency shift $\omega_{\textrm{S}}$, the pure dephasing rate $\gamma_{\varphi_{\textrm{eff}}}$, and the relaxation/excitation rates $\gamma_{\downarrow \uparrow}$ are all time-dependent in general due to their dependence on the cavity state, as shown in Fig.\ref{fig: GammaPhiVsTime}. In particular, all of these dephasing and transition rates can be temporarily negative. However, the temporary negativity of the dephasing and transition rate are not unphysical, but rather capture the time-dependent two-way exchange of information between the spin and the cavity mode as well as the two-way exchange of energy between the spin and the drive, as will be discussed later in this subsection. In the literature, such negative qubit dephasing rates are sometimes considered as manifestation of the non-Markovianity of the dynamics of the reduced spin system~\cite{Breuer2009, Megier2017, Rivas2010, Shrikant2018}.

For the set of parameters in Eq.(\ref{Eq: dispersive parameters}), the spin QME coefficients defined in Eq.(\ref{Eq: omega_S}-\ref{Eq: gamma_jump}) of Subsection \ref{Subsec: bath effects} are dominated by the following contributions:
\begin{equation}
\label{Eq: spin_ME_parameters}
\begin{split}
&
\omega_{\textrm{S}}(t)
\approx
-
\frac{2g_{ac}^{2} \omega_{a}}{\omega_{d}^{2} - \omega_{a}^{2}}
\Big(
1
+
2 \textrm{Re}\big[ \alpha_{g}(t) \alpha_{e}^{\ast}(t) \big]
\Big)\\
&
\gamma_{\varphi_{\textrm{eff}}}(t)
\approx
-
\frac{4g_{ac}^{2} \omega_{a}}{\omega_{d}^{2} - \omega_{a}^{2}} \textrm{Im}\big[ \alpha_{g}(t) \alpha_{e}^{\ast}(t) \big]\\
&
\gamma_{\downarrow}(t)
\approx
\frac{4\kappa_{c} g_{ac}^{2}\omega_{a}\omega_{d}}{(\omega_{d}^{2}-\omega_{a}^{2})^{2}}
-
\frac{4\epsilon_{d}g_{ac}^{2}\big(\omega_{d}^{2}+\omega_{a}^{2}\big)}{\big(\omega_{d}^{2}-\omega_{a}^{2}\big)^{2}} \textrm{Im}\big[\alpha_{e}(t)\big]\\
&
\gamma_{\uparrow}(t)
\approx
-
\frac{4\epsilon_{d}g_{ac}^{2}\big(\omega_{d}^{2}+\omega_{a}^{2}\big)}{\big(\omega_{d}^{2}-\omega_{a}^{2}\big)^{2}} \textrm{Im}\big[\alpha_{g}(t)\big]
\end{split}
\end{equation}

In particular, we notice that on average, $\omega_{\textrm{S}}(t)$ and $\gamma_{\varphi_{\textrm{eff}}}(t)$ are on the same order of magnitude for more than one photon in the cavity, and they are both much greater than the drive-induced spin transition rates $\gamma_{\downarrow \uparrow}$.


Therefore, the spin dynamics take place on two distinct time scales: a short dephasing time $T_{2}$ determined by the total dephasing rate $\gamma_{\phi} \equiv \gamma_{\varphi_{\textrm{eff}}} + \frac{1}{2}(\gamma_{\downarrow}+\gamma_{\uparrow})$, and a relatively long relaxation/excitation time $T_{1}$ determined by $\gamma_{\uparrow\downarrow}$ alone. If we define the dephasing time $T_{2}$ during readout by the relation
\begin{equation}
\begin{split}
\int_{0}^{T_{2}} dt \cdot \gamma_{\phi}(t) = 1
\end{split}
\end{equation}
then for parameters on the same order of magnitude as those in Eq.(\ref{Eq: dispersive parameters}), we have
\begin{equation}
\begin{split}
T_{2}
\approx&
\sqrt{\frac{\sqrt{2}}{ \epsilon_{d} \abs{\chi} }}
+
\frac{1}{12}\sqrt{\frac{\abs{\chi}}{\sqrt{2}\epsilon_{d}^{3}}}
+
\frac{7\kappa_{c}}{30\sqrt{2}\abs{\chi} \epsilon_{d}}\\
&
+
\frac{59\kappa_{c}^{2}}{1800\cdot 2^{\frac{1}{4}} (\abs{\chi} \epsilon_{d})^{\frac{3}{2}}}
\end{split}
\end{equation}
where the first term is the dominant contribution and is indeed on the same order of magnitude as information acquisition time in Eq.(\ref{Eq: tau_info}).
The spin relaxation/excitation time scale $T_{1}$, on the other hand, can be characterized by
\begin{equation}
\label{Eq:spinmodelrates}
\begin{split}
T_{1}
\equiv&
\frac{1}{\gamma_{\downarrow}(\infty) + \gamma_{\uparrow}(\infty)}\\
\approx&
\frac{(\omega_{d}^{2}-\omega_{a}^{2})^{2}}{4\kappa_{c}g_{ac}^{2}}
\Big[
\omega_{a}\omega_{d}
+
\frac{\epsilon_{d}^{2} \big(\omega_{d}^{2}+\omega_{a}^{2}\big)}{\abs{\chi}^{2} + (\frac{\kappa_{c}}{2})^{2}}
\Big]^{-1}
\end{split}
\end{equation}
since the cavity state stabilizes over a much smaller time scale $\kappa_{c}^{-1} \ll T_{1} \sim \gamma_{\uparrow\downarrow}^{-1}$.
In terms of the (semi-classical) steady-state cavity occupation number
\begin{equation}
\begin{split}
n_{c}^{\textrm{ss}}
\equiv
\lim_{\quad t \gg \kappa_{c}^{-1}}
\langle n_{c}(t) \rangle
\approx
\frac{
\abs{\epsilon_{d}}^{2}
}{
\chi^{2}
+
\frac{\kappa_{c}^{2}}{4}
}
,
\end{split}
\end{equation}
the inverse $T_{1}$ time can in general be written as a power series
\begin{equation}
\label{Eq: InvT1_expansion}
\begin{split}
T_{1}^{-1}
=
\kappa_{s}
\big(
1
+
c_{1} n_{c}^{\textrm{ss}}
+
c_{2} (n_{c}^{\textrm{ss}})^{2}
+
\cdots
\big)
\end{split}
\end{equation}
where performing TCG up to the 5th order gives us
\begin{equation}
\begin{split}
c_{1}
\approx
\frac{ \omega_{d}^{2} + \omega_{a}^{2} }{ \omega_{d}\omega_{a} }
-
\frac{2g_{ac}^{2}}{\omega_{a}\omega_{d}}
\Big[
\frac{8\omega_{a}^{2}(\omega_{a}^{2}+\omega_{d}^{2})}{(\omega_{d}^{2}-\omega_{a}^{2})^{2}}
-
1
\Big]
\end{split}
\end{equation}
\begin{equation}
\begin{split}
c_{2}
\approx
-
\frac{
2g_{ac}^{2} \big(
7 \omega_{a}^{4}
+
10 \omega_{a}^{2} \omega_{d}^{2}
-
\omega_{d}^{4}
\big)
}{
\omega_{a}\omega_{d}\big(\omega_{d}^{2} - \omega_{a}^{2}\big)^{2}
}
\end{split}
\end{equation}
We emphasize that the dependence of $T_{1}$ on the drive strength $\epsilon_{d}$ is due to the emergent incoherent spin level transitions in the effective EQME, which has not been captured by analytical models before. 

\begin{figure}[!h]
\centering
\captionsetup{justification=Justified, font=footnotesize}
\includegraphics[width=0.48\textwidth]{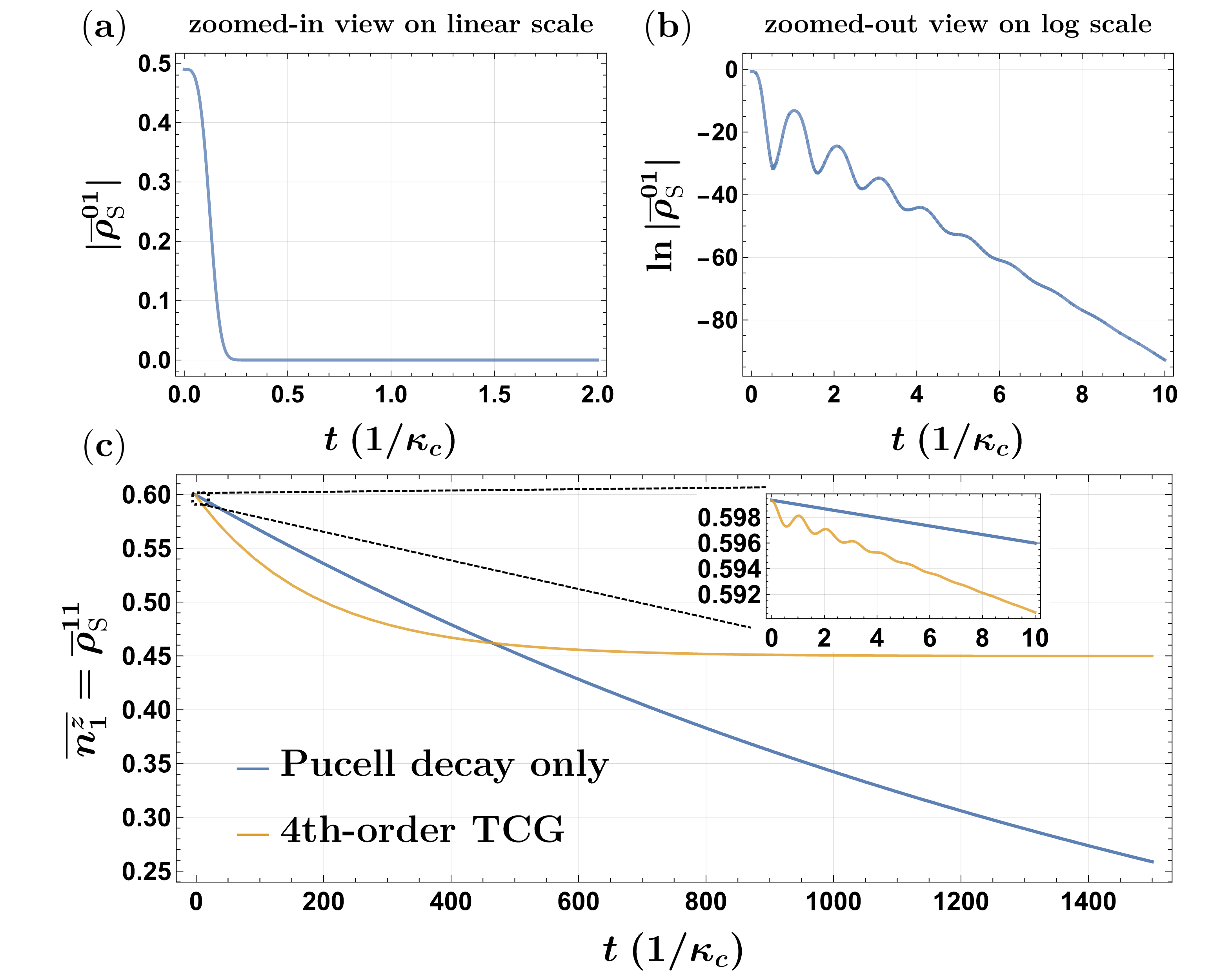}
\caption{The numerically simulated spin dynamics.}
\label{fig: SpinDynamics}
\end{figure}

As can be seen in Fig.\ref{fig: GammaPhiVsTime} and Fig.\ref{fig: SpinDynamics}, while the temporarily negative $\gamma_{\phi}(t)$ and $\gamma_{\uparrow\downarrow}(t)$ lead to revival of spin coherence and reversed level transitions during the corresponding periods in time, they are nonetheless limited to either a short transient period in the early dynamics or longer periods where their effects are rendered negligible by the corresponding dynamics. When the initial condition for the coarse-grained density matrix is taken from a physical domain (which is a subset of the density matrix space and depends on the coarse-graining time scale), the temporarily negative rates do not produce any unphysical states. In fact, throughout the time evolution, there is still strong directionality in the exchange of information and energy although they are two-way in principle.


In addition, we see that the pure dephasing rate $\gamma_{\varphi_{\textrm{eff}}}$ and the drive-induced transition rates $\gamma_{\uparrow\downarrow}$ are responsible for physics at two very distinct time scales: the decay of the off-diagonal matrix element $\abs{\overline{\rho}_{\textrm{S}}^{01}(t)} \propto e^{-\int_{0}^{t} dt^{\prime} \cdot \gamma_{\phi}(t^{\prime})}$ is controlled by $\gamma_{\phi}$ and takes about $100\textrm{ns}$ whereas the drive-induced transitions of the diagonal matrix elements $\overline{\rho}_{\textrm{S}}^{00}$ and $\overline{\rho}_{\textrm{S}}^{11}$ occur over about $100\mu\textrm{s}$. However, they can be obtained in a unified fashion as corrections that occur at different orders in the TCG perturbative expansion. Usually, the higher-order corrections are responsible for phenomenology over longer time periods. Lastly, we note that this analysis does not include additional loss channels acting locally on the qubit itself, which in many applications can not be ignored. 
\begin{figure}[!h]
\centering
\captionsetup{justification=Justified, font=footnotesize}
\includegraphics[width=0.48\textwidth]{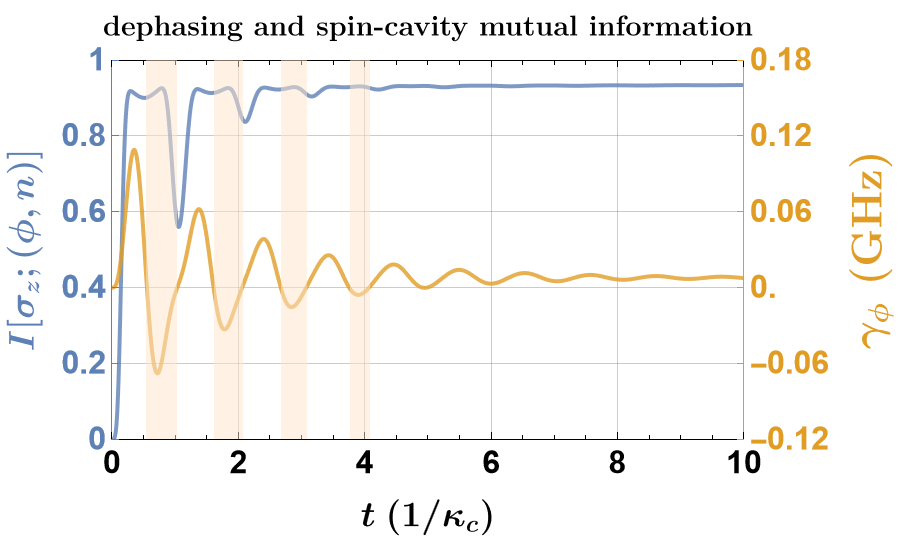}
\caption{Time evolution of the dephasing rate $\gamma_{\phi}(t)$ and the mutual information $I\big( \sigma_{z}, (\phi,n) \big)$.}
\label{fig: dephasing_and_MIZ}
\end{figure}

From another perspective, the dependence of $\gamma_{\varphi_{\textrm{eff}}}$ and $\gamma_{\uparrow\downarrow}$ on $\alpha_{g/e}(t)$ can be regarded as representing channels of information/energy exchange that are controlled by the entire history of the readout drive (resolved up to the coarse-graining time scale $\tau$). In fact, as mentioned in the previous subsection, the time-dependent spin dephasing is modulating the exchange between quantum information of the spin and the mutual information $I\big( \sigma_{z}, (\phi,n) \big)$, as shown in Fig.\ref{fig: dephasing_and_MIZ} where we see that the time periods of negative $\gamma_{\phi}(t)$ correspond to the periods of decreasing mutual information as well as increasing spin coherence (since $\partial_{t} \abs{\overline{\rho}_{\textrm{S}}^{01}(t)} = - \gamma_{\phi}(t) \abs{\overline{\rho}_{\textrm{S}}^{01}(t)}$). Now, if we are free to give the coefficient $\epsilon_{d}$ of the readout drive some engineered time dependence that is slow compared to the coarse-graining time scale $\tau$, then we expect the EQME to be still valid if we replace $\epsilon_{d}$ by $\epsilon_{d}(t)$ or its complex conjugate, and it would follow from Eq.(\ref{Eq: spin_ME_parameters}) that the flow of information and energy due to $\gamma_{\phi}(t)$ and $\gamma_{\uparrow\downarrow}$ may be controlled to some degree. For example, one can slowly modulate the frequency of the drive by replacing $\epsilon_{d}$ with $\abs{\epsilon_{d}}e^{i\theta(t)}$ for some slowly-varying phase $\theta(t)$ (slow in comparison to $\tau$).
\begin{figure}[h]
\centering
\captionsetup{justification=Justified, font=footnotesize}
\includegraphics[width=0.475\textwidth]{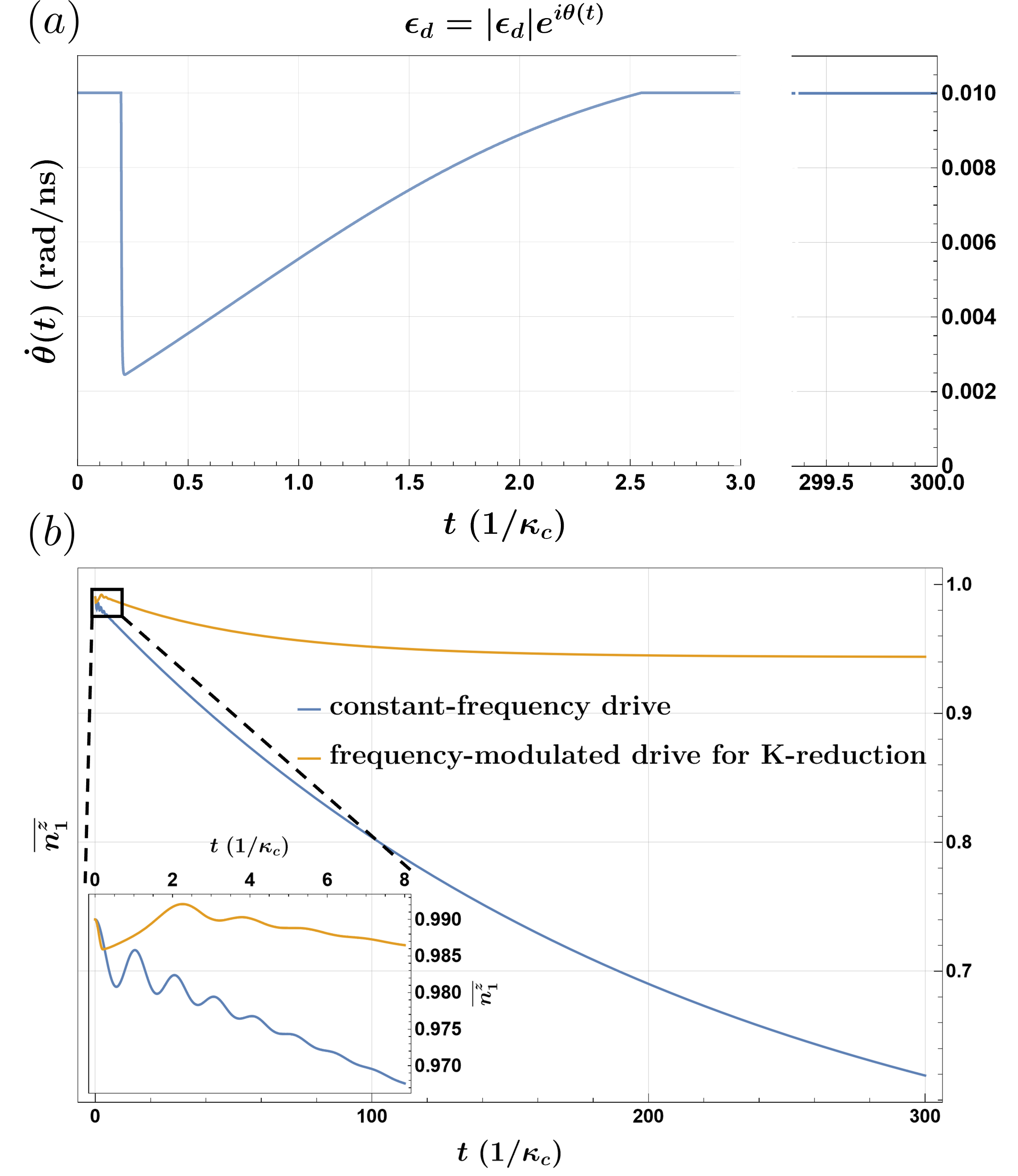}
\caption{The fine-tuned time dependence of the frequency of the readout drive shown in panel (a) is capable of suppressing the drive-induced spin relaxation, as shown in panel (b) with a spin initialized mostly in its excited state.}
\label{fig: SC_frequency_modulation}
\end{figure}

In particular, if one adopts the $\theta(t)$ defined by panel (a) of Fig.\ref{fig: SC_frequency_modulation}, then in theory one can significantly reduce the drive-induced relaxation of the spin by minimizing $\abs{\textrm{Im}\big[\alpha_{e}(t)\big]}$ while at the same time maintaining approximately the same distance between the cavity pointer states in the phase space. However, this scheme requires very precise control of the fine-tuned frequency of the readout drive which may not be easily achievable in readout experiments.

Finally, we note that, apart from the Purcell decay, both spin dephasing and the incoherent energy level transitions can be viewed as drive-induced phenomena, since $\alpha_{g/e}$ is proportional to $\epsilon_{d}$. If one increases the drive strength, then $\gamma_{\varphi_{\textrm{eff}}}(t)$, $\gamma_{\downarrow}$, and $\gamma_{\uparrow}$ will all increase as $\abs{\epsilon_{d}}^{2}$, leading to faster spin dephasing as well as more frequent incoherent spin  level transitions, which has been widely observed in readout experiments~\cite{Devoret_March_Meeting, Minev_Nature}. In view of an accurate analysis of modern readout experiments, however, the multi-level structure of the transmon must be taken into account, which we do in \Sec{Sec: the readout problem}.

\subsection{The stochastic master equation for the spin-cavity model}
\label{Subsec: spin-cavity SME}

When relaxation into the bath modes are constantly monitored by homodyne detection, stochastic terms appear in the master equation to account for the backaction of the continuous measurement. Consider the field-quadrature measurement schemes discussed in \cite{Wiseman_Milburn_1993} for example. If we imagine that the homodyne measurement has infinite time resolution ($\tau = 0$), then the time evolution of the conditional density matrix $\rho_{c}(t;\tau=0)$ can be written as
\begin{equation}
\begin{split}
\partial_{t} \rho_{c}
=&
-
i \left[ H_{I}(t), \rho_{c} \right]
+
\sqrt{\kappa_{c}} \xi(t;\tau=0)\\
&\cdot
\Big(
c \rho_{c}
+
\rho_{c} c^{\dagger}
-
\textrm{Tr}\left[
\rho_{c} (c + c^{\dagger})
\right] \rho_{c}
\Big)
\end{split}
\end{equation}
where $H_{I}(t)$ is given in Eq.(\ref{Eq: spin-cavity HI}) while $\xi(t;\tau=0)$ is the standard Gaussian white noise.
If instead we work in the $\tau \gg \abs{\omega_{d}-\omega_{a}}^{-1}$ limit, then performing TCG expansion to the third order gives rise to the following SME:

\cornerlineR{0.4pt}{6pt}{0pt}
\onecolumngrid

\begin{equation}
\label{Eq: third order spin-cavity SME}
\begin{split}
\partial_{t} \overline{\rho}_{\textrm{c}}(t;\tau)
=&
-
i\left[
\epsilon_{d} \left( 1+\frac{\lambda^{2}\sigma_{z}}{2} \right)\left( c + c^{\dagger} \right)
-
\chi \frac{\sigma_{z}}{2}\left( c^{\dagger} c + \frac{1}{2}
\right)
,
\overline{\rho}_{\textrm{c}}(t;\tau)
\right]\\
&
+
\kappa_{c} D_{(1+\lambda^{2}\sigma_{z}/2)c, (1+\lambda^{2}\sigma_{z}/2)c^{\dagger}} \overline{\rho}_{\textrm{c}}(t;\tau)
+
\kappa_{s} D_{\sigma_{-}, \sigma_{+}} \overline{\rho}_{\textrm{c}}(t;\tau)\\
&
+
\left[
i \kappa^{-}_{d}
\left(
D_{\sigma_{-},\sigma_{+}c}
+
D_{\sigma_{+}c,\sigma_{-}}
\right) \overline{\rho}_{\textrm{c}}(t;\tau)
+
i \kappa^{+}_{d}
\left(
D_{\sigma_{-}c,\sigma_{+}}
+
D_{\sigma_{+},\sigma_{-}c}
\right) \overline{\rho}_{\textrm{c}}(t;\tau)
+
h.c.
\right]\\
&
+
\sqrt{\kappa_{c}} \xi(t;\tau)
\Big(
(1+\lambda^{2}\sigma_{z}/2)c \overline{\rho}_{\textrm{c}}(t;\tau)
+
\overline{\rho}_{\textrm{c}}(t;\tau) (1+\lambda^{2}\sigma_{z}/2)c^{\dagger}\\
&\qquad\qquad\;\;
-
\textrm{Tr}\left[ \overline{\rho}_{\textrm{c}}(t;\tau) ((1+\lambda^{2}\sigma_{z}/2)c+(1+\lambda^{2}\sigma_{z}/2)c^{\dagger}) \right] \overline{\rho}_{\textrm{c}}(t;\tau)
\Big)
\end{split}
\end{equation}
\twocolumngrid
\cornerlineL{0.4pt}{0pt}{6pt}
with
\begin{equation}
\kappa^{\pm}_{d}
\equiv
\frac{\epsilon_{d} g_{ac}^{2}}{(\omega_{d}\pm\omega_{a})^{2}}
,\qquad
\chi
\equiv
2g_{ac}^{2}
\left(
\frac{1}{\omega_{d} - \omega_{a}}
-
\frac{1}{\omega_{d} + \omega_{a}}
\right)
\end{equation}
whereas $\xi(t;\tau)$ now represents a stationary colored Gaussian noise process determined by the coarse-graining window function $f(t;\tau)$. In particular, for the gaussian window function $f(t;\tau) = \frac{1}{\sqrt{2\pi}\tau} e^{-\frac{t^{2}}{2\tau^{2}}}$ we have
\begin{equation}
\begin{split}
\textrm{Cov}_{\textrm{gs}}
\left[
\xi(t;\tau),
\xi(t+\Delta t;\tau)
\right]
=
\frac{1}{2\sqrt{\pi} \tau} e^{-\frac{\Delta t^{2}}{4\tau^{2}}}
.
\end{split}
\end{equation}
As shown in the appendix, it follows from Eq.(\ref{Eq: third order spin-cavity SME}) that the population of state $|1\rangle$ evolves according to Eq.(\ref{Eq: p1 from spin-cavity SME}) where $\langle x\rangle_{n}$ and $\langle y\rangle_{n}$ are the conditional expectation values of $a+a^{\dagger}$ and $\langle-i(a-a^{\dagger})\rangle$ respectively for the spin state $|n\rangle$. Similar to Eq.(13) in \cite{Mabuchi_Wiseman_1999}, the stochastic term in the last line may be interpreted as the random fluctuations in the recorded photocurrent noise $\xi(t;\tau)$ retroactively ``causing'' quantum jumps in the spin state.

However, with finite time resolution $\tau$, one not only replaces the Wiener white noise by the colored noise $\xi(t;\tau)$, but also needs to modify the smooth evolution and the quadrature variable expectation values coupled to $\xi(t;\tau)$ in order for the underlying relation between the statistics and backaction of the continuous measurement to self-consistent.
In fact, Eq.(\ref{Eq: p1 from spin-cavity SME}) is derived in a limit where most of the terms are already in the ``IR'' limit and not sensitive to the value of $\tau$. One can easily consider models where the smooth part of the time evolution depends much more conspicuously on the color (i.e. time resolution and coarse-graining window) of $\xi(t;\tau)$.

\cornerlineR{0.4pt}{6pt}{0pt}
\onecolumngrid
\begin{equation}
\label{Eq: p1 from spin-cavity SME}
\begin{split}
\partial_{t} p_{1}(t;\tau)
=&
-
\kappa_{s} p_{1}(t;\tau)
+
2 ( \kappa^{-}_{d} + \kappa^{+}_{d} )
\left(
-
\frac{4\epsilon_{d} \kappa_{c}\left(p_{1}(t;\tau)-\frac{1}{2}\right)}{\kappa_{c}^{2}+\chi^{2}}
-
\frac{\langle y \rangle_{0}}{2}
+
\frac{p_{1}(t;\tau) \left( \langle y \rangle_{1} + \langle y \rangle_{0} \right)}{2}
\right)\\
&
+
\sqrt{\kappa_{c}} \xi(t;\tau)
p_{1}(t;\tau) \left(1-p_{1}(t;\tau)\right)
\left(
\langle x \rangle_{1}
-
\langle x \rangle_{0}
+
\frac{\lambda^{2}}{2}
(\langle x \rangle_{1} + \langle x \rangle_{0})
\right)
\end{split}
\end{equation}
\twocolumngrid
\cornerlineR{0.4pt}{0pt}{6pt}

\section{The transmon readout problem}
\label{Sec: the readout problem}

\begin{figure}[h!]
\centering
\captionsetup{justification=Justified, font=footnotesize}
\includegraphics[scale=0.6]{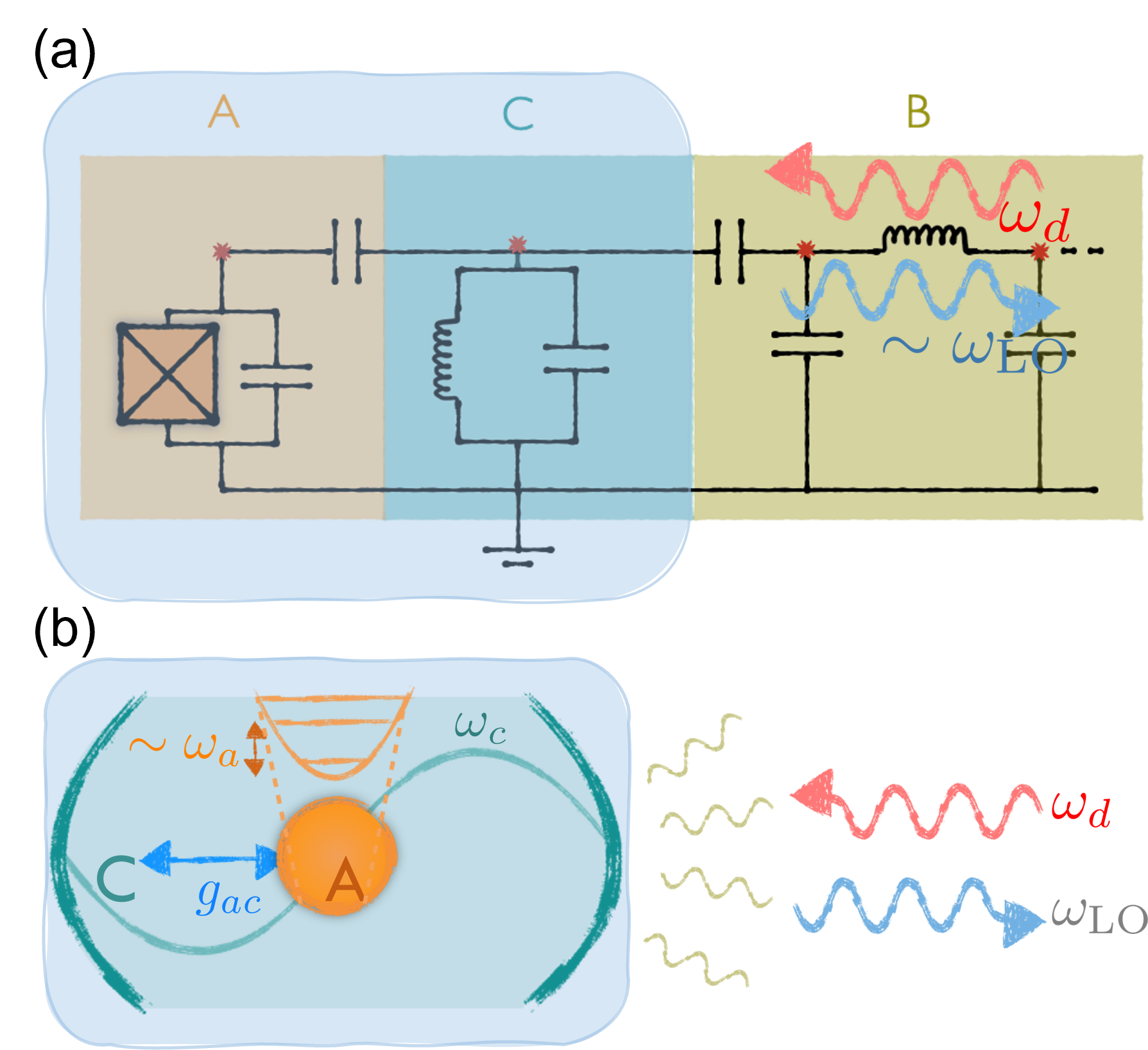}
\captionof{figure}[]{The Hamiltonian in Eq.(\ref{Eq: S-pic H0},\ref{Int Hamiltonian}) can be used to model either a superconducting circuit as shown in (a), or a driven atom-cavity system as shown in (b). Therefore, we use ``the transmon'' and ``the atom'' interchangeably, and also use ``the (linear) resonator'' and ``the cavity'' interchangeably. The bath is modeled by the bosonic modes supported by a transmission line, and we refer to the transmon (atom) and the resonator (cavity mode) as ``the system''.}
\label{fig:circuit}
\end{figure}

In this section we consider the readout dynamics for a more accurate model of the qubit that is being measured. To this end we consider the dispersive readout of a transmon qubit, modeled by its Josephson potential, monitored through a detuned linear resonator to which it is capacitively coupled (see Fig.\ref{fig:circuit} for the schematic circuit diagram). During the measurement, a readout drive close to the resonator frequency $\omega_{c}$ is applied to the resonator to perform heterodyne detection of its quadrature variables. Since different energy levels of the transmon shift the resonator frequency by different amounts, the steady states of the resonator have previously been shown to function as pointer states for the different qubit excitation levels\cite{Blais_etal_dispersive_cQED, Blais_qubit_cloaking}.

Rapid qubit readout requires a large SNR which can be achieved through a strong probe pulse, i.e. a large $\epsilon_{d}$. But increasing the amplitude of the probe pulse results in undesired transitions to other states of the qubit than the one started out, limiting the readout fidelity through misassignment errors. In the two-level model of the qubit in the previous section, such transition are limited to transitions between the two spin levels (as found e.g. in \Eq{Eq:spinmodelrates}) however such transitions can also be significant to levels outside the computational subspace, as has been reported in recent experiments~\cite{Devoret_March_Meeting, Minev_Nature} and analyzed in early theoretical work~\cite{Sank2016, MPT_I, MPT_II, Blais_etal_transmon_ionization, Hanai_McDonald_Clerk, Blais_etal_dispersive_cQED,Lescanne2019}. Such transitions can easily be perceived in experiments as leading to a reduction in the effective spin-$T_1$ time, as concluded in Figure S2 of~\cite{Minev_Nature}. To distinguish these drive-induced transitions from each other, the occupation of other levels have to be simultaneously monitored in well-calibrated readout experiments~\cite{Pop_et_al_readout_with_large_photon_number, Zhou2021}.  

We analyze this situation as an informative case study in this section. We shall not be concerned here with the effects of an additional qubit-bath which can be important~\cite{thorbeck2023}, nor are we interested in the question of engineering qubits to suppress such readout errors~\cite{Lledo2023, Martinis2021}. The goal here is to provide the perspective offered by the TCG approach, analyze the effectiveness of a TCG-based numerical analysis and the interpretive power provided by the availability of high-order analytical expressions.

	
We demonstrate that time-coarse graining the system dynamics down to below the GHz range would:
\begin{itemize}
	\item give rise to explicit formula for the dispersive shift of the resonator frequency as a function of the transmon energy level;
	\item give rise to effective direct coupling between the transmon and the bath, from which we derive the Purcell decay rate for different transmon energy levels;
	\item reveal \emph{drive-induced incoherent transitions} among the transmon energy levels which can significantly impact the lifetimes of all transmon levels as well as their populations in non-equilibrium steady states.
\end{itemize}
	
\subsection{The Model for Dispersive Transmon Qubit Readout}
\label{Subsec: the model}

In the rest of this paper, we investigate an experimentally-relevant model for the dispersive readout of a transmon qubit, where we suppose that the system both dissipates to and is driven through a bath of transmission line modes capacitively coupled to the readout resonator (see \Fig{fig:circuit}). We also assume the bath to be at zero temperature, although generalization to finite temperatures is straightforward. In addition, as motivated in Appendix A of the Supplementary Materials\cite{supplement} through input-output theory, we suppose that the recorded signal at the end of the measurement chain is some time average (with resolution $\tau > 2 \textrm{ns}$) of the output signal carried by the transmission line after down-converting it by a local oscillator (LO) at frequency $\omega_{\textrm{LO}}$. See Appendix H of the Supplementary Materials\cite{supplement} for further details. For simplicity, we assume that the local oscillator frequency $\omega_{\textrm{LO}}$ is the same as the readout drive frequency $\omega_{d}$, since small deviations from this value would not qualitatively change the system dynamics. The free and interaction-picture Hamiltonians can be written as
\begin{equation}
\label{Eq: S-pic H0}
\begin{split}
\hat{H}_{0}
=&
\omega_{\textrm{LO}} c^{\dagger}c
+
\omega_{a} a^{\dagger}a
-
\frac{\epsilon\omega_{a}}{48} \big( a + a^{\dagger} \big)^{4}
+
\int \frac{d\omega \mathcal{D}_{\omega} \omega}{2\pi} B_{\omega}^{\dagger} B_{\omega}
\end{split}
\end{equation}
and
\begin{equation}
\label{Int Hamiltonian}
\begin{split}
H_{I}(t)
=&
H_{AC}(t)
+
H_{A3C}(t)
+
H_{D}(t)
+
H_{\delta}
+
H_{CB}(t)
\end{split}
\end{equation}
respectively, with
\begin{equation}
\label{Int Hamiltonian terms}
\begin{split}
&H_{AC}(t)
=
g_{ac} e^{ i( \omega_{a}^{\prime}(n_{a}) - \omega_{d} )t } \Big( 1 - \frac{\epsilon n_{a}}{8} \Big) ca^{\dagger}\\
&\qquad\qquad\;
-
g_{ac} ca \Big( 1 - \frac{\epsilon n_{a}}{8} \Big) e^{ -i( \omega_{a}^{\prime}(n_{a}) + \omega_{d} )t }
+
h.c.;\\
&
H_{A^{3}C}(t)
=
-
\frac{\epsilon g_{ac}}{16} e^{ i( 3\omega_{a}^{\prime}(n_{a}-1) - \omega_{d} )t } ca^{\dagger 3}\\
&\qquad\qquad\;\;\;
+
\frac{\epsilon g_{ac}}{16}ca^{3} e^{ -i( 3\omega_{a}^{\prime}(n_{a}-1) + \omega_{d} )t }
+
h.c.;\\
&H_{D}(t)
=
\big( \epsilon_{d}^{\ast} c - \epsilon_{d} e^{-2i\omega_{d}t}c \big)
+
h.c.;\\
&
H_{\delta}
=
-
(\omega_{d}-\omega_{c}) c^{\dagger}c;\\
&H_{CB}(t)
=
\int \frac{d\omega}{2\pi} \mathcal{D}_{\omega} g_{\omega}
\big(
e^{i( \omega - \omega_{d} ) t} c B_{\omega}^{\dagger} - e^{-i( \omega + \omega_{d} ) t} c B_{\omega}
\big)\\
&\qquad\qquad\;
+
h.c.
\end{split}
\end{equation}
where the operator-valued function
\begin{equation}
\begin{split}
\omega_{a}^{\prime}(n_{a})
:=
\omega_{a}\big( 1-\frac{\epsilon}{4} n_{a} \big)
\end{split}
\end{equation}
gives us the phase frequency of the transmon quadrature variable $a$ depending on the energy level $n_{a}$, with the dependence being a consequence of transmon anharmonicity. In addition to variation in the transmon transition energies, a nonzero anharmonicity $\epsilon$ also gives rise to the nonlinear atom-resonator couplings in $H_{AC}$ and $H_{A^{3}C}$.
	
Finally, we emphasize that we have chosen to include the anharmonic transmon potential into the free Hamiltonian $\hat{H}_{0}$ in order to simplify the interaction Hamiltonian and more accurately calculate the dispersive shift and Purcell rates at low orders in perturbative TCG. The price we pay, however, is that the modified transmon frequency $\omega^{\prime}_{a}(n_{a})$ in the interaction picture now depends the energy level $n_{a}$. This also leads to the inconvenient fact that the frequencies to be used in the TCG calculations would be operator-valued and do not commute with an arbitrary density matrix in general. However, since we are interested in the coarse-grained density matrix $\overline{\rho}(t)$ with sub-GHz resolution, the diagonal matrix elements in the transmon energy eigenbasis (i.e., the ``pointer basis'') are expected to decouple from the off-diagonal ones below the fifth order in TCG, because the only operator products of no more than four Hamiltonian terms in Eq.(\ref{Int Hamiltonian terms}) whose total frequencies can be resolved by $\tau$ would contain equal numbers of $a$ and $a^{\dagger}$. In fact, we can verify a posteriori from the TCG master equation obtained in Subsection \ref{Subsec: TCG transmon readout} that there are no significant superoperators coupling the diagonal density matrix elements in the transmon pointer basis to the off-diagonal ones up to the fourth-order EQME. Therefore, in the derivation of the TCG master equation, one can adopt the rule of thumb that $n_{a}$ commutes with $\overline{\rho}$ and understand the resulting TCG superoperators as acting on those diagonal matrix elements in the transmon pointer basis. Furthermore, when we obtain a superoperator whose coefficient depends on $n_{a}$, we adopt the convention that the operator-valued coefficient acts on the diagonal elements of $\overline{\rho}$ before any action of the dissipator, i.e.,
\begin{equation}
\begin{split}
\kappa(n_{a}) D_{A,B} \overline{\rho}
\equiv
A \kappa(n_{a}) \overline{\rho} B - \frac{ B A \kappa(n_{a}) \overline{\rho} + \kappa(n_{a}) \overline{\rho} B A }{ 2 }
\end{split}
\end{equation}
with the coefficient $\kappa(n_{a})$ being any function of $n_{a}$.

\subsection{The TCG master equation and drive-induced dynamics}
\label{Subsec: TCG transmon readout}

To explicitly calculate the TCG superoperators, we assume that the coarse graining time scale $\tau$ is much greater than $\frac{1}{\omega_{c}}$, $\frac{1}{\omega_{a}^{\prime}(n_{a})}$, and $\frac{1}{\abs{\omega_{c}-\omega_{a}^{\prime}(n_{a})}}$ (which are typically below the nanosecond range) while allowing $\tau$ to be comparable or smaller than $\frac{1}{\abs{\omega_{d}-\omega_{c}}}$. In summary, we make the following assumptions in this section:
\begin{equation}
\label{assumptions}
\begin{split}
&
\frac{1}{\omega_{c}}, \frac{1}{\omega_{a}^{\prime}(n_{a})}, \frac{1}{\abs{\omega_{c}-\omega_{a}^{\prime}(n_{a})}}
\ll
\tau
\lesssim
\frac{1}{\abs{\omega_{d}-\omega_{c}}}, \frac{1}{J(\omega_{c})};\\
&
\frac{g}{\abs{\omega_{c}-\omega_{a}^{\prime}(n_{a})}} \ll 1;
\quad
\epsilon < 1;
\quad
\omega_{d} \sim \omega_{c}.
\end{split}
\end{equation}
With these conditions, we calculate the EQME up to the fourth order while omitting superoperators that are exponentially suppressed by factors smaller than or equal to $e^{-\frac{(\omega_{c}-\omega_{a}^{\prime}(n_{a}))^{2}\tau^{2}}{2}}$. For all the numerical simulations performed in this section, $\tau$ is assumed to be a few nanoseconds so that the coefficients of the omitted superoperators are many orders of magnitude smaller than the smallest of those to be presented in the remainder of this subsection. In addition, for the superoperators up to the fourth-order, $\tau > 2 \textrm{ns}$ would be safely in the IR limit, so the approximate (super)operator coefficients to be presented are all independent of $\tau$. The bath modes are traced out at zero temperature from the master equation according to the formalism in \Sec{Sec: TCG open quantum systems}.

In addition, as detailed in Appendix I of the Supplementary Materials\cite{supplement}, we use the TCG master equation to numerically solve for the dynamics of the transmon pointer state populations $p_{n}(t)$ together with those of the conditional resonator cumulant variables for each transmon level $n$. For that purpose, we would be using the following set of parameters that are not atypical of an experimental readout scenario:
\begin{equation}
\label{Eq: transmon parameters}
\begin{split}
&
\epsilon = 0.2
\qquad\quad\;\;\;\,
\kappa_{c} = 0.48 \textrm{MHz}
\qquad\quad\;\;\;\,
\frac{\omega_{c}}{2\pi} = 7 \textrm{GHz}\\
&
\omega_{01} \equiv \frac{\omega_{a}\big( 1 - \frac{\epsilon}{4} \big)}{2\pi} = 5 \textrm{GHz}
\qquad\qquad\;
\frac{g_{ac}}{2\pi} = 48.0 \textrm{MHz}
\end{split}
\end{equation}
where $\omega_{01}$ is the (approximate) transition energy between the ground state and the first excited state of the transmon.

The transmon readout model described by the Hamiltonian in Eq.(\ref{Int Hamiltonian},\ref{Int Hamiltonian terms}) is in its essence similar to the toy model analyzed in the previous section, the Hamiltonian in Eq.(\ref{Eq: spin-cavity HI}). In fact, the types of effective corrections obtained at each order in the TCG master equations are exactly the same for both models, and therefore we only present the fourth-order TCG master equation here with corrections from all orders lumped together. As in the general formalism, the fourth-order TCG master equation can be written as
\begin{equation}
\begin{split}
\partial_{t} \overline{\rho}_{ac}(t)
=
-i \Big[ H_{\textrm{eff}}, \overline{\rho}_{ac}(t) \Big]
+
D_{\textrm{eff}} \overline{\rho}_{ac}(t)
\end{split}
\end{equation}
where the effective Hamiltonian contains the following corrections
\begin{equation}
\begin{split}
H_{\textrm{eff}}
=&
h_{0}(n_{a})
+
h_{1}(n_{a}) n_{c}
+
h_{2}(n_{a}) n_{c}^{2}
+
\Big[
h^{(-1)}_{0}(n_{a}) c\\
&
+
h^{(-1)}_{1}(n_{a}) c^{\dagger} c^{2}\
+
h^{(-2)}_{0}(n_{a}) c^{2}
+
h.c.
\Big]
\end{split}
\end{equation}
while the effective dissipators can be split into three terms
\begin{equation}
\begin{split}
D_{\textrm{eff}}
=&
D^{(0)}_{\textrm{eff}}
+
D^{(-)}_{\textrm{eff}}
+
D^{(+)}_{\textrm{eff}}
\end{split}
\end{equation}
with $D^{(0)}_{\textrm{eff}}$, $D^{(-)}_{\textrm{eff}}$, and $D^{(+)}_{\textrm{eff}}$ defined as the collections of dissipators that keep the transmon level invariant, decrease it by one, and increase it by one respectively. More explicitly, we have
\begin{fleqn}
\begin{equation}
\begin{split}
D^{(0)}_{\textrm{eff}}
=&
d^{(-1,-1)}_{0,0}(n_{a}) D_{c,c^{\dagger}}
+
\Big[
d^{(-2,-1)}_{0,0}(n_{a}) D_{c^{2},c^{\dagger}}\\
&
+
d^{(1,2)}_{0,0}(n_{a}) D_{c^{\dagger},c^{2}}
+
h.c.
\Big]
\end{split}
\end{equation}
\end{fleqn}
\begin{fleqn}
\begin{equation}
\begin{split}
D^{(-)}_{\textrm{eff}}
=&
\kappa^{(0,0)}_{0,0}(n_{a}) D_{a,a^{\dagger}}
+
\tilde{\kappa}^{(0,0)}_{0,0}(n_{a}) D_{a^{3},a^{\dagger 3}}\\
&
+
\Big[
\kappa^{(-1,-1)}_{(1,0)}(n_{a}) D_{ac^{\dagger}c^{2},a^{\dagger}c^{\dagger}}
+
\kappa^{(1,1)}_{(1,0)}(n_{a}) D_{ac^{\dagger 2}c,a^{\dagger}c}\\
&
+
\kappa^{(0,0)}_{0,1}(n_{a}) D_{a,a^{\dagger}c^{\dagger}c}
+
\kappa^{(-1,0)}_{0,0}(n_{a}) D_{ac,a^{\dagger}}\\
&
+
\kappa^{(-1,0)}_{0,1}(n_{a}) D_{ac,a^{\dagger}c^{\dagger}c}
+
\kappa^{(0,1)}_{0,0}(n_{a}) D_{a,a^{\dagger}c}\\
&
+
\kappa^{(0,1)}_{1,0}(n_{a}) D_{ac^{\dagger}c,a^{\dagger}c}
+
h.c.
\Big]
\end{split}
\end{equation}
\end{fleqn}
\begin{fleqn}
\begin{equation}
\begin{split}
D^{(+)}_{\textrm{eff}}
=&
\gamma^{(-1,-1)}_{(1,0)}(n_{a}) D_{a^{\dagger}c^{\dagger}c^{2},ac^{\dagger}}
+
\gamma^{(1,1)}_{(1,0)}(n_{a}) D_{a^{\dagger}c^{\dagger 2}c,ac}\\
&
+
\gamma^{(0,0)}_{0,1}(n_{a}) D_{a^{\dagger},ac^{\dagger}c}
+
\gamma^{(-1,0)}_{0,0}(n_{a}) D_{a^{\dagger}c,a}\\
&
+
\gamma^{(-1,0)}_{0,1}(n_{a}) D_{a^{\dagger}c,ac^{\dagger}c}
+
\gamma^{(0,1)}_{0,0}(n_{a}) D_{a^{\dagger},ac}\\
&
+
\gamma^{(0,1)}_{1,0}(n_{a}) D_{a^{\dagger}c^{\dagger}c,ac}
+
h.c.
\end{split}
\end{equation}
\end{fleqn}
where we have again used the notation for generalized dissipators defined in Eq.(\ref{Eq: diss_def}). In particular, we comment on the following corrections which appear to have the most significant impact on observable dynamics at the coarse-graining time $\tau$:
\begin{itemize}
\item The emergent longitudinal coupling $h_{1}(n_{a}) n_{c}$ between the transmon and the resonator gives rise to the dispersive shift of the cavity frequency
\begin{flalign*}
&&\chi_{ac}
\equiv&
h_{1}(0) - h_{1}(1)
\approx
\frac{\epsilon g_{ac}^{2}\big( \omega_{a}^{\prime}(1) + \omega_{a}^{\prime}(2) \big)\omega_{d}^{2}}{\big( \omega_{d}^{2} - \omega_{a}^{\prime}(1)^{2} \big)\big( \omega_{d}^{2} - \omega_{a}^{\prime}(2)^{2} \big)}
\end{flalign*}
which can be easily observed by heterodyne detection and is the basis for dispersive readout of the transmon state. From the structure of the contraction coefficients in Eq.(\ref{Eq: contraction coefficient}), we know that $\chi_{ac}$ receives contributions from virtual transitions between transmon levels $1 \leftrightarrow 2$ as well as $0 \leftrightarrow 1$. If the artificial atom has nonzero matrix elements $\langle 0 |(a-a^{\dagger})| n\rangle$ or $\langle 1 |(a-a^{\dagger})| n\rangle$ for some higher levels $n>2$, then virtual transitions to those higher levels will also contribute to $\chi_{ac}$ according to the loop diagrams introduced in \Sec{Sec: TCG perturbation theory}. In addition, the well-known Lamb shift
\begin{flalign*}
\begin{split}
\Delta_{\textrm{Lamb}}
=
-
\int_{0}^{\infty} \frac{d\omega}{2\pi} J(\omega)
\Big( \frac{1}{\omega-\omega_{c}} + \frac{1}{\omega+\omega_{c}} \Big)
\end{split}
\end{flalign*}
is also found to be an important contribution to the constant part of $h_{1}(n_{a})$.
    
\item The resonator decay rate $\kappa_{c} \equiv d^{(-1,-1)}_{0,0}(n_{a}) = J(\omega_{d})$ is found by tracing out bath modes at zero temperature. When the explicit form of the spectral density $J(\omega)$ is needed in this section, we assume that
\begin{flalign*}
\begin{split}
J(\omega) = \frac{\alpha \omega}{ 1 + \frac{\omega^{2}}{\Lambda^{2}} }
\end{split}
\end{flalign*}
as suggested by the asymptotic behavior of the power spectral density studied in \cite{Cutoff_Free} where the bath is modeled by the modes of a transmission-line coupled to semi-infinite waveguides at both ends. For the purpose of numerical simulation, we would use the cutoff value $\Lambda = 2\pi \cdot 100 \textrm{GHz}$. Note that $J(\omega) \approx \alpha \omega$ is approximately Ohmic and insensitive to the cutoff $\Lambda$ as long as $\omega$ is comparable to system transition frequencies; on the contrary, the Lamb shift $\Delta_{\textrm{Lamb}}$ only converges if $J(\omega)$ is properly regularized at high frequencies, because TCG \emph{only filters out direct transitions} with large energy differences but not any virtual transitions which are responsible for the Lamb shift.
    
\item The one- and three-photon Purcell decay rates are found to be
\begin{flalign*}
\begin{split}
\kappa^{(0,0)}_{0,0}(n_{a})
=
J\big( \omega_{a}^{\prime}(n_{a}) \big) \Big[ \frac{ 2 g_{ac} \big( 1 - \frac{\epsilon n_{a}}{8} \big) \omega_{d}  }{ \omega_{d}^{2}-\omega_{a}^{\prime}(n_{a})^{2} } \Big]^{2}
\end{split}
\end{flalign*}
and
\begin{flalign*}
\begin{split}
\tilde{\kappa}^{(0,0)}_{0,0}(n_{a})
=
\frac{ \epsilon^{2} g_{ac}^{2} \omega_{d}^{2} J\big( 3\omega_{a}^{\prime}(n_{a}-1) \big) }{ 64 \big(9\omega_{a}^{\prime}(n_{a}-1)^{2} - \omega_{d}^{2}\big)^{2} }
\end{split}
\end{flalign*}
respectively. These Purcell decay processes are responsible for the drive-independent part of the transmon relaxation. However, since the three-photon decay rate $\tilde{\kappa}^{(0,0)}_{0,0}(n_{a})$ is at order $\epsilon^{2}$ and much smaller than $\kappa^{(0,0)}_{0,0}(n_{a})$ for the parameters in Eq.(\ref{Eq: transmon parameters}), we ignore it in the effective dissipator $D_{\textrm{eff}}$ during numerical simulations.

\item The effective dissipators
\begin{flalign*}
\begin{split}
&\qquad\qquad
\kappa^{(-1,0)}_{0,0}(n_{a}) D_{ac,a^{\dagger}},\qquad
\kappa^{(0,1)}_{0,0}(n_{a}) D_{a,a^{\dagger}c},\qquad\\
&\qquad\qquad
\gamma^{(-1,0)}_{0,0}(n_{a}) D_{a^{\dagger}c,a},\qquad
\gamma^{(0,1)}_{0,0}(n_{a}) D_{a^{\dagger},ac},
\end{split}
\end{flalign*}
and their hermitian conjugates are responsible for drive-induced transitions among the transmon levels. These dissipators are analogous to $D_{\textrm{TCG}}^{(3)}$ in Eq.(\ref{Eq: spin D3}) for the spin-cavity toy model, and their explicit forms are presented in Eq.(H27) of Appendix H in the Supplementary Materials\cite{supplement}. Tracing over the resonator states, we find the corresponding relaxation rate from level $n$ to level $(n-1)$ of the transmon to be
\begin{flalign*}
\begin{split}
&\qquad
K(n)
=
2 \textrm{Re}\Big[ \big( \delta^{(3)} \kappa^{(-1,0)}_{0,0}(n)
+
\delta^{(3)} \kappa^{(0,1)}_{0,0}(n) \big) \cdot \langle c \rangle_{n} \Big]
\end{split}
\end{flalign*}
whereas the corresponding excitation rate from level $n$ to level $(n+1)$ is
\begin{flalign*}
\begin{split}
&\qquad
\Gamma(n)
=
2 \textrm{Re}\Big[ \big( \delta^{(3)} \gamma^{(-1,0)}_{0,0}(n)
+
\delta^{(3)} \gamma^{(0,1)}_{0,0}(n) \big) \cdot \langle c \rangle_{n} \Big].
\end{split}
\end{flalign*}

\end{itemize}

\begin{figure}[!h]
\centering
\captionsetup{justification=Justified, font=footnotesize}
\includegraphics[width=0.48\textwidth]{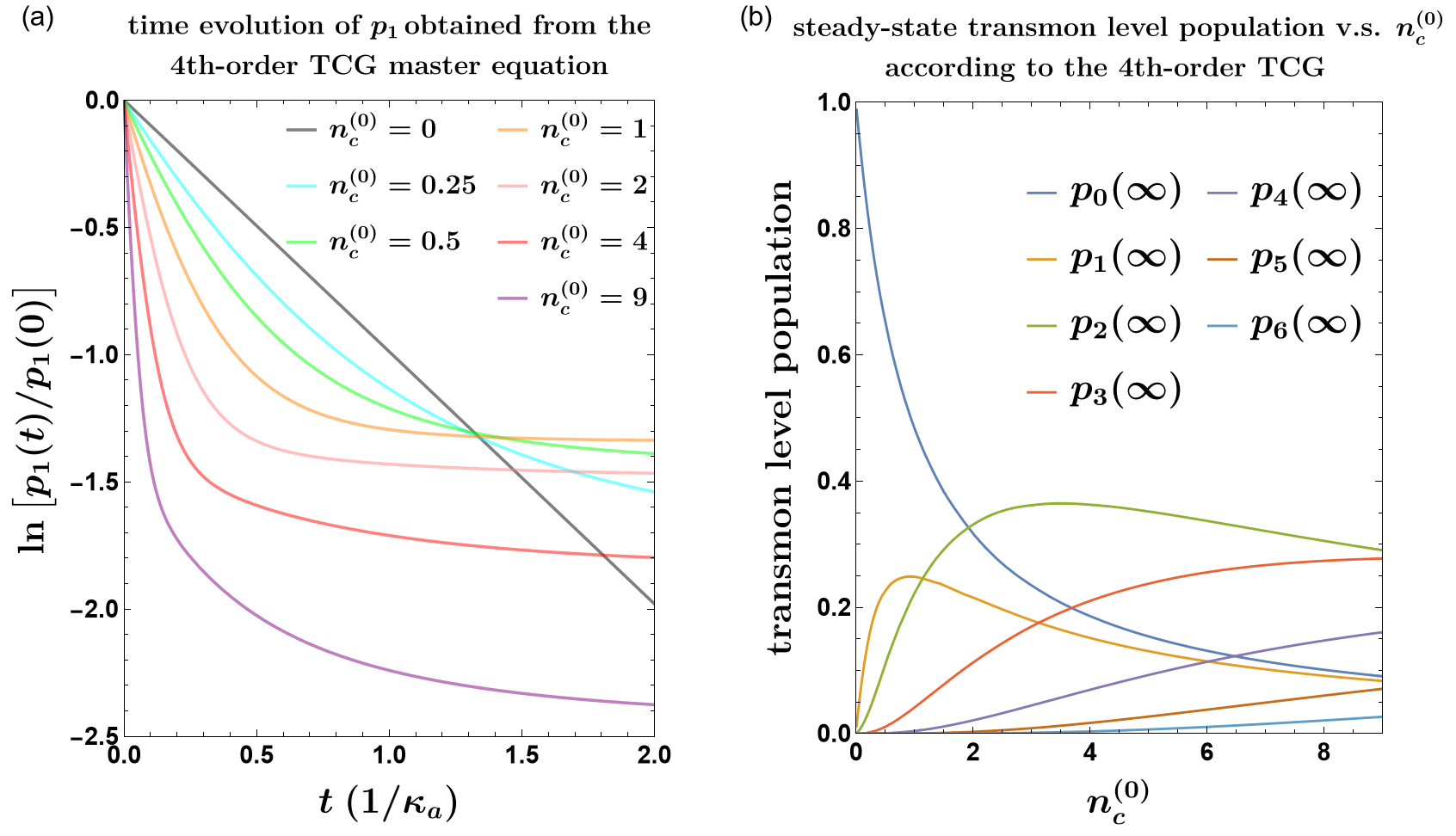}
\caption{(a) Time evolution of the first excited state population $p_{1}$ of the transmon according to the 4th-order TCG master equation. The drive induced transitions are significant compared to the Purcell decay rate for a moderately driven readout resonator. The curves level off towards the end of the time evolution as the system approach a non-equilibrium steady state. The horizontal axis is in units of the zero-temperature Purcell decay time $\kappa_{a}^{-1} = 2.36 \textrm{ms}$ as defined in Eq.(H20) of Appendix H in the Supplementary Materials(\cite{supplement}). (b) The non-equilibrium steady-state transmon population $p_{n}$ at level $n$ plotted as a function the drive strength measured in $n_{c}^{(0)}$, the steady-state resonator photon number when the transmon is in its ground state. All the simulations were obtained from the TCG QEM assuming that $\tau = 2\textrm{ns} = 25(\omega_{c}-\omega_{a}^{\prime}(1))^{-1} = 8.5\times 10^{-7} \kappa_{a}^{-1}$.}
\label{fig:4th TCG p1 and SS atom population}
\end{figure}

With more detailed discussion at each order left to Appendix H of the Supplementary Materials\cite{supplement}, we focus on the phenomenological predictions of the TCG master equation. In particular, we are interested in the dynamics of the population $p_{n}(t)$ of the transmon level $n$. Using the cumulant expansion method introduced in Appendix I of the Supplementary Materials\cite{supplement}, we solve for numerical solutions of the 4th-order EQME for different drive strengths (measured in terms of the steady-state resonator photon number $n_{c}^{(0)} := \langle c^{\dagger}c \rangle_{n_{a}=0}$ when the transmon is in the ground state); for all the numerical simulations, the initial condition is taken to be the direct product of the pure resonator vacuum state and the mixed transmon state with the following $n$th-level populations $p_{n}(0)$:
\begin{equation*}
\begin{split}
p_{0}
=
0.04
;\quad
p_{1}
=
0.95
;\quad
p_{n}
=
9\times 10^{-(n+1)}
\;\;\forall\;\;
n\ge 2
\end{split}
\end{equation*}

As shown in panel (a) of Fig.\ref{fig:4th TCG p1 and SS atom population}, the lifetime of the first excited state of the transmon decreases by more than a factor of two for relatively small resonator photon occupancy ($n_{c}^{(0)} \sim 1$) if the intrinsic dissipation of the transmon is negligible in comparison with the Purcell decay rate. However, we emphasize that the EQME suggests different drive-induced transition rates between all neighboring levels of the transmon, so the relaxation of the transmon cannot be adequately described by $K(1)+\Gamma(1)$ alone.

In addition, the steady-state population $p_{n}(\infty)$ is nonzero for any transmon level $n$ due to the drive-induced transitions, as shown in panel (b) of Fig.\ref{fig:4th TCG p1 and SS atom population} (it has been numerically verified that the steady-state populations are independent of the initial condition.). Moreover, for sufficiently large drive strength, the steady-state population of a higher transmon level will exceed that of a lower level, making it impossible to assign a positive effective temperature to the transmon in its steady-state~\cite{Blais_et_al_chaos_in_transmon, MPT_II, Devoret_March_Meeting}. Even for smaller drive strengths, an effective-temperature description of the steady state is only possible if we combine all the excited states of the transmon into a single level for statistical purposes.

\begin{figure}[h]
\centering
\captionsetup{justification=Justified, font=footnotesize}
\includegraphics[width=0.45\textwidth]{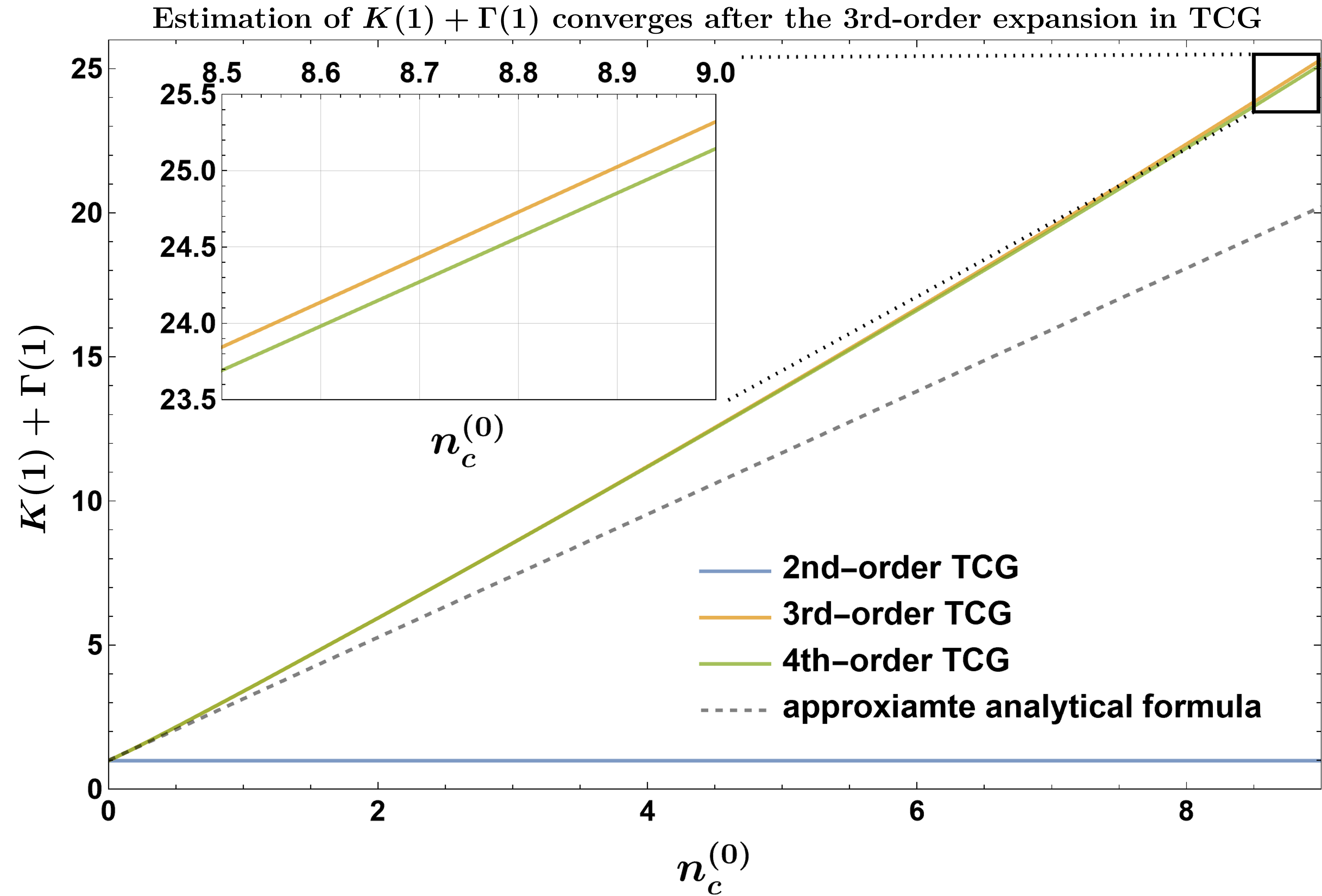}
\caption{The predicted level-1 decay rate $K(1)+\Gamma(1)$ as a function of the drive strength measured in $n_{c}^{(0)}$ is plotted at each order in the TCG perturbation theory. Numerically, $K(1)+\Gamma(1)$ is estimated by exponential curve fitting during the time period where $0.78< p_{1} < 1$. The dashed curve, on the other hand, represents the estimation given by the approximate analytical formula in Eq.(\ref{Eq: K over kappa}). We observe good convergence of the theoretical predictions starting at the third order in perturbative TCG.}
\label{fig:InvT1 vs nc}
\end{figure}

In order to estimate the drive-induced transition rates $K(n)$ and $\Gamma(n)$, we may approximate the conditional resonator state by a coherent state with
\begin{equation}
\langle c \rangle_{n}
\approx
\frac{\epsilon_{d}}{ -\big( \tilde{\omega}_{c}(n) - \omega_{d} \big) + \frac{\kappa_{c}}{2}i}
\end{equation}
during all but the very initial period of the transmon time evolution, where $\tilde{\omega}_{c}(n) \equiv \omega_{c} + h_{1}(n)$ is the modified resonator frequency when the transmon is at level $n$. This approximation is justified since the lifetime of the resonator is much shorter than that of the transmon in practical applications of dispersive readout.
Using this approximation of $\langle c \rangle_{n}$ in the general formulas for $K(n)$ and $\Gamma(n)$, we find $K(n)$ and $\Gamma(n)$ to be on the same order of magnitude with the corresponding ratios between them and the Purcell decay rate $\kappa_{a}(n)$ being
\begin{equation}
\label{Eq: K over kappa}
\begin{split}
&
\frac{K(n)}{\kappa_{a}(n)}
=
\frac{ \abs{\epsilon_{d}}^{2}\big( 1 - \frac{\epsilon n}{4} \big) }{ \big( \tilde{\omega}_{c}(n) - \omega_{d} \big)^{2} + \frac{\kappa_{c}^{2}}{4} } \frac{ \omega_{d}^{2} + \omega^{\prime}_{a}(n)^{2} }{ 2\omega_{d}\omega^{\prime}_{a}(n) \big( 1 - \frac{\epsilon n}{8} \big)^{2} }\\
&\qquad\;\;\,\approx
n_{c}^{(n)} \frac{ \omega_{d}^{2} + \omega^{\prime}_{a}(n)^{2} }{ 2\omega_{d}\omega^{\prime}_{a}(n) }\\
&
\frac{\Gamma(n)}{\kappa_{a}(n)}
\approx
n_{c}^{(n)} \frac{ \omega_{d}^{2} + \omega^{\prime}_{a}(n+1)^{2} }{ 2\omega_{d}\omega^{\prime}_{a}(n+1) }
\end{split}
\end{equation}
where $n_{c}^{(n)}$ is the quasi-steady state expectation value of $n_{c}\equiv c^{\dagger}c$ when the transmon is at level $n$. Analogous to Eq.(\ref{Eq: InvT1_expansion}) for the spin-cavity model, the drive-induced transition rates can in general  also be written as a power series in $n_{c}^{(n)}$:
\begin{equation}
\begin{split}
&
\frac{K(n)}{\kappa_{a}(n)}
=
c^{-}_{1}(n) n_{c}^{(n)}
+
c^{-}_{2}(n) (n_{c}^{(n)})^{2}
+
\cdots\\
&
\frac{\Gamma(n)}{\kappa_{a}(n)}
=
c^{+}_{1}(n) n_{c}^{(n)}
+
c^{+}_{2}(n) (n_{c}^{(n)})^{2}
+
\cdots
\end{split}
\end{equation}
where higher-order terms are necessary for higher readout drive power.

Since $\omega_{d}$ and $\omega^{\prime}_{a}(n)$ are both in the GHz range, the ratio $K(n)/\kappa_{a}(n)$ is on the same order of magnitude as $n_{c}^{(n)}$. If Purcell decay is the dominant decay channel of the transmon, then we expect the readout drive to significantly decrease the $T_{1}$ lifetime of the transmon qubit initially prepared in its first excited state as soon as the drive is strong enough to maintain a few photons in the readout resonator. This is consistent with observations from recent experimental studies\cite{Devoret_March_Meeting, Minev_npj}. In fact, our analysis here show that the ratio between drive-induced transition rates and the Purcell decay rate cannot be easily reduced by tuning the transmon and resonator frequencies or adjusting the transmon-resonator coupling strengths. To minimize the relative strength of drive-induced transitions, one has to fundamentally change the type of atom-resonator couplings (employing longitudinal couplings for example) or engineer the functional form of the spectral density.

On the other hand, we can also estimate the drive-induced transition rate $K(1)+\Gamma(1)$ by fitting the numerical solution of $p_{1}(t)$ to an exponential curve.
The resulting estimates for $K(1)+\Gamma(1)$ obtained with TCG master equations at different orders can be found in Fig.\ref{fig:InvT1 vs nc}.
In particular, we see that the drive-induced transitions kick in at the third order in the TCG expansion, and higher-order corrections only provide small quantitative modifications to those transitions rates.
\onecolumngrid

\begin{figure}[!h]
\centering
\captionsetup{justification=Justified, font=footnotesize}
\includegraphics[width=0.95\textwidth]{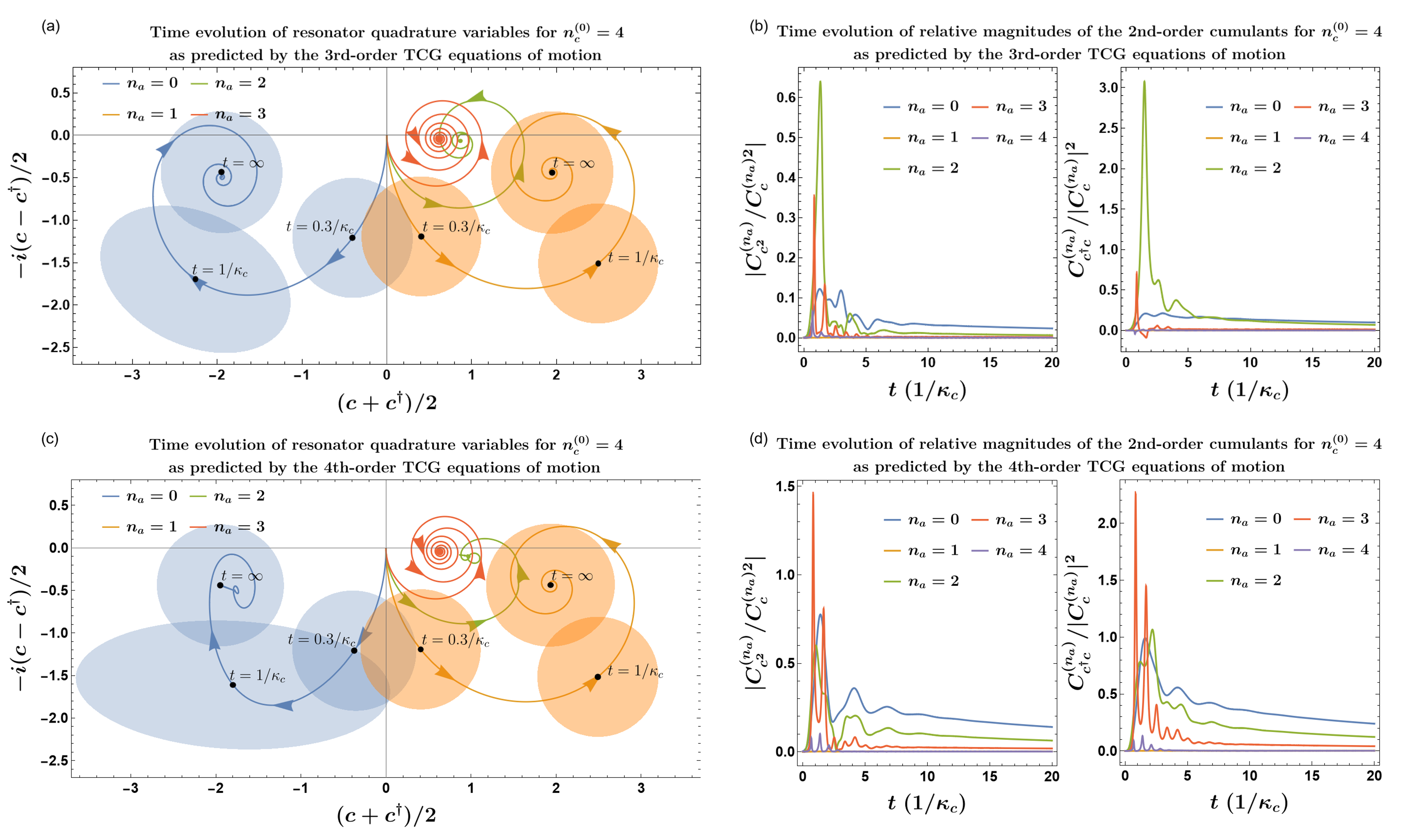}
\caption{(a) The transient time evolution of resonator quadrature variables conditioned on the energy level of the transmon, as predicted by the 3rd-order TCG equations of motion. The expectation values are plotted in solid curves whereas the second-order cumulants are represented by the shaded ellipses that correspond to the one-sigma region of equivalent gaussian Husimi functions (i.e., the gaussian Husimi functions have the same second-order cumulants as the corresponding cavity states, even though the cavity states are not necessarily gaussian). (b) The relative magnitude of the 2nd-order cumulants in the numerical solutions to the 3rd-order TCG equations of motion. (c) The transient time evolution of resonator quadrature variables conditioned on the energy level of the transmon, as predicted by the 4th-order TCG equations of motion. (d) The relative magnitude of the 2nd-order cumulants in the numerical solutions to the 4th-order TCG equations of motion.}
\label{fig:3rd order transient}
\end{figure}
\twocolumngrid
In addition, we notice that the approximate analytical formula in Eq.(\ref{Eq: K over kappa}) is accurate for small drive strengths, but deviates from the curve-fit values at high drive power. This is mainly due to the fact that the resonator state sees greater deviation from our coherent state ansatz with stronger drive. This deviation can be characterized by the higher-order cumulant variables introduced in Appendix I of the Supplementary Materials\cite{supplement} and shown in Fig.\ref{fig:3rd order transient}.

More precisely, the second-order cumulant $C_{c^{\dagger}c}^{(n)}(t)$ measures the average increase in the variance of $c$ from its Heisenberg limit along all directions in the phase space, whereas $C_{c^{2}}^{(n)}(t)$ reflects the amount and direction of squeezing. The main effect of the fourth-order TCG superoperators appears to be increasing the second-order cumulants during transient evolution of the resonator.

For example, with the resonator empty at $t=0$ and $p_{1}(0) = 0.95$, numerical solutions to the third- and fourth-order TCG equations of motion show noticeably different transient time evolution of the resonator quadrature variables, as one can see by comparing the upper panels with the lower ones in Fig.\ref{fig:3rd order transient}. More precisely, the fourth-order TCG superoperators generate additional deviations from coherent states before the resonator settles into its steady state.

Furthermore, in both the third- and the fourth-order simulations, we find that the resonator only deviates from coherent states during an intermediate transient time period where $t$ is on the same order of magnitude as $1/\kappa_{c}$ (see the left panels of Fig.\ref{fig:3rd order transient}). For both early- and late-time evolution, coherent states can be considered good approximations to the conditional resonator state for each transmon energy level. We emphasize that this fact agrees with and provides theoretical support for the experimentally successful practice of using resonator states as pointer states to indicate the transmon energy level.

As a sanity check, we also present the relative magnitudes of the second-order cumulants in the right panels of Fig.\ref{fig:3rd order transient} where $C_{c^{2}}^{(n_{a})}$ and $C_{c^{\dagger}c}^{(n_{a})}$ are weighed against the squared absolute value of the first-order cumulant $C_{c}^{(n_{a})}$. Since cumulant expansion can be considered as a semi-classical expansion around the coherent state, we may confidently truncate the cumulant expansion only if the ratios $\vert C_{c^{2}}^{(n_{a})} / C_{c}^{(n_{a})2} \vert$ and $C_{c^{\dagger}c}^{(n_{a})} / \vert C_{c}^{(n_{a})} \vert^{2}$ remain below $1$ for the vast majority of time and never increase much beyond $1$. This is indeed the case according to the right panels of Fig.\ref{fig:3rd order transient}. And we have also verified that the transmon dynamics as well as the steady-state means and variances of the resonator quadrature variables is very insensitive to variations in the initial state, which further increases our confidence in the validity of truncating the TCG master equation.

\section{Summary and outlook}
\label{Sec: Summary}

To summarize, we have developed the measurement-adapted time-coarse graining (MaTCG) as a systematic perturbation theory with both closed-form combinatorial formulas for the superoperators generated at each order and a diagrammatic representation for each term in the corresponding coefficients. In addition, we have given a prescription for modeling the time-coarse grained dynamics of open quantum systems, which starts from the standard system+bath formalism and treats the system and bath degrees of freedom on the same footing in deriving the time-coarse grained effective master equation before the bath degrees of freedom are traced out. This approach not only directly relates the measurement time resolution $\tau$ to the validity of Markov and secular approximations, but also suggests that the effective incoherent processes need to be treated on the same footing as the coherent ones in a measurement-informed system+bath formalism, which opens up the possibility for new types of bath-induced effective dynamics of the reduced system density matrix observed with finite time resolution. In particular, we have demonstrated through the examples in this paper that the interplay among the bath degrees of freedom, the external drives, and the system nonlinearity can lead to intricate forms of non-Markovian incoherent transitions which have not been captured by any effective Hamiltonian method to the best of our knowledge. To the contrary, such transitions are natural in the MaTCG framework since we acknowledge the fact that for any nonlinear system, information can flow across time scales and time coarse-graining arguments are unreliable if the framework concerns pure states only. Furthermore, when a quantum system is subject to continuous measurement with finite time resolution, we have shown that MaTCG also provides a natural prescription for deriving effective stochastic master equations where the relation between the noise statistics and its backaction on the quantum system are consistently related in the time-coarse grained picture.

As a demonstration for the method, we have applied the MaTCG perturbation theory to an experimentally relevant model for the dispersive readout of a transmon qubit, where simple analytical expressions for the drive-induced relaxation and excitation rates have been found for each transmon energy level, along with other corrections which contain all the known effective dynamics in the literature as a subset. The drive-induced transition rates have been found to be surpass the Purcell decay rates for moderate drive strengths ($\langle n_{c} \rangle \sim 1$), for which there has been at least indirect evidence from recent experimental analysis \cite{Devoret_March_Meeting, Minev_npj}. Noticeably, the most dominant contributions to the drive-induced transitions are not sensitive to the anharmonicity $\epsilon$, and we expect similar phenomena both for coupled linear quantum oscillators and for fluxonium-resonator systems. In addition, with cumulant expansion in the resonator quadrature variables, we have derived effective equations of motion for the relevant transmon and resonator variables with sub-GHz time resolution. The limited number of relevant variables and the absence of fast dynamics in those equations of motion make numerical simulation much easier than in conventional approaches, and we have been able to solve for the transient resonator dynamics and the late-time steady state of the transmon in a single simulation. Furthermore, we have also been able to solve for higher moments of the resonator quadrature variables in a time-efficient way using the TCG equations of motion.

Similar lines of thought have also motivated the recently developed perturbation method for computing static effective Hamiltonians of rapidly driven nonlinear quantum systems \cite{Devoret_Kamiltonian, Devoret_Lindbladian}. This method offers another view of the effective dynamics by canonically transforming to a frame with time-independent Hamiltonian instead of direct time-coarse graining. In fact, the Hamiltonian part of the TCG master equation exactly reproduces the static effective Hamiltonian for the examples calculated in \cite{Devoret_Kamiltonian}. However, unlike the canonically transformed density matrix $\varrho(t)$, the time-coarse grained $\overline{\rho}(t)$ does not undergo purely unitary time evolution generated by an effective Hamiltonian in general, and the TCG dissipators become important in certain situations as we have seen in the analysis of the transmon readout problem. As a general rule of thumb, the TCG dissipators account for the incoherent transfer of information/energy across different degrees of freedom in the system; while such incoherent processes are usual weak in comparison with the coherent ones, they assume great importance for the long-time dynamics of the system when external drives and/or dissipative environments induce directional flow of information/energy throughout the system in a far-from equilibrium scenario. Furthermore, the choice of frame in the MaTCG method is based on the readout channel rather than some presupposed steady state, which makes results from MaTCG more reliable regarding the transient dynamics. From a practical point of view, the effective model given by MaTCG is fully informed by the choice of the observation channel and time resolution in experiments, and the resulting EQMEs describe variables that can be directly observed in principle, without needing to perform any extra transformation of the experimental observables. Finally, we note that although the explicit analytical results in this paper are derived assuming gaussian window functions, our general formulas in Eq.(\ref{Eq: Lk}) and Eq.(\ref{Eq: contraction coefficient}) make no assumption about the particular form of the window function $f(t;\tau)$, and more complicated measurement chains may be effectively modeled by some more sophisticated $f(t;\tau)$ through which information about the quantum system is aggregated by the measurement devices.

Considered in a broader context, MaTCG bears some significant resemblance to the philosophy behind Wilsonian renormalization~\cite{Wilson1971_1, Wilson1971_2,Wilson_1983}. Indeed, as an important theoretical tool and conceptual bedrock for quantum field theories, Wilsonian renormalization allows one to obtain effective theories at low-momentum/large scale from a microscopic theory. It filters out the small-scale fluctuations from a model, while keeping track of their contributions to the dynamics at a certain large length scale that can be understood as the spatial resolution of the resulting effective theory. In the same spirit, MaTCG postulates that details of the high-frequency transitions in a quantum theory should be irrelevant when a system is prepared and observed with finite time-resolution through a certain measurement channel, and it is only their impact on the long-time dynamics of the resolvable degrees of freedom, the ``corrections from renormalization'', that should be incorporated into the effective theory. However, as pointed out in the Introduction, there are also important conceptual difference between Wilsonian renormalization and MaTCG, in particular since the MaTCG does not explicitly truncate the Hilbert space. In fact, MaTCG only validates the truncation of the Hilbert space if the initial condition effectively limits the relevant degrees of freedom to a particular section of the Hilbert space. This feature renders MaTCG more similar to the flow equation method by Wilson and Wegner~\cite{Wilson_flowEq,WEGNER2000141,Kehrein_flowEq}, which has been shown to be effective for solving non-equilibrium dynamics of quantum systems. An additional important feature of MaTCG is that the integrated high-frequency transitions are with respect to a finite frequency $\omega_0$ depending on the measurement channel; from that point of view, MaTCG implements not a low-pass filter in the frequency domain of the many-body dynamics (as in the flow-equation method) but a band-pass filter. In addition, since here we have mostly concerned ourselves with the measurable quantum dynamics of finite dimensional dimensional non-linear quantum systems embedded in the electromagnetic continuum, ``integrating out'' the fast transitions in MaTCG would generate relevant superoperators in the effective master equation that are new and qualitatively different from those in the microscopic theory, in contrast with high-dimensional quantum field theories where only a handful of interactions remain relevant in the IR limit. Inspired by the explanatory power of Wilsonian renormalization and the flow equation method, we think it is reasonable to expect that MaTCG would be useful in a wide range of studies, from parametrically driven qubits and quantum-limited amplifiers in $0+1$ dimensions, to higher-dimensional field theories with well-defined measurement channels.

Potential future work includes combining TCG with spatial coarse-graining in a congruent way so that effective field theories may be developed to efficiently describe dissipative and/or non-equilibrium quantum system in (d+1) dimensions. This may not only provide a theoretical framework complementary to the Schwinger-Keldysh formalism, but also establish a more natural and experimentally relevant connection between field theories and their lattice counterparts described in the discrete exterior calculus (DEC) formalism where variables defined on the simplicial complex of a lattice are interpreted as spatially and temporally integrated degrees of freedom that constitute the physical observables in the theory\cite{Dung2023}.

\section{Acknowledgments}
\label{Sec: Acknowledgments}
We would like to thank Kanupriya Sinha and Moein Malekakhlagh for their insightful discussions and suggestions regarding this work. We acknowledge support from the US Department of Energy, Office of Basic Energy Sciences, Division of Materials Sciences and Engineering, under Award No. DESC0016011. The simulations presented in this article were performed on computational resources managed and supported by Princeton Research Computing, a consortium of groups including the Princeton Institute for Computational Science and Engineering (PICSciE) and the Office of Information Technology's High-Performance Computing Center and Visualization Laboratory at Princeton University.

\section{Data
Availability Statement}
The Julia code used in this article is openly available on GitHub \cite{quantumgraining-github}.

\bibliography{refs}

\end{document}